# New Resonant Bivacuum Mediated
# Interaction, as a Possible Explanation of Psi Phenomena

Alex Kaivarainen

University of Turku, Department of physics
Vesilinnantie 5, FIN-20014, Turku, Finland
H2o@karelia.ru
http://www.karelia.ru/~alexk/new_articles/index.html

## CONTENTS

# SUMMARY

The coherent physical theory of Psi phenomena, like remote vision, telepathy, telekinesis, remote healing, clairvoyance - is absent till now due to its high complexity and multilateral character.

The mechanism of Bivacuum mediated Psi - phenomena is proposed in this work. It is based on many stages of my long term efforts, including creation of few new theories:

1) Unified theory of Bivacuum, rest mass and charge origination, fusion of elementary particles (electrons, protons, neutrons, photons, etc.) from certain number of sub-elementary fermions and dynamic mechanism of their corpuscle-wave [C - W] duality (http://arxiv.org/abs/physics/0207027);

2) Quantitative Hierarchic theory of liquids and solids, verified on examples of water and ice by special, theory based, computer program (http://arxiv.org/abs/physics/0102086);

3) Hierarchic model of consciousness: from mesoscopic Bose condensation (mBC) to synaptic reorganization (http://arxiv.org/abs/physics/0003045);

4) Theory of primary Virtual Replica (**VR**) of material objects in Bivacuum and **VR** Multiplication: **VRM (r,t)**. The **VR** represents a three-dimensional (3D) superposition of Bivacuum virtual standing waves $\mathbf{VPW}_m^{\pm}$ and $\mathbf{VirSW}_m^{\pm 1/2}$, modulated by $[\mathbf{C} \rightleftharpoons \mathbf{W}]$ pulsation of elementary particles and translational and librational de Broglie waves of molecules of macroscopic object (http://arxiv.org/abs/physics/0207027). The infinitive multiplication of *primary* **VR** in space in form of 3D packets of virtual standing waves: **VRM(r)**, is a result of interference of all pervading external coherent basic *reference waves* - Bivacuum Virtual Pressure Waves ($\mathbf{VPW}_{q=1}^{+}$) and Virtual Spin Waves ($\mathbf{VirSW}_{q=1}^{\pm 1/2}$) with similar waves, forming primary **VR**. This phenomena may stand for *remote vision* of psychic. The ability of enough complex system of **VRM(t)** to self-organization in nonequilibrium conditions, make it possible multiplication of VR not only in space but as well, in time in both time direction - positive (evolution) and negative (devolution). The feedback reaction between most probable/stable **VRM(t)** and nerve system of psychic, including visual centers of brain, can by responsible for *clairvoyance*;

5) Theory of *nonlocal* Virtual Guides (**VirG**$_{SME}$) of spin, momentum and energy, representing virtual microtubules with properties of quasi one-dimensional virtual Bose condensate, constructed from 'head-to-tail' polymerized Bivacuum bosons (**BVB**$^{\pm}$) or Cooper pairs of Bivacuum fermions (**BVF**$^{\updownarrow}$). The bundles of **VirG**$_{SME}$, connecting coherent atoms of Sender (S) and Receiver (S), as well as nonlocal component of **VRM(r,t)**, determined by interference pattern of Virtual Spin Waves, are responsible for nonlocal weak interaction, telekinesis, telepathy and remote healing;

6) Theory of *Bivacuum Mediated Interaction* (**BMI**) as a new fundamental interaction due to superposition of Virtual replicas of Sender and Receiver, because of **VRM(r,t)** mechanism, and connection of their coherent atoms via **VirG**$_{SME}$ bundles. Just **BMI** is responsible for remote ultraweak nonlocal interaction and different psi-phenomena. The system: [S + R] should be in nonequilibrium state.

The correctness of our approach follows from its ability to explain a lot of unconventional experimental data, like Kozyrev ones, remote genetic transmutation, remote vision, mind-matter interaction, etc. without contradictions with fundamental laws of nature. For details see: http://arxiv.org/abs/physics/0103031.

# Abbreviations and Definitions*

- $(\mathbf{V}^+)$ and $(\mathbf{V}^-)$ are correlated actual torus and complementary antitorus (pair of 'donuts') of Bivacuum of the opposite energy, charge and magnetic moment, formed by collective excitations of non mixing subquantum particles and antiparticles of opposite angular momentums;

- $(\mathbf{BVF}^\uparrow = \mathbf{V}^+ \uparrow\uparrow \mathbf{V}^-)^i$ and $(\mathbf{BVF}^\downarrow = \mathbf{V}^+ \downarrow\downarrow \mathbf{V}^-)^i$ are virtual dipoles of three opposite poles: actual (inertial) and complementary (inertialess) mass, positive and negative charge, positive and negative magnetic moments, separated by energetic gap, named Bivacuum fermions and Bivacuum antifermions. The opposite half integer spin $S = \pm\frac{1}{2}\hbar$ of $(\mathbf{BVF}^\updownarrow)^i$, notated as ($\uparrow$ and $\downarrow$), depends on direction of clockwise or anticlockwise in-phase rotation of pairs of [torus $(\mathbf{V}^+)$ + antitorus $(\mathbf{V}^-)$], forming them. The index: $i = e, \mu, \tau$ define the energy and Compton radiuses of $(\mathbf{BVF}^\updownarrow)^i$ of three electron generations;

- $(\mathbf{BVB}^\pm = \mathbf{V}^+ \Updownarrow \mathbf{V}^-)^i$ are Bivacuum bosons, representing the intermediate transition state between Bivacuum fermions of opposite spins: $\mathbf{BVF}^\uparrow \rightleftharpoons \mathbf{BVB}^\pm \rightleftharpoons \mathbf{BVF}^\downarrow$;

- $|\mathbf{m}_V^+|\mathbf{c}^2$ and $|\mathbf{m}_V^-|\mathbf{c}^2$ are the energies of torus and antitorus of Bivacuum dipoles: $\left[\mathbf{BVF}^\updownarrow\right]_{j,k}^i$ and $\left[\mathbf{BVB}^\pm\right]_{j,k}^i$;

- $(\mathbf{VC}_{j,k}^+ \sim \mathbf{V}_j^+ - \mathbf{V}_k^+)^i$ and $(\mathbf{VC}_{j,k}^- \sim \mathbf{V}_j^- - \mathbf{V}_k^-)^i$ are virtual clouds and anticlouds, composed from subquantum particles and antiparticles, correspondingly. Virtual clouds and anticlouds emission/absorption accompany the correlated transitions between different excitation energy states ($j$ and $k$) of torus $(\mathbf{V}_{j,k}^+)^i$ and antitorus $(\mathbf{V}_{j,k}^-)^i$ of Bivacuum dipoles: $\left[\mathbf{BVF}^\updownarrow\right]_{j,k}^i$ and $\left[\mathbf{BVB}^\pm\right]_{j,k}^i$;

- $\mathbf{VirP}^\pm$ is *virtual pressure,* resulted from the process of subquantum particles density oscillation, accompanied the virtual clouds $(\mathbf{VC}_{j,k}^\pm)$ emission and absorption in the process of torus and antitorus transitions between their $j$ and $k$ states;

- $\mathbf{\Delta VirP}_{j,k}^\pm = |\mathbf{VirP}^+ - \mathbf{VirP}^-|_{j,k..} \sim \|\mathbf{m}_V^+| - |\mathbf{m}_V^-|\mathbf{c}^2 \geq 0$ means the excessive virtual pressure, being the consequence of Bivacuum dipoles asymmetry. It determines the *kinetic energy* of Bivacuum, which can be positive or zero;

- $\sum \mathbf{VirP}_{j,k}^\pm = |\mathbf{VirP}^+ + \mathbf{VirP}^-|_{j,k} \sim \|\mathbf{m}_V^+| + |\mathbf{m}_V^-|\mathbf{c}^2 > 0$ is the total virtual pressure. It determines the potential energy of Bivacuum and always is positive;

- $\mathbf{VPW}_{q=1,2..}^+$ and $\mathbf{VPW}_{q=1,2..}^-$ are the *positive and negative virtual pressure waves,* related with oscillations of $\mathbf{VirP}_{j,k}^\pm$. In symmetric primordial Bivacuum the energy of these oscillations compensate each other;

- $\mathbf{F}_\updownarrow^+$ and $\mathbf{F}_\updownarrow^-$ are sub-elementary *fermions and antifermions* of the opposite charge (+/-) and energy. They emerge due to stable symmetry shift of the *mass and charge* between the *actual* $(\mathbf{V}^+)$ and *complementary* $(\mathbf{V}^-)$ torus of $\mathbf{BVF}^\updownarrow$ dipoles, providing the rest mass and charge origination: $[\mathbf{m}_V^+ - \mathbf{m}_V^-]^\phi = \pm\mathbf{m}_0$ and $[\mathbf{e}_V^+ - \mathbf{e}_V^-]^\phi = \pm\mathbf{e}_0$ to the left or right, correspondingly. Their stabilization and fusion to triplets, represented by electrons and protons, is accompanied by big energy release, determined by mass defect, occur when the velocity of rotation of Cooper pairs $[\mathbf{BVF}^\uparrow \bowtie \mathbf{BVF}^\downarrow]$ around the common axis corresponds to Golden mean: $(\mathbf{v/c})^2 = 0.618$;

- *Hidden Harmony* condition means the equality of the internal and external group and phase velocities of Bivacuum fermions and Bivacuum bosons: $\mathbf{v}_{gr}^{in} = \mathbf{v}_{gr}^{ext}$; $\mathbf{v}_{ph}^{in} = \mathbf{v}_{ph}^{ext} = \mathbf{v}$. It is proved that this condition is a natural background of Golden mean realization in physical systems: $\phi = (\mathbf{v}^2/\mathbf{c}^2)^{ext,in} = 0.6180339887$;

- $\langle[\mathbf{F}_\updownarrow^+ \bowtie \mathbf{F}_\updownarrow^-] + \mathbf{F}_\updownarrow^\pm\rangle^{e^-,p^+}$ are the coherent triplets of fused sub-elementary fermions and antifermions of $\mu$ and $\tau$ generations, representing the electron/positron or proton/antiproton. In the latter case a sub-elementary fermions and antifermions corresponds to $u$ and $d$ quarks;

- $(\mathbf{CVC}^+$ and $\mathbf{CVC}^-)$ are the *cumulative virtual clouds* of subquantum particles and antiparticles, standing for [W] phase of sub-elementary fermions and antifermions, correspondingly. The reversible quantum beats $[\mathbf{C} \rightleftharpoons \mathbf{W}]$ between asymmetric torus and antitorus of sub-elementary fermions are accompanied by [emission $\rightleftharpoons$ absorption] of $\mathbf{CVC}^\pm$. The stability of triplets of leptons and partons is determined by the resonant interaction of sub-elementary fermions and antifermions by $\mathbf{CVC}^\pm$ exchange in the process of [Corpuscle $\rightleftharpoons$ Wave] pulsations. The virtual pairs $[\mathbf{CVC}^+ \bowtie \mathbf{CVC}^-]_{e,p,n}$ display the gluons (bosons) properties, stabilizing the electrons, protons and neutrons;

- **VirBC** means *virtual Bose condensation* of Cooper - like pairs $[\mathbf{BVF}^{\uparrow} \bowtie \mathbf{BVF}^{\downarrow}]$ and/or $[\mathbf{BVB}^{\pm}]$ with external translational momentum close to zero: $\mathbf{p} \simeq \mathbf{0}$ and corresponding de Broglie wave length close to infinity: $\lambda_B = (\mathbf{h}/\mathbf{p}) \simeq \infty$, providing the nonlocal properties of huge Bivacuum domains;

- **TE** and **TF** are *Tuning Energy and Tuning Force* of Bivacuum, driving the matter to Golden Mean conditions and slowing down (cooling) the thermal dynamics of particles, driving their mass to the rest mass value. Such Bivacuum - matter interaction is responsible for realization of principle of Least action, 2nd and 3d laws of thermodynamics;

- **VirSW**$^{\pm 1/2}$ are the *Virtual spin waves*, excited as a consequence of angular momentums of cumulative virtual clouds (**CVC**$^{\pm}$) of sub-elementary particles in triplets $\langle[\mathbf{F}_{\uparrow}^{-} \bowtie \mathbf{F}_{\downarrow}^{+}] + \mathbf{F}_{\updownarrow}^{\pm}\rangle$ due to angular momentum conservation law. The **VirSW**$^{\pm 1/2}$ are highly anisotropic, depending on orientation of triplets in space and their rotational/librational dynamics, being the physical background of torsion field;

- **VirG**$_{\mathbf{SME}}^{S}$ is the nonlocal virtual spin-momentum-energy guide (quasi-1D virtual microtubule), formed primarily by standing **VirSW**$_{S}^{S=+1/2}$ $\underset{\mathbf{BVF}^{\uparrow} \bowtie \mathbf{BVF}^{\downarrow}}{\overset{\mathbf{BVB}^{\pm}}{=\!\bigcirc\!=}}$ **VirSW**$_{R}^{S=-1/2}$ of opposite spins and induced self-assembly of Bivacuum bosons $(\mathbf{BVB}^{\pm})^i$ or Cooper pairs of $[\mathbf{BVF}^{\uparrow} \bowtie \mathbf{BVF}^{\downarrow}]^i$, representing quasi one-dimensional Bose condensate. The bundles of **VirG**$_{\mathbf{SME}}^{S}$ connect the remote coherent triplets $\langle[\mathbf{F}_{\uparrow}^{-} \bowtie \mathbf{F}_{\downarrow}^{+}] + \mathbf{F}_{\updownarrow}^{\pm}\rangle^{e,p}$, representing elementary particles, like protons and electrons in free state or in composition of atoms or their coherent groups, providing remote nonlocal interaction - microscopic and macroscopic ones;

- **(mBC)** means *mesoscopic molecular Bose condensate* in the volume of condensed matter with dimensions, determined by the length of 3D standing de Broglie waves of molecules, related to their librations and translations;

- **VR** means three-dimensional (3D) *Virtual Replica* of elementary, particles, atoms, molecules and macroscopic objects, including living organisms. The primary **VR** of object represents a complex superposition of Bivacuum virtual standing waves **VPW**$_m^{\pm}$ and **VirSW**$_m^{\pm 1/2}$, modulated by $[\mathbf{C} \rightleftharpoons \mathbf{W}]$ pulsation of elementary particles and translational and librational de Broglie waves of molecules of macroscopic object. The **VR** of elementary particles coincide with notion of their *anchor site*, representing two-three Cooper pairs $3[\mathbf{BVF}^{\uparrow} \bowtie \mathbf{BVF}^{\downarrow}]_{as}^{i}$ of asymmetric Bivacuum fermions. The stochastic jumps of **CVC**$^{\pm}$ of [W] phase of particle from one anchor site to another and the ability of interference of single particle with its own anchor site explains two slit experiment;

- **VRM(r,t)** means the primary **VR** multiplication in space and time. The infinitive multiplication of primary **VR** in space in form of 3D packets of virtual standing waves is a result of interference of all pervading external coherent basic *reference waves* - Bivacuum Virtual Pressure Waves (**VPW**$_{q=1}^{\pm}$) and Virtual Spin Waves (**VirSW**$_{q=1}^{\pm 1/2}$) with similar kinds of modulated standing waves, like that, forming the primary **VR**. The latter has a properties of the *object waves* in terms of holography. Consequently, the **VRM** can be named **Holoiteration** by analogy with hologram (in Greece *'holo'* means the 'whole' or 'total'). The spatial **VRM(r)** may stand for *remote vision* of psychic. The ability of enough complex system of **VRM(t)** to self-organization in nonequilibrium conditions, make it possible multiplication of VR not only in space but as well, in time in both time direction - positive (evolution) and negative (devolution). The feedback reaction between most probable/stable **VRM(t)** and nerve system of psychic, including visual centers of brain, can be responsible for *clairvoyance;*

- **Psi − channels** are multiple correlated bundles of **VirG**$_{\mathbf{SME}}^{ext}$, connecting coherent particles of nerve cells of [S]- *psychic* and [R] - *target* in superimposed **VRM(r,t)**$_S$ $\bowtie$ **VRM(r,t)**$_R$. This combination of Bivacuum mediated interactions (BMI), providing the transmission of not only information, but as well the momentum and energy, can be responsible for *telekinesis and remote healing;*

- **BMI** is a new fundamental *Bivacuum Mediated Interaction* due to superposition of Virtual replicas of Sender [S] and Receiver [R] in nonequilibrium state, provided by **VRM(r,t)** mechanism and connection of coherent atoms of [S] and [R] via **VirG**$_{\mathbf{SME}}$ bundles. Just this **BMI** is responsible for remote ultraweak nonlocal interaction and psi-phenomena.
*************************************************************************
*The abbreviations are not in alphabetic, but in logical order to make this glossary more clear for perception of new notions, introduced in Unified theory.*

## 1. Introduction

The Dirac's equation points to equal probability of positive and negative energy (Dirac, 1947). In asymmetric Dirac's vacuum its realm of negative energy is saturated with infinitive number of electrons. However, it was assumed that these electrons, following Pauli principle, have not any gravitational or viscosity effects. Positrons and electron in his model represent the 'holes', originated as a result of the electrons jumps in realm of positive energy over the energetic gap: $\Delta = \mathbf{2m_0}c^2$. Currently it becomes clear, that the Dirac type model of vacuum is not general enough to explain all known experimental data, for example, the bosons emergency. The model of Bivacuum, presented in this paper and in other works (Kaivarainen, 1995; 2000; 2004; 2005a,b: http://arxiv.org/abs/physics/0207027) looks to be more advanced. However, it use the same starting point of equal probability of positive and negative energy, confined in each of Bivacuum cell-dipole in the absence of these dipoles symmetry shift.

Aspden (2003) introduced in his aether theory the basic unit, named Quon, as a pair of virtual muons of opposite charges, i.e. [muon + antimuon]. This idea has some common with our model of Bivacuum dipoles of torus + antitorus of opposite energy/mass, charge and magnetic moments with Compton radiuses, corresponding to electron, muon and tauon (Kaivarainen, 2004, 2005).

The superfluid model of vacuum, composed from pairs of fermions of opposite spins and charge where discussed earlier by Sinha et. al., (1976; 1976a; 1978) and also by Boldyreva and Sotina (1999).

In 1957 Bohm published a book: Causality and Chance in Modern Physics. Later he comes to conclusion, that Universe has a properties of giant, flowing hologram. Taking into account its dynamic nature, he prefer to use term: **holomovement**. In his book: Wholeness and the Implicate Order (1980) he develops an idea that our *explicated unfolded reality is a product of enfolded (implicated) or hidden order of existence. He consider the manifestation of all forms in the universe, as a result of enfolding and unfolding exchange between two orders, determined by super quantum potential.*

In book, written by D. Bohm and B. Hiley (1993): "THE UNDIVIDED UNIVERSE. An ontological interpretation of quantum theory" the electron is considered, as a particle with well- defined position and momentum which are, however, under influence of special wave (quantum potential). Elementary particle, in accordance with these authors, is a *sequence of incoming and outgoing waves*, which are very close to each other. However, particle itself does not have a wave nature. Interference pattern in double slit experiment after Bohm is a result of periodically "bunched" character of quantum potential.

After Bohm, the manifestation of corpuscle - wave duality of particle is dependent on the way, which observer interacts with a system. He claims that both of this properties are always enfolded in particle. *It is a basic difference with our model of duality, assuming that the wave and corpuscle phase are realized alternatively with high frequency during two different semiperiods of sub-elementary particles, forming particles in the process of quantum beats between sublevels of positive (actual) and negative (complementary) energy. This frequency is amplitude and phase modulated by experimentally revealed de Broglie wave of particles.* The important point of Bohmian philosophy, coinciding with our concept, is that everything in the Universe is a part of dynamic continuum. Neurophysiologist Karl Pribram does made the next step in the same direction as Bohm: *"The brain is a hologram enfolded in a holographic Universe".*

Sidharth (1998, 1999) considered *elementary particle as a relativistic vortex of Compton radius, from which he recovered its mass and quantized spin.* He pictured a particle as a fluid vortex steadily circulating with light velocity along a 2D ring or spherical 3D shell with radius

$$L = \frac{\hbar}{2mc} \qquad\qquad 1$$

Inside such vortex the notions of negative energy, superluminal velocities and nonlocality are acceptable without contradiction with conventional theory.

He treated also a charged Dirac fermions, as a Kerr-Newman black holes. Within the region of Compton vortex the superluminal velocity and negative energy are possible after Sidharth. If measurements are averaged over time $t \sim mc^2/\hbar$ and over space $L \sim \hbar/mc$, the imaginary part of particle's position disappears and we are back in usual Physics (Sidharth, 1998).

## 2. New Hierarchical Model of Bivacuum, as a Superfluid Multi-Dipole Structure

### 2.1. Properties of Bivacuum dipoles - Bivacuum fermions and Bivacuum bosons

New concept of Bivacuum is elaborated, as a dynamic superfluid matrix of the Universe with large domains of virtual Bose condensation, standing for their nonlocal properties. Bivacuum is represented by continuum of *subquantum particles and antiparticles* of the opposite energies, with properties of quantum liquids, separated by energy gap. The collective excitations of such quantum liquid, form the quantized vortical structures in Bivacuum - strongly interrelated *donuts:* toruses $\mathbf{V}^+$ and antitoruses $\mathbf{V}^-$ of the opposite energies with Compton radiuses $\mathbf{L}_0^i = \hbar/\mathbf{m}_0^i \mathbf{c}$ of three electron's generation ($\mathbf{i} = \mathbf{e}, \boldsymbol{\mu}, \boldsymbol{\tau}$). The pairs of these in-phase clockwise or anticlockwise rotating toruses and antitoruses (cell-dipoles), form Bivacuum fermions ($\mathbf{BVF}^\uparrow = \mathbf{V}^+\Uparrow \mathbf{V}^-)^i$ and antifermions ($\mathbf{BVF}^\downarrow = \mathbf{V}^+\Downarrow \mathbf{V}^-)^i$ of opposite spins. The intermediate state between Bivacuum fermions of opposite spins, named Bivacuum bosons, has two possible polarization: ($\mathbf{BVB}^+ = \mathbf{V}^+\uparrow\downarrow \mathbf{V}^-)^i$ and ($\mathbf{BVB}^- = \mathbf{V}^+\downarrow\uparrow \mathbf{V}^-)^i$. Two Bivacuum fermions of opposite spins may form Cooper pair: $[\mathbf{BVF}^\uparrow \bowtie \mathbf{BVF}^\downarrow]$. The correlated *actual torus* ($\mathbf{V}^+$) and *complementary antitorus* ($\mathbf{V}^-$) have the opposite quantized energy, mass, charges and magnetic moments, which compensate each other in symmetric primordial Bivacuum (in the absence of matter and fields) and do not compensate in secondary Bivacuum.

### 2.2 The basic (carrying) Virtual Pressure Waves (VPW$_q^\pm$) and Virtual spin waves (VirSW$_q^{\pm 1/2}$) of Bivacuum

The emission and absorption of Virtual clouds $(\mathbf{VC}_{j,k}^+)^i$ and anticlouds $(\mathbf{VC}_{j,k}^-)^i$ in primordial Bivacuum, i.e. in the absence of matter and fields or where their influence on symmetry of Bivacuum is negligible, are the result of correlated transitions between different excitation states $(j,k)$ of torus $(\mathbf{V}_{j,k}^+)^i$ and antitoruses $(\mathbf{V}_{j,k}^-)^i$, forming symmetric $[\mathbf{BVF}^\updownarrow]^i$ and $[\mathbf{BVB}^\pm]^i$, corresponding to three lepton generations ($i = e, \mu, \tau$) :

$$(\mathbf{VC}_q^+)^i \equiv [\mathbf{V}_j^+ - \mathbf{V}_k^+]^i \; - \; virtual \; cloud \qquad\qquad 2.1$$

$$(\mathbf{VC}_q^-)^i \equiv [\mathbf{V}_{\bar{j}}^- - \mathbf{V}_{\bar{k}}^-]^i \; - \; virtual \; anticloud \qquad\qquad 2.1a$$

where: $j > k$ are the integer quantum numbers of torus and antitorus excitation states; $q = j - k$.

The virtual clouds: $(\mathbf{VC}_q^+)^i$ and $(\mathbf{VC}_q^-)^i$ exist in form of collective excitation of *subquantum* particles and antiparticles of opposite energies, correspondingly. They can be considered as 'drops' of virtual Bose condensation of subquantum particles of positive and negative energy and similar in case of $[\mathbf{BVF}^\updownarrow]^i$ and opposite in case of $[\mathbf{BVB}^\pm]^i$ angular momentums.

The process of [*emission* $\rightleftharpoons$ *absorption*] of virtual clouds should be accompanied by oscillation of *virtual pressure (*$\mathbf{VirP}^{\pm}$*) and excitation of positive and negative virtual pressure waves:* $\mathbf{VPW}_q^+$ *and* $\mathbf{VPW}_q^-$. In primordial Bivacuum the energies of opposite virtual pressure waves totally compensate each other: $\mathbf{VPW}_q^+ + \mathbf{VPW}_q^- = 0$. However, in asymmetric secondary Bivacuum, in presence of matter and fields, the total compensation is absent and the resulting virtual pressure is nonzero (Kaivarainen, 2005): $(\Delta \mathbf{VirP}^{\pm} = |\mathbf{VirP}^+| - |\mathbf{VirP}^-| = \mathbf{T}_k) > 0$. This difference in absolute values of virtual pressure determines the kinetic energy of bivacuum ($\mathbf{T}_k$), in contrast to their sum, responsible for Bivacuum potential energy: $|\mathbf{VPW}_q^+| + |\mathbf{VPW}_q^-| \simeq 2|\mathbf{VPW}_q^{\pm}| = \mathbf{V}_p$.

In accordance with our model of Bivacuum, virtual particles and antiparticles represent the asymmetric Bivacuum dipoles $(\mathbf{BVF}^{\updownarrow})^{as}$ and $(\mathbf{BVB}^{\pm})^{as}$ of three electron generations ($i = e, \mu, \tau$) in unstable state, not corresponding to Golden mean conditions. Virtual particles and antiparticles are the result of correlated and opposite Bivacuum dipole symmetry fluctuations. Virtual particles, like the real sub-elementary particles, may exist in Corpuscular and Wave phases. The Corpuscular [C]- phase, represents the correlated pairs of asymmetric torus ($\mathbf{V}^+$) and antitorus ($\mathbf{V}^-$) of two different by absolute values energies. The Wave [W]- phase, results from quantum beats between these states, which are accompanied by emission or absorption of Cumulative Virtual Clouds (CVC$^+$ or CVC$^-$), formed by subquantum particles.

Virtual particles have a mass, charge, spin, etc., but they differs from real sub-elementary ones by their lower stability (short life-time) and inability for fusion to stable triplets. They are a singlets or very unstable triplets or other clusters of Bivacuum dipoles $(\mathbf{BVF}^{\updownarrow})^{as}$ in contrast to real stable fermions-triplets or bosons, containing the integer number of sub-elementary fermion and antifermion pairs.

For Virtual Clouds ($\mathbf{VC}^{\pm}$) and virtual pressure waves ($\mathbf{VPW}^{\pm}$) excited by them, the relativistic mechanics is not valid. *Consequently, the causality principle also does not work in a system (interference pattern) of* $\mathbf{VPW}^{\pm}$.

The quantized energies of positive and negative $\mathbf{VPW}_q^+$ and $\mathbf{VPW}_q^-$ and corresponding virtual clouds and anticlouds, emitted $\rightleftharpoons$ absorbed by $(\mathbf{BVF})^i$ and $(\mathbf{BVB})^i$, as a result of their transitions between $\mathbf{n} = \mathbf{j}$ and $\mathbf{n} = \mathbf{k}$ states can be presented as:

$$\mathbf{E}_{\mathbf{VPW}_q^+}^i = \hbar \omega_0^i (\mathbf{j} - \mathbf{k}) = \mathbf{m}_0^i \mathbf{c}^2 (\mathbf{j} - \mathbf{k}) \qquad 2.2$$

$$\mathbf{E}_{\mathbf{VPW}_q^-}^i = -\hbar \omega_0^i (\mathbf{j} - \mathbf{k}) = -\mathbf{m}_0^i \mathbf{c}^2 (\mathbf{j} - \mathbf{k}) \qquad 2.2a$$

The quantized fundamental Compton frequency of $\mathbf{VPW}_q^{\pm}$:

$$\mathbf{q}\,\omega_0^i = \mathbf{q}\,\mathbf{m}_0^i \mathbf{c}^2 / \hbar \qquad 2.3$$

where: $\mathbf{q} = (\mathbf{j} - \mathbf{k}) = \mathbf{1}, \mathbf{2}, \mathbf{3}..$ is the quantization number of $\mathbf{VPW}_{j,k}^{\pm}$ energy;

In symmetric primordial Bivacuum the total compensation of positive and negative Virtual Pressure Waves takes a place:

$$\mathbf{q}\mathbf{E}_{\mathbf{VPW}_{j,k}^+}^i = -\mathbf{q}\mathbf{E}_{\mathbf{VPW}_{j,k}^-}^i = \mathbf{q}\,\hbar\omega_0^i \qquad 2.4$$

The density oscillation of $\mathbf{VC}_{j,k}^+$ and $\mathbf{VC}_{j,k}^-$ and virtual particles and antiparticles represent *positive and negative virtual pressure waves* ($\mathbf{VPW}_{j,k}^+$ *and* $\mathbf{VPW}_{j,k}^-$). The symmetric excitation of positive and negative energies/masses of torus and antitorus means increasing of primordial Bivacuum potential energy, corresponding to increasing of energy gap between them:

$$[\mathbf{A}_{BVF} = \mathbf{E}_{\mathbf{V}^+} - (-\mathbf{E}_{\mathbf{V}^-}) = \hbar\boldsymbol{\omega}_0(1+2\mathbf{n})]^i = \mathbf{m}_0^i \mathbf{c}^2(1+2\mathbf{n}) = \frac{h\mathbf{c}}{[\mathbf{d}_{\mathbf{V}^+ \Updownarrow \mathbf{V}^-}]_n^i} \qquad 2.5$$

where the characteristic distance between torus $(\mathbf{V}^+)^i$ and antitorus $(\mathbf{V}^-)^i$ of Bivacuum dipoles *(gap dimension)* is a quantized parameter:

$$[\mathbf{d}_{\mathbf{V}^+ \Updownarrow \mathbf{V}^-}]_n^i = \frac{h}{\mathbf{m}_0^i \mathbf{c}(1+2\mathbf{n})} \qquad 2.5a$$

The symmetric transitions/beats between the excited and basic states of torus and antitorus are accompanied by virtual pressure waves excitation of corresponding frequency (2.2 and 2.2a).

The correlated *virtual Cooper pairs* of adjacent Bivacuum fermions $(\mathbf{BVF}_{S=\pm1/2}^{\updownarrow})$, rotating in opposite direction with resulting spin, equal to zero and Boson properties, can be presented as:

$$[\mathbf{BVF}_{S=1/2}^{\uparrow} \bowtie \mathbf{BVF}_{S=-1/2}^{\downarrow}]_{S=0} \equiv [(\mathbf{V}^+ \Uparrow \mathbf{V}^-) \bowtie (\mathbf{V}^+ \Downarrow \mathbf{V}^-)]_{S=0} \qquad 2.6$$

Such a pairs, as well as Bivacuum bosons $(\mathbf{BVB}^{\pm})$ in conditions of ideal equilibrium, like the *Goldstone bosons,* have zero mass and spin: $S = 0$. The virtual clouds $(\mathbf{VC}_q^{\pm})$, emitted and absorbed in a course of correlated transitions of $[\mathbf{BVF}^{\uparrow} \bowtie \mathbf{BVF}^{\downarrow}]_{S=0}^{j,k}$ between (j) and (k) sublevels: $q = j - k$, excite the virtual pressure waves $\mathbf{VPW}_q^+$ and $\mathbf{VPW}_q^-$, carrying the opposite angular momentums. They compensate the energy and momentums of each other totally in primordial Bivacuum and partly in *secondary Bivacuum* - in presence of matter and fields.

*The nonlocal virtual spin waves* $(\mathbf{VirSW}_{j,k}^{\pm1/2})$, with properties of massless collective Nambu-Goldstone modes, like a real spin waves, represent the oscillation of angular momentum equilibrium of Bivacuum fermions with opposite spins via "flip-flop" mechanism, accompanied by origination of intermediate states - Bivacuum bosons $(\mathbf{BVB}^{\pm})$:

$$\mathbf{VirSW}_{j,k}^{\pm1/2} \sim \left[\mathbf{BVF}^{\uparrow}(\mathbf{V}^+ \Uparrow \mathbf{V}^-) \rightleftharpoons \mathbf{BVB}^{\pm}(\mathbf{V}^+ \Updownarrow \mathbf{V}^-) \rightleftharpoons \mathbf{BVF}^{\downarrow}(\mathbf{V}^+ \Downarrow \mathbf{V}^-)\right] \qquad 2.7$$

The $\mathbf{VirSW}_{j,k}^{+1/2}$ and $\mathbf{VirSW}_{j,k}^{-1/2}$ are excited by $(\mathbf{VC}_q^{\pm})_{S=1/2}^{\circlearrowleft}$ and $(\mathbf{VC}_q^{\pm})_{S=-1/2}^{\circlearrowleft}$ of opposite angular momentums, $S_{\pm1/2} = \pm\frac{1}{2}\hbar = \pm\frac{1}{2}\mathbf{L}_0\mathbf{m}_0\mathbf{c}$ and frequency, equal to $\mathbf{VPW}_q^{\pm}$:

$$\mathbf{q}\boldsymbol{\omega}_{\mathbf{VirSW}^{\pm1/2}}^i = \mathbf{q}\boldsymbol{\omega}_{\mathbf{VPW}^{\pm}}^i = \mathbf{q}\mathbf{m}_0^i\mathbf{c}^2/\hbar = \mathbf{q}\boldsymbol{\omega}_0^i \qquad 2.8$$

The most probable basic virtual pressure waves $\mathbf{VPW}_0^{\pm}$ and virtual spin waves $\mathbf{VirSW}_0^{\pm1/2}$ correspond to minimum quantum number $\mathbf{q} = (\mathbf{j} - \mathbf{k}) = \mathbf{1}$. The $\mathbf{VirSW}_{j,k}^{\pm1/2}$, like so-called torsion field, can serve as a carrier of the phase/spin (angular momentum) and information - *qubits*, but not the energy.

The Bose-Einstein statistics of energy distribution, valid for system of weakly interacting bosons (ideal gas), do not work for Bivacuum due to strong coupling of pairs $[\mathbf{BVF}^{\uparrow} \bowtie \mathbf{BVF}^{\downarrow}]_{S=0}$ and $(\mathbf{BVB}^{\pm})$, forming virtual Bose condensate $(\mathbf{VirBC})$ with nonlocal properties.

### 2.3 *Virtual Bose condensation (VirBC), as a base of Bivacuum nonlocality*

**Nonlocality**, as the independence of potential on the distance from its source in the volume or filaments of virtual or real Bose condensate, follows from application of the Virial theorem to systems of Cooper pairs of Bivacuum fermions $[\mathbf{BVF}^{\uparrow} \bowtie \mathbf{BVF}^{\downarrow}]_{S=0}$ and Bivacuum bosons ($\mathbf{BVB}^\pm$) (Kaivarainen: http://arxiv.org/abs/physics/0103031).

It follows from our model of Bivacuum, that the infinite number of Cooper pairs of Bivacuum fermions $[\mathbf{BVF}^\uparrow \bowtie \mathbf{BVF}^\downarrow]_{S=0}^i$ and their intermediate states - Bivacuum bosons ($\mathbf{BVB}^\pm)^i$, as elements of Bivacuum, have zero or very small (in presence of fields and matter) translational momentum: $\mathbf{p}_{\mathbf{BVF}^\uparrow \bowtie \mathbf{BVF}^\downarrow}^i = \mathbf{p}_{\mathbf{BVB}}^i \to 0$ and corresponding de Broglie wave length tending to infinity: $\boldsymbol{\lambda}_{\mathbf{VirBC}}^i = \mathbf{h}/\mathbf{p}_{\mathbf{BVF}^\uparrow \bowtie \mathbf{BVF}^\downarrow, \mathbf{BVB}}^i \to \infty$. It leads to origination of 3D net of virtual adjacent pairs of virtual microtubules from Cooper pairs $[\mathbf{BVF}^\uparrow \bowtie \mathbf{BVF}^\downarrow]_{S=0}$, rotating in opposite direction and $(\mathbf{BVB}^\pm)_{S=0}$, which may form single microtubules, with resulting angular momentum, equal to zero. These twin and single microtubules, termed Virtual Guides ($\mathbf{VirG}^{\mathbf{BVF}^\uparrow \bowtie \mathbf{BVF}^\downarrow}$ and $\mathbf{VirG}^{\mathbf{BVB}^\pm}$), represent a quasi one-dimensional Bose condensate with nonlocal properties close to that of 'wormholes'. Their radiuses are determined by the Compton radiuses of the electrons, muons and tauons and state of corresponding Bivacuum dipoles excitation. Their length is limited only by decoherence effects. In symmetric Bivacuum, unperturbed by matter and fields, the length of $\mathbf{VirG}$ bundles with nonlocal properties may have the order of stars and galactics separation.

### 3. Three conservation rules for Bivacuum fermions $(\mathbf{BVF}^\ddagger)_{as}$ and Bivacuum bosons $(\mathbf{BVB}^\pm)_{as}$

There are three basic postulates/rules in our theory, interrelated with each other:

**I.** The absolute values of internal rotational kinetic energies of torus and antitorus are equal to each other and to the half of the rest mass energy of the electrons of corresponding lepton generation, independently on the external group velocity ($\mathbf{v}$), turning the symmetric Bivacuum fermions ($\mathbf{BVF}^\ddagger$) to asymmetric ones:

$$[\mathbf{I}] : \quad \left( \tfrac{1}{2}\mathbf{m}_V^+(\mathbf{v}_{gr}^{in})^2 = \tfrac{1}{2}|-\mathbf{m}_V^-|(\mathbf{v}_{ph}^{in})^2 = \tfrac{1}{2}\mathbf{m}_0\mathbf{c}^2 = \mathbf{const} \right)_{in}^i \qquad 3.1$$

where: $\mathbf{m}_V^+$ and $\mathbf{m}_V^-$ are the 'actual' - inertial and 'complementary' - inertialess masses of torus ($\mathbf{V}^+$) and antitorus ($\mathbf{V}^-$); the $\mathbf{v}_{gr}^{in}$ and $\mathbf{v}_{ph}^{in}$ are the *internal* angular group and phase velocities of subquantum particles and antiparticles, forming torus and antitorus, correspondingly. In symmetric conditions of *primordial* Bivacuum and its virtual dipoles, when the influence of matter and fields is absent: $\mathbf{v}_{gr}^{in} = \mathbf{v}_{ph}^{in} = \mathbf{c}$ and $\mathbf{m}_V^+ = \mathbf{m}_V^- = \mathbf{m}_0$.

It is proved, that the above condition means the infinitive life-time of torus and antitorus of $\mathbf{BVF}^\ddagger$ and $\mathbf{BVB}^\pm$ (Kaivarainen, 2005).

**II.** The internal magnetic moments of torus ($\mathbf{V}^+$) and antitorus ($\mathbf{V}^-$) of asymmetric Bivacuum fermions $\mathbf{BVF}_{as}^\uparrow = [\mathbf{V}^+ \uparrow\uparrow \mathbf{V}^-]$ and antifermions: $\mathbf{BVF}_{as}^\downarrow = [\mathbf{V}^+ \downarrow\downarrow \mathbf{V}^-]$, when $\mathbf{v}_{gr}^{in} \neq \mathbf{v}_{ph}^{in}$, $\mathbf{m}_V^+ \neq |-\mathbf{m}_V^-|$ and $|\mathbf{e}_+| \neq |\mathbf{e}_-|$, are equal to each other and to that of Bohr magneton: $[\boldsymbol{\mu}_B = \boldsymbol{\mu}_0 = \tfrac{1}{2}|\mathbf{e}_0|\frac{h}{\mathbf{m}_0\mathbf{c}}]$, independently on their external translational velocity ($\mathbf{v} > 0$) and symmetry shift. In contrast to permanent magnetic moments of $\mathbf{V}^+$ and $\mathbf{V}^-$, their actual and complementary masses $\mathbf{m}_V^+$ and $|-\mathbf{m}_V^-|$, internal angular velocities ($\mathbf{v}_{gr}^{in}$ and $\mathbf{v}_{ph}^{in}$) and electric charges $|\mathbf{e}_+|$ and $|\mathbf{e}_-|$, are dependent on ($\mathbf{v}$), however, they compensate each other variations:

$$[\mathbf{II}] : \quad \left( \begin{array}{c} |\pm\boldsymbol{\mu}_+| = \tfrac{1}{2}|\mathbf{e}_+|\frac{|\pm h|}{|\mathbf{m}_V^+(\mathbf{v}_{gr}^{in})_{rot}|} = |\pm\boldsymbol{\mu}_-| = \tfrac{1}{2}|-\mathbf{e}_-|\frac{|\pm h|}{|-\mathbf{m}_V^-|(\mathbf{v}_{ph}^{in})_{rot}} = \\ = \boldsymbol{\mu}_0 = \tfrac{1}{2}|\mathbf{e}_0|\frac{h}{\mathbf{m}_0\mathbf{c}} = \mathbf{const} \end{array} \right)^i \qquad 3.2$$

This postulate reflects the condition of the invariance of the spin value, with respect to the external velocity of Bivacuum fermions.

**III.** The equality of Coulomb interaction between torus and antitorus $\mathbf{V}^+ \Updownarrow \mathbf{V}^-$ of primordial Bivacuum dipoles of all three generations $i = e, \mu, \tau$ (electrons, muons and tauons), providing uniform electric energy density distribution in Bivacuum:

$$[\text{III}] : \quad \mathbf{F}_0^i = \left( \frac{\mathbf{e}_0^2}{[\mathbf{d}_{\mathbf{V}^+ \Diamond \mathbf{V}^-}^2]_n} \right)^e = \left( \frac{\mathbf{e}_0^2}{[\mathbf{d}_{\mathbf{V}^+ \Diamond \mathbf{V}^-}^2]_n} \right)^\mu = \left( \frac{\mathbf{e}_0^2}{[\mathbf{d}_{\mathbf{V}^+ \Diamond \mathbf{V}^-}^2]_n} \right)^\tau \qquad 3.2a$$

where: $[\mathbf{d}_{\mathbf{V}^+ \Diamond \mathbf{V}^-}]_n = \frac{h}{\mathbf{m}_0^i \mathbf{c}(1+2\mathbf{n})}$ is the separation between torus and antitorus of Bivacuum three pole dipoles (1.4) at the same state of excitation ($n$). A similar condition is valid as well for opposite magnetic poles interaction; $|\mathbf{e}_+| \, |\mathbf{e}_-| = \mathbf{e}_0^2$.

The important consequences of postulate **III** are the following equalities:

$$(\mathbf{e}_0 \mathbf{m}_0)^e = (\mathbf{e}_0 \mathbf{m}_0)^\mu = (\mathbf{e}_0 \mathbf{m}_0)^\tau = \sqrt{|\mathbf{e}_+ \mathbf{e}_-| \, |\mathbf{m}_V^+ \mathbf{m}_V^-|} = const \qquad 3.2b$$

It means that the toruses and antitoruses of symmetric Bivacuum dipoles of generations with bigger mass: $\mathbf{m}_0^\mu = 206,7 \, \mathbf{m}_0^e; \; \mathbf{m}_0^\tau = 3487,28 \, \mathbf{m}_0^e$ have correspondingly smaller charges:

$$\mathbf{e}_0^\mu = \mathbf{e}_0^e(\mathbf{m}_0^e/\mathbf{m}_0^\mu); \quad \mathbf{e}_0^\tau = \mathbf{e}_0^e(\mathbf{m}_0^e/\mathbf{m}_0^\tau) \qquad 3.2c$$

As is shown in the next section, just these conditions provide *the same charge symmetry shift* of Bivacuum fermions of three generations ($i = e, \mu$) at the different mass symmetry shift between corresponding torus and antitorus, determined by Golden mean.

It follows from second postulate, that the resulting magnetic moment of sub-elementary fermion or antifermion ($\mathbf{\mu}^\pm$), equal to the Bohr's magneton, is interrelated with the actual spin of Bivacuum fermion or antifermion as:

$$\mathbf{\mu}^\pm = (|\pm \mathbf{\mu}_+| \, |\pm \mathbf{\mu}_-|)^{1/2} = \mathbf{\mu}_B = \pm \frac{1}{2} \hbar \frac{\mathbf{e}_0}{\mathbf{m}_0 \mathbf{c}} = \mathbf{S} \frac{\mathbf{e}_0}{\mathbf{m}_0 \mathbf{c}} \qquad 3.3$$

where: $\mathbf{e}_0/\mathbf{m}_0 \mathbf{c}$ is gyromagnetic ratio of Bivacuum fermion, equal to that of the electron.

One may see from (3.3), that the spin of the actual torus, equal to that of the resulting spin of Bivacuum fermion (symmetric or asymmetric), is:

$$\mathbf{S} = \pm \frac{1}{2} \hbar \qquad 3.4$$

Consequently, the permanent absolute value of spin of torus and antitorus is a consequence of 2nd postulate.

The dependence of the *actual inertial* mass ($\mathbf{m}_V^+$) of torus $\mathbf{V}^+$ of asymmetric Bivacuum fermions ($\mathbf{BVF}_{as}^\uparrow = \mathbf{V}^+ \uparrow\uparrow \mathbf{V}^-$) on the external translational group velocity ($\mathbf{v}$) follows relativistic mechanics:

$$\mathbf{m}_V^+ = \mathbf{m}_0/\sqrt{1 - (\mathbf{v}/\mathbf{c})^2} = \mathbf{m} \qquad 3.5$$

while the *complementary inertialess* mass ($\mathbf{m}_V^-$) of antitorus $\mathbf{V}^-$ has the reverse velocity dependence:

$$-\mathbf{m}_V^- = i^2 \mathbf{m}_V^- = -\mathbf{m}_0 \sqrt{1 - (\mathbf{v}/\mathbf{c})^2} \qquad 3.6$$

where $i^2 = -1; \; i = \sqrt{-1}$ and complementary mass in terms of mathematics is the imaginary parameter.

For Bivacuum antifermions $\mathbf{BVF}_{as}^\downarrow = \mathbf{V}^+ \downarrow\downarrow \mathbf{V}^-$ the relativistic dependences of positive and negative mass are opposite to those described by (3.5) and (3.6) for Bivacuum fermions and the notions of actual and complementary parameters change place. Corresponding symmetry shifts between torus and antitorus in secondary Bivacuum take a place in presence of matter and fields. The product of actual (inertial) and complementary (inertialess) mass is a constant, equal to the rest mass of particle of corresponding generation squared:

$$(|\mathbf{m}_V^+ \mathbf{m}_V^-| = \mathbf{m}_0^2)^i \qquad\qquad 3.7$$

The difference of total energies of torus and antitorus is equal to doubled kinetic energy of particle:

$$(\mathbf{m}_V^+ - \mathbf{m}_V^-)\mathbf{c}^2 = \mathbf{m}_V^+\mathbf{v}^2 = 2\mathbf{T}_k = \frac{\mathbf{m}_0\mathbf{v}^2}{\sqrt{1 - (\mathbf{v}/\mathbf{c})^2}} \qquad\qquad 3.8$$

and their ratio:

$$\frac{|-\mathbf{m}_V^-|}{\mathbf{m}_V^+} = \frac{\mathbf{m}_0^2}{(\mathbf{m}_V^+)^2} = 1 - \left(\frac{\mathbf{v}}{\mathbf{c}}\right)^2 \qquad\qquad 3.9$$

The asymmetry between torus and antitorus is zero in the absence of the external motion of Bivacuum dipoles ($\mathbf{v} = \mathbf{0}$).

The product of the *internal* group and phase velocities of positive and negative subquantum particles, forming torus and antitorus, correspondingly, is equal to product of the external velocities:

$$\mathbf{v}_{gr}^{in} \, \mathbf{v}_{ph}^{in} = \mathbf{v}_{gr}^{ext} \, \mathbf{v}_{ph}^{ext} = \mathbf{c}^2 \qquad\qquad 3.8$$

A similar symmetry rule reflects the *charge compensation principle, following from (3.2)*:

$$|\mathbf{e}_+| \, |\mathbf{e}_-| = \mathbf{e}_0^2 \qquad\qquad 3.9$$

For Bivacuum antifermions $\mathbf{BVF}_{as}^\downarrow = \mathbf{V}^+ \downarrow\downarrow \mathbf{V}^-$ the relativistic dependences of positive and negative charge, like the positive and negative masses of torus and antitorus are opposite to that of Bivacuum fermions $\mathbf{BVF}_{as}^\downarrow = \mathbf{V}^+ \uparrow\uparrow \mathbf{V}^-$. The symmetry of Bivacuum bosons ($\mathbf{BVB}^{\pm} = \mathbf{V}^+\uparrow\downarrow \mathbf{V}^-)^i$ of each electron's generation ($i = e, \mu, \tau$) can be ideal and independent on external velocity, due to opposite relativistic effects of their torus and antitorus, compensating each other.

The opposite shift of symmetry between $\mathbf{V}^+$ and $\mathbf{V}^-$ of two Bivacuum fermions of opposite spins occur due to relativistic effects, accompanied their rotation *side-by-side* as a Cooper pairs $[\mathbf{BVF}^\uparrow \bowtie \mathbf{BVF}^\downarrow]_{as}$ around common axe. In this case the quantum beats between $\mathbf{V}^+$ and $\mathbf{V}^-$ of $\mathbf{BVF}^\uparrow \bowtie \mathbf{BVF}^\downarrow$ can occur in the same phase.

### 3.1 The rest mass and charge origination of sub-elementary fermions

The important interrelations between the internal and external velocities of torus and antitorus and their masses, charges and radiuses can be obtained from our three conservation rules (3.1-3.2):

$$\left(\frac{\mathbf{m}_V^+}{\mathbf{m}_V^-}\right)^{1/2} = \frac{\mathbf{m}_+^+\mathbf{c}^2}{\mathbf{m}_0\mathbf{c}^2} = \frac{\mathbf{v}_{ph}^{in}}{\mathbf{v}_{gr}^{in}} = \left(\frac{\mathbf{c}}{\mathbf{v}_{gr}^{in}}\right)^2 = \qquad 3.10$$

$$= \frac{\mathbf{L}_V^-}{\mathbf{L}_V^+} = \frac{\mathbf{L}_0^2}{(\mathbf{L}_V^+)^2} = \frac{|\mathbf{e}_+|}{|\mathbf{e}_-|} = \left(\frac{\mathbf{e}_+}{\mathbf{e}_0}\right)^2 = \frac{1}{[1 - (\mathbf{v}^2/\mathbf{c}^2)^{ext}]^{1/2}} \qquad 3.10a$$

where:

$$\mathbf{L}_V^+ = \hbar/(\mathbf{m}_V^+\mathbf{v}_{gr}^{in}) = \mathbf{L}_0[1 - (\mathbf{v}^2/\mathbf{c}^2)^{ext}]^{1/4} \qquad 3.11$$

$$\mathbf{L}_V^- = \hbar/(\mathbf{m}_V^-\mathbf{v}_{ph}^{in}) = \frac{\mathbf{L}_0^2}{\mathbf{L}_V^+} = \frac{\mathbf{L}_0}{[1 - (\mathbf{v}^2/\mathbf{c}^2)^{ext}]^{1/4}}$$

$$\mathbf{L}_0 = (\mathbf{L}_V^+\mathbf{L}_V^-)^{1/2} = \hbar/\mathbf{m}_0\mathbf{c} \ - \ Compton \ radius \qquad 3.11a$$

are the radii of torus ($\mathbf{V}^+$), antitorus ($\mathbf{V}^-$) and the resulting radius of $\mathbf{BVF}_{as}^\updownarrow = [\mathbf{V}^+ \Updownarrow \mathbf{V}^-]$, equal to Compton radius, correspondingly.

When the external velocity ($\mathbf{v}$) of the external rotation of pair $[\mathbf{BVF}^\uparrow \bowtie \mathbf{BVF}^\downarrow]_{as}$ reach the Golden mean (GM) condition ($\mathbf{v}^2/\mathbf{c}^2 = \phi = 0,618$), this results in origination of *the rest mass*: $\mathbf{m}_0 = |\mathbf{m}_V^+ - \mathbf{m}_V^-|^\phi$ and *elementary charge*: $\mathbf{e}^\phi = |\mathbf{e}_+ - \mathbf{e}_-|$ of opposite sign for sub-elementary fermion: $\left(\mathbf{BVF}_{as}^\uparrow\right)^\phi = \mathbf{F}_\uparrow^+$ and sub-elementary antifermion $\left(\mathbf{BVF}_{as}^\downarrow\right)^\phi = \mathbf{F}_\downarrow^-$ with spatial image of pair of truncated cone of opposite symmetry. The resulting mass/energy, charge and spin of Cooper pairs $[\mathbf{F}_\uparrow^+ \bowtie \mathbf{F}_\downarrow^-]$ is zero because of compensation effects.

On the other hand, two adjacent asymmetric Bivacuum fermions and antifermions of similar direction of rotation and spin can not rotate 'side-by-side' in opposite direction $[\mathbf{BVF}^\uparrow \bowtie \mathbf{BVF}^\downarrow]_{as}$, but only 'head-to-tail' in the same: clockwise or anticlockwise directions: $\mathbf{N}^+[\mathbf{BVF}^\uparrow + \mathbf{BVF}^\downarrow]_{as}$ or $\mathbf{N}^-[\mathbf{BVF}^\downarrow + \mathbf{BVF}^\downarrow]_{as}$. The Pauli allows spatial compatibility of two $\mathbf{BVF}^\uparrow$ and $\mathbf{BVF}^\uparrow$ only in the case if quantum beats between $\mathbf{V}^+$ and $\mathbf{V}^-$ occur in the counterphase manner. In such configuration, corresponding to integer spin, the uncompensated energy/ mass, charge and half-integer spin, provided by their $\mathbf{V}^+$ and $\mathbf{V}^-$ symmetry shift, are the additive values.

The positive and negative energy, charge and integer spin of such pairs determines their polarization. As far in primordial Bivacuum the average mass/energy, charge and spin should be zero, it means that the number of 'head-to-tail' pairs of Bivacuum fermions is equal to similar pairs of Bivacuum antifermions: $\mathbf{N}^+ = \mathbf{N}^-$.

**The absence of magnetic monopole** - spatially localized magnetic charge, is one of the important consequences of our model of elementary particles, as far the magnetic moments symmetry shift of $\mathbf{BVF}_{as}^\updownarrow$ is independent on velocity $(\mathbf{v})^{in,ext}$ and always zero, in contrast to mass and charge symmetry shift:

$$\Delta\boldsymbol{\mu}^\pm = \boldsymbol{\mu}_V^+ - \boldsymbol{\mu}_V^- = 0 \qquad 3.12$$

## 4. The fusion of sub-elementary fermions to elementary fermions, like electrons and protons at Golden Mean conditions

*The fusion* of sub-elementary fermions and antifermions to stable triplets $< [\mathbf{F}_\uparrow^+ \bowtie \mathbf{F}_\downarrow^-]_{x,y} + \mathbf{F}_\updownarrow^\pm >_z^i$ also becomes possible at the Golden mean velocity $\mathbf{v}_{rot} = \mathbf{c}\phi^{1/2}$ after the gyration radius of pair $[\mathbf{F}_\uparrow^+ \bowtie \mathbf{F}_\downarrow^-]_{x,y}$ around common axe at GM conditions declines to Compton radius:

$$\left[ \mathbf{L}_0 = \frac{\hbar}{(\mathbf{m}_V^+)^{\phi}(\mathbf{v}_{rot}^2)^{\phi}} = \frac{\hbar}{\mathbf{m}_0 \mathbf{c}} = \frac{\mathbf{c}}{\boldsymbol{\omega}_{rot}} \right]^i \qquad 4.1$$

where: $(\mathbf{m}_V^+)^{\phi} = \mathbf{m}_0/\phi$ is the actual mass of torus at GM conditions; $(\mathbf{v}_{rot}^2)^{\phi} = \mathbf{c}^2 \phi$ is the external group velocity at GM conditions. At this conditions, the triplets, representing electrons, positrons, protons and anti protons, are stabilized by the resonance exchange interaction of unpaired sub-elementary fermion $\mathbf{F}_{\updownarrow}^{\pm} >^i$ and paired $(\mathbf{F}_{\uparrow}^+$ and $\mathbf{F}_{\downarrow}^-)^i$ with Bivacuum Virtual Pressure Waves $(\mathbf{VPW}_q^{\pm})^i$ with fundamental quantized frequency $(\boldsymbol{\omega}_{\mathbf{VPW}_q} = \mathbf{q}\boldsymbol{\omega}_0)^i$. This interaction occur in the process of quantum beats between the actual and complementary states of sub-elementary fermions, representing their $[\mathbf{Corpusle(C)} \rightleftharpoons \mathbf{Wave(W)}]$ pulsations with fundamental Compton frequency:

$$\left[ \boldsymbol{\omega}_{\mathbf{C} \rightleftharpoons \mathbf{W}} = \boldsymbol{\omega}_{\mathbf{VPW}_q} = \mathbf{q}\boldsymbol{\omega}_0 = \mathbf{q}(\mathbf{m}_0 \mathbf{c}^2)/\hbar \right]^i \qquad 4.2$$

where: $\mathbf{m}_0 \mathbf{c}^2$ is a rest-mass energy of sub-elementary fermion of $i$-generation.

The *1st stage* of matter creation in form of sub-elementary fermions or antifermions is a result of cells-dipoles symmetry shift towards the positive or negative energy, correspondingly, accompanied by uncompensated *mass and charge origination*. The *2nd stage* of matter formation is fusion of triplets $< [\mathbf{F}_{\uparrow}^+ \bowtie \mathbf{F}_{\downarrow}^-] + \mathbf{F}_{\updownarrow}^{\pm} >^i$ from sub-elementary fermions and antifermions. Both of stages occur at Golden mean (GM) conditions: $(\mathbf{v}/\mathbf{c})^2 = 0.618$ and results in elementary particles and antiparticles (electrons, protons and neutrons) origination.

At the Golden Mean (GM) conditions: $(\mathbf{v}/\mathbf{c})^2 = \phi = 0.618$, the Cooper pairs of asymmetric Bivacuum fermions, rotating in opposite direction around the common axis of vorticity, turns to pair of sub-elementary fermion and antifermion with ratio of radiuses of torus and antitorus: $\mathbf{L}^+/\mathbf{L}^- = \pi(\mathbf{L}^+)^2/\pi\mathbf{L}_0^2 = \mathbf{S}^+/\mathbf{S}_0 = \phi$ :

$$[\mathbf{F}_{\uparrow}^+ \bowtie \mathbf{F}_{\downarrow}^-] \equiv [\mathbf{BVF}_{as}^{\uparrow} \bowtie \mathbf{BVF}_{as}^{\downarrow}]^{\phi} \qquad 4.3$$

of opposite charge, spin and energy with common Compton radius. The spatial image of pair $[\mathbf{F}_{\uparrow}^+ \bowtie \mathbf{F}_{\downarrow}^-]$ is two identical *truncated cones* of the opposite orientation of planes rotating without slip around common rotation axis (Fig.1).

**Model of the electron, as a triplet of
rotating sub-elementary fermions:**
$$< [\mathbf{F}_\uparrow^+ \bowtie \mathbf{F}_\downarrow^-] + \mathbf{F}_\downarrow^- >$$

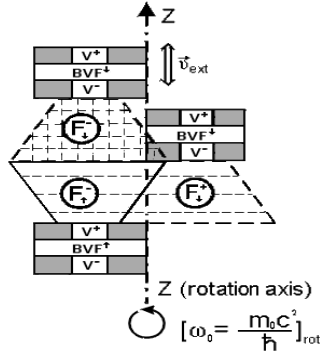

**The total energy of each sub-elementary fermion:**

$$\mathbf{E}_{tot} = \mathbf{mc}^2 = \sqrt{1-(\mathbf{v/c})^2}\,(\mathbf{m}_0\omega_0^2\mathbf{L}^2)_{rot}^{in} + \left(\frac{\mathbf{h}^2}{\mathbf{m}\lambda_B^2}\right)_{tr}^{ext}$$

$$or:\ \ \mathbf{E}_{tot} = \sqrt{1-(\mathbf{v/c})^2}\,\hbar\omega_0^{in} + \hbar\omega^{ext}; \qquad \lambda_B = \mathbf{h/mv}_{tr}^{ext}$$

**Fig. 1.** Model of the electron, as a triplets $< [\mathbf{F}_\uparrow^+ \bowtie \mathbf{F}_\downarrow^-] + \mathbf{F}_\downarrow^\pm >^i$ , resulting from fusion of third sub-elementary antifermion $[\mathbf{F}_\downarrow^-]$ to sub-elementary antifermion $[\mathbf{F}_\uparrow^-]$ with opposite spin in rotating pair $[\mathbf{F}_\uparrow^+ \bowtie \mathbf{F}_\downarrow^-]$. The velocity of rotation of unpaired sub-elementary $[\mathbf{F}_\downarrow^-]$ around the same axis of common rotation axis of pair provide the similar rest mass ($\mathbf{m}_0$) and absolute charge $|\mathbf{e}^\pm|$, as have the paired $[\mathbf{F}_\uparrow^+$ and $\mathbf{F}_\downarrow^-]$. Three effective anchor $(\mathbf{BVF}^\updownarrow = [\mathbf{V}^+ \Updownarrow \mathbf{V}^-])_{anc}$ in the vicinity of sub-elementary particles base, participate in recoil effects, accompanied their $[\mathbf{C} \rightleftharpoons \mathbf{W}]$ pulsation and modulation of Bivacuum pressure waves $(\mathbf{VPW}_q^\pm)$. The recoil effects of paired $[\mathbf{F}_\uparrow^+ \bowtie \mathbf{F}_\downarrow^-]$ totally compensate each other and the relativistic mass change of triplets is determined only by the anchor Bivacuum fermion $(\mathbf{BVF}^\updownarrow)_{anc}$ of the unpaired sub-elementary fermion $\mathbf{F}_\updownarrow^\pm >$.

We suppose, that regular electrons and positrons are the result of fusion of three $\mu$ −electrons/positrons. For the other hand, the protons and antiprotons are resulted from fusion of three $\tau$ −electrons/positrons.

The fusion of asymmetric sub-elementary fermions and antifermions of $\mu$ and $\tau$ generations $\left[ \mathbf{F}_\updownarrow^\pm \equiv \left(\mathbf{BVF}_{as}^\updownarrow\right)^\phi \right]^{\mu,\tau}$ (Fig.1) to triplets of electrons/positrons, protons/antiprotons and neutrons/antineutrons

$$< [\mathbf{F}_\uparrow^+ \bowtie \mathbf{F}_\downarrow^-]_{x,y} + \mathbf{F}_z^\pm >_z^{e,p} \qquad\qquad 4.4$$

becomes possible also at the Golden mean (GM) conditions. It is accompanied by energy release and electronic and hadronic $e, h$ −gluons origination, equal in sum to the mass defect. It was demonstrated theoretically, that the vortical structures at certain conditions self-organizes into vortex crystals (Jin and Dubin, 2000).

The fusion threshold overcoming is due to ’switching on’ the resonant exchange interaction of $\mathbf{CVC}^\pm$ with Bivacuum virtual pressure waves $\mathbf{VPW}^\pm$ of fundamental frequency ($\omega_0 = \mathbf{m}_0\mathbf{c}^2/\hbar)^{e,\mu,\tau}$ in the process of $[\mathbf{corpuscle(C)} \rightleftharpoons \mathbf{wave(W)}]$ transitions of elementary particles. The *triplets* of elementary particles and antiparticles formation (Fig.2) is a result of conjugation of third sub-elementary fermion (antifermion) $[\mathbf{F}_\updownarrow^\pm]$ to one of sub-elementary fermion (antifermion) of rotating pair $[\mathbf{F}_\uparrow^+ \bowtie \mathbf{F}_\downarrow^-]$ of the opposite spins. The opposite spins means that their $[\mathbf{C} \rightleftharpoons \mathbf{W}]$ pulsations are counterphase and these two sub-elementary particles are spatially compatible (see ). The velocity of rotation of unpaired sub-elementary fermion $[\mathbf{F}_\downarrow^-]$ around the same axis of common rotation axis of pair (Fig.1) provide the similar mass and charge $|e^{\pm}|$, as have the paired $[\mathbf{F}_{\uparrow}^{+}$ and $\mathbf{F}_{\downarrow}^{-}]$ because of similar symmetry shift.

Let us consider the rotational dynamics of unpaired $\mathbf{F}_{\updownarrow}^{\pm} >^{e,\mu,\tau} = [\mathbf{V}^{+} \Updownarrow \mathbf{V}^{-}]^{as}$ in triplets (Fig.2) just after fusion to triplet at GM conditions in the absence of the external translational motion of triplet.

Its properties are the result of participation in two rotational process simultaneously:

1) rotation of asymmetric $\mathbf{F}_{\updownarrow}^{\pm} >^{e,\mu,\tau}$ around its own axis (Fig.1) with spatial image of truncated cone with resulting radius:

$$\mathbf{L}_{\mathbf{BVF}_{as}}^{\phi} = \hbar/|\mathbf{m}_{V}^{+} + \mathbf{m}_{\overline{V}}^{-}|^{\phi}\mathbf{c} = \hbar/[\mathbf{m}_0(1/\phi + \phi)\mathbf{c}] = \hbar/2.236\,\mathbf{m}_0\mathbf{c} = \mathbf{L}_0/2.23 \qquad 4.5$$

2) rolling of this truncated cone of $\mathbf{F}_{\updownarrow}^{\pm} >^{e,\mu,\tau}$ around the another axis, common for pair of sub-elementary particles $[\mathbf{F}_{\uparrow}^{+} \bowtie \mathbf{F}_{\downarrow}^{-}]$ (Fig.2) inside of a larger vorticity with bigger radius, equal to *Compton radius*:

$$\mathbf{L}_{\mathbf{BVF}_{as}^{\uparrow} \bowtie \mathbf{BVF}_{as}^{\downarrow}}^{\phi} = \hbar/|\mathbf{m}_{V}^{+} - \mathbf{m}_{\overline{V}}^{-}|^{\phi}\mathbf{c} = \hbar/\mathbf{m}_0\mathbf{c} = \mathbf{L}_0 \qquad 4.6$$

The ratio of radius of $\left(\mathbf{BVF}_{as}^{\uparrow}\right)^{\phi} \equiv \mathbf{F}_{\updownarrow}^{\pm} >$ to radius of pairs $[\mathbf{F}_{\uparrow}^{+} \bowtie \mathbf{F}_{\downarrow}^{-}]$ at GM conditions is equal to the ratio of potential energy ($\mathbf{V}$) to kinetic energy ($\mathbf{T}_k$) of relativistic de Broglie wave (wave B) at GM conditions. This ratio is the same, as in known formula for relativistic wave B $\left(\frac{\mathbf{V}}{\mathbf{T}_k} = 2\frac{\mathbf{v}_{ph}}{\mathbf{v}_{gr}} - 1\right)$:

$$\frac{\mathbf{L}_{\mathbf{BVF}_{as}^{\uparrow} \bowtie \mathbf{BVF}_{as}^{\downarrow}}^{\phi}}{\mathbf{L}_{\mathbf{BVF}_{as}}^{\phi}} = \frac{|\mathbf{m}_{V}^{+} + \mathbf{m}_{\overline{V}}^{-}|^{\phi}\mathbf{c}^2}{|\mathbf{m}_{V}^{+} - \mathbf{m}_{\overline{V}}^{-}|^{\phi}\mathbf{c}^2} = \left(\frac{\mathbf{V}}{\mathbf{T}_k}\right)^{\phi} = 2\left(\frac{\mathbf{v}_{ph}}{\mathbf{v}_{gr}}\right)^{\phi} - 1 = 2,236 \qquad 4.7$$

This result is a good evidence in proof of our expressions for total energy of sub-elementary particle, as sum of internal potential energy and rotational-translational kinetic energy (see http://arxiv.org/abs/physics/0103031):

$$\mathbf{V} = |\mathbf{m}_{V}^{+} + \mathbf{m}_{\overline{V}}^{-}|\mathbf{c}^2 \qquad 4.7a$$

$$\mathbf{T}_k = |\mathbf{m}_{V}^{+} - \mathbf{m}_{\overline{V}}^{-}|\mathbf{c}^2 \qquad 4.7b$$

*The triplets of the regular electrons and positrons* of the same or opposite spin state are the result of fusion of sub-elementary particles of $\mu-$ electrons (muons) generation:

$$\mathbf{e}^{-} \equiv < [\mathbf{F}_{\uparrow}^{-} \bowtie \mathbf{F}_{\uparrow}^{+}] + \mathbf{F}_{\updownarrow}^{-} >^{e} \qquad 4.8$$

$$\mathbf{e}^{+} \equiv < [\mathbf{F}_{\uparrow}^{-} \bowtie \mathbf{F}_{\downarrow}^{+}] + \mathbf{F}_{\updownarrow}^{+} >^{e} \qquad 4.8a$$

with mass, charge and spins, determined by uncompensated/unpaired sub-elementary particle: $\mathbf{F}_{\updownarrow}^{+} >^{e}$.

**The fusion of triplets** (electrons and protons) is accompanied by release of huge amount of kinetic energy and the *electronic* ($e$) − *gluons* and *hadronic* ($h$) − *gluons* origination (i.e. strong interaction, providing stability of triplets).

The mass of *muon* is 105.69 MeV and the rest mass of the electron is 0.511 MeV. We suppose, that like in the case of protons, approximately the same energy as the latter one becomes involved in ($e$) − *gluons* formation. At this conditions the energy of mass defect about: $\mathbf{\Delta mc^2 \simeq 105.69\,MeV - (2 \cdot 0.511)MeV \simeq 104\,MeV}$ should be released as a result of each electron or positron fusion.

In the case of proton:

$$\mathbf{p} = < [\mathbf{F}^+_\uparrow \bowtie \mathbf{F}^-_\downarrow]_{x,y} + \mathbf{F}^+_\updownarrow >^\tau_z \qquad 4.9$$

the sub-elementary fermions and antiferms, resulting from fusion of $\boldsymbol{\tau}$ - electrons to stable triplets, have a properties of quarks and antiquarks. In our model of hadrons we do not need the hypothesis of fractional quark charge, because of compensation of integer charges of paired sub-elementary fermion and antifermion in triplets.

The mass of tau-electron is 1782 MeV and mass of proton: 938.28 MeV. The difference between them is about 844 MeV. This corresponds to known data, that the energy of 8 gluons is about 50% of energy/mass of quarks and antiquarks ($[\mathbf{q}^+ + \mathbf{q}^-]_{S=0,1}$): the fused *tauons* and *antitauons*. These gluons, in accordance to our model of elementary particles, are represented by 8 different superposition of $[\mathbf{CVC}^+ \bowtie \mathbf{CVC}^-]$ in two spin states ($\pm 1/2$) of triplets of $(\mathbf{p})$, emitted and absorbed with in-phase $[\mathbf{C} \rightleftarrows \mathbf{W}]$ pulsation of pair of quark + antiquark:

$$[\mathbf{F}^+_\uparrow \bowtie \mathbf{F}^-_\downarrow]^p_{S=0,1} = [\mathbf{q}^+ + \mathbf{q}^-]_{S=0,1} \qquad 4.10$$

of triplets (Kaivarainen: http://arxiv.org/abs/physics/0207027).

The pairs of quark ($\mathbf{q}^+ \sim \boldsymbol{\tau}^+$) and antiquark ($\mathbf{q}^- \sim \boldsymbol{\tau}^-$) represents *mesons* with neutral bosons properties.

## 5. The new formulas for total energy of de Broglie wave

The total energy of sub-elementary fermions, composing the triplets of the electrons or protons $< [\mathbf{F}^-_\uparrow \bowtie \mathbf{F}^+_\downarrow]_{S=0} + (\mathbf{F}^\pm_\updownarrow)_{S=\pm 1/2} >^{e,p}$, equal in both - Corpuscular and Wave phase, can be presented in three modes, as a sum of their potential $\mathbf{V}_{tot}$ and kinetic $\mathbf{T}_{tot}$ energies, including internal and external contributions:

$$\mathbf{E}_{tot} = \mathbf{V}_{tot} + \mathbf{T}_{tot} = \hbar\omega_B = \frac{1}{2}(\mathbf{m}^+_V + \mathbf{m}^-_{\bar V})\mathbf{c}^2 + \frac{1}{2}(\mathbf{m}^+_V - \mathbf{m}^-_{\bar V})\mathbf{c}^2 \qquad 5.1$$

$$\mathbf{E}_{tot} = \mathbf{m}^+_V \mathbf{c}^2 = \frac{1}{2}\mathbf{m}^+_V(2\mathbf{c}^2 - \mathbf{v}^2) + \frac{1}{2}\mathbf{m}^+_V \mathbf{v}^2 \qquad 5.1a$$

$$\mathbf{E}_{tot} = 2\mathbf{T}_k(\mathbf{v}/\mathbf{c})^2 = \frac{1}{2}\mathbf{m}^+_V \mathbf{c}^2[1 + \mathbf{R}^2] + \frac{1}{2}\mathbf{m}^+_V \mathbf{v}^2 \qquad 5.1b$$

where: $\mathbf{R} = \mathbf{m}_0/\mathbf{m}^+_V = \sqrt{1 - (\mathbf{v}/\mathbf{c})^2}$ is the dimensionless relativistic factor; $\mathbf{v}$ is external translational - rotational velocity of particle; $\mathbf{m}^+_V$ and $\mathbf{m}^-_{\bar V}$ are the *absolute* masses of torus and antitorus of Bivacuum dipoles.

One may see, that $\mathbf{E}_{tot} \to \mathbf{m}_0\mathbf{c}^2$ at $\mathbf{v} \to \mathbf{0}$ and $\mathbf{m}^+_V \to \mathbf{m}_0$.

Taking into account that the total kinetic energy of dipoles is

$$\mathbf{T}^W_{tot} = \frac{1}{2}(\mathbf{m}^+_V - \mathbf{m}^-_{\bar V})\mathbf{c}^2 = \frac{1}{2}\mathbf{m}^+_V \mathbf{v}^2 = \mathbf{T}^C_{tot} \qquad 5.2$$

$\mathbf{T}^W_{tot} = \frac{1}{2}(\mathbf{m}^+_V - \mathbf{m}^-_{\bar V})\mathbf{c}^2 = \mathbf{T}^C_{tot} = \frac{1}{2}\mathbf{m}^+_V \mathbf{v}^2$ and $\mathbf{c}^2 = \mathbf{v}_{gr}\mathbf{v}_{ph}$, where $\mathbf{v}_{gr} \equiv \mathbf{v}$, then dividing the left and right parts of (5.1 and 5.1a) by $\frac{1}{2}\mathbf{m}^+_V \mathbf{v}^2$, we get formula, similar to (4.7):

$$2\frac{\mathbf{c}^2}{\mathbf{v}^2} - 1 = 2\frac{\mathbf{v}_{ph}}{\mathbf{v}_{gr}} - 1 = \frac{(\mathbf{m}^+_V + \mathbf{m}^-_{\bar V})\mathbf{c}^2}{\mathbf{m}^+_V \mathbf{v}^2} = \frac{\mathbf{m}^+_V + \mathbf{m}^-_{\bar V}}{\mathbf{m}^+_V - \mathbf{m}^-_{\bar V}} \qquad 5.3$$

Comparing formula (5.3) with known relation for relativistic de Broglie wave for ratio of its potential and kinetic energy (Grawford, 1973), we get the confirmation of our definitions of potential and kinetic energies of elementary particle:

$$2\frac{\mathbf{v}_{ph}}{\mathbf{v}_{gr}} - 1 = \frac{\mathbf{V}_{tot}}{\mathbf{T}_{tot}} = \frac{(\mathbf{m}_{\bar{V}}^+ + \mathbf{m}_{\bar{V}}^-)\mathbf{c}^2}{(\mathbf{m}_{\bar{V}}^+ - \mathbf{m}_{\bar{V}}^-)\mathbf{c}^2} \qquad 5.4$$

In Golden mean conditions, necessary for triplet fusion, the ratio $(\mathbf{V}_{tot}/\mathbf{T}_{tot})^\phi = (1/\phi + \phi) = 2.236$.

The well known Dirac equation for total/real energy of a free relativistic particle, following also from Einstein relativistic formula, can be easily derived from (5.1a), multiplying its left and right part on $\mathbf{m}_{\bar{V}}^+\mathbf{c}^2$ and using introduced mass compensation principle ($|\mathbf{m}_{\bar{V}}^+\mathbf{m}_{\bar{V}}^-| = \mathbf{m}_0^2$). The same formula for energy of *torus ($\mathbf{V}^+$)* can be obtained directly from Lorentz relativistic equation (3.5):

$$\mathbf{E}_{tot}^2 = (\mathbf{m}_{\bar{V}}^+\mathbf{c}^2)^2 = (\mathbf{m}_0\mathbf{c}^2)^2 + (\mathbf{m}_{\bar{V}}^+)^2\mathbf{v}^2\mathbf{c}^2 \qquad 5.5$$

where: the actual *inertial* mass of torus of unpaired sub-elementary fermion in triplets is equal to regular mass of particle: $\mathbf{m}_{\bar{V}}^+ = \mathbf{m}_0/\sqrt{1-(\mathbf{v}/\mathbf{c})^2}$.

From the formula (3.6), describing dependence of *inertialess* mass $\mathbf{m}_{\bar{V}}^-$ of antitorus ($\mathbf{V}^-$) on the external velocity of Bivacuum dipole or unpaired sub-elementary fermion in triplets $\mathbf{m}_{\bar{V}}^- = \mathbf{m}_0\sqrt{1-(\mathbf{v}/\mathbf{c})^2}$, we get:

$$(\mathbf{m}_{\bar{V}}^-\mathbf{c}^2)^2 = (\mathbf{m}_0\mathbf{c}^2)^2 - \mathbf{m}_0^2\mathbf{v}^2\mathbf{c}^2 \qquad 5.6$$

The difference between 5.5 and 5.6 can be easily transformed to:

$$(\mathbf{m}_{\bar{V}}^+\mathbf{c}^2)^2 - (\mathbf{m}_{\bar{V}}^-\mathbf{c}^2)^2 = [(\mathbf{m}_{\bar{V}}^+)^2 + \mathbf{m}_0^2]\mathbf{v}^2\mathbf{c}^2 \qquad 5.7$$

$$(\mathbf{m}_{\bar{V}}^+\mathbf{c}^2 - \mathbf{m}_{\bar{V}}^-\mathbf{c}^2)(\mathbf{m}_{\bar{V}}^+\mathbf{c}^2 + \mathbf{m}_{\bar{V}}^-\mathbf{c}^2) = [(\mathbf{m}_{\bar{V}}^+)^2 + \mathbf{m}_0^2]\mathbf{v}^2\mathbf{c}^2 \qquad 5.7a$$

$$\mathbf{T}_k\mathbf{V} = \frac{1}{4}[(\mathbf{m}_{\bar{V}}^+)^2 + \mathbf{m}_0^2]\mathbf{v}^2\mathbf{c}^2 \qquad 5.7b$$

We got the new important formula, expressing the product of kinetic and potential energy of asymmetric Bivacuum dipole or unpaired sub-elementary fermion in triplets ($\mathbf{T}_k\mathbf{V}$) via its actual inertial ($\mathbf{m}_{\bar{V}}^+$), the rest mass ($\mathbf{m}_0$) and external velocity ($\mathbf{v}$). As far the kinetic energy of asymmetric dipole like the unpaired sub-elementary fermion of triplet is $\mathbf{T}_k = \mathbf{m}_{\bar{V}}^+\mathbf{v}^2/2$, the potential energy from 5.7b can be calculated from the known empirical data:

$$\mathbf{V} = \frac{1}{2}[\mathbf{m}_{\bar{V}}^+ + \mathbf{m}_0^2/\mathbf{m}_{\bar{V}}^+]\mathbf{c}^2 \qquad 5.8$$

## 6. The Dynamic Mechanism of Corpuscle-Wave Duality

It is shown, that the [corpuscle (C) $\rightleftharpoons$ wave (W)] duality represents the *modulation* of quantum beats between the asymmetric 'actual' (torus) and 'complementary' (antitorus) states of sub-elementary fermions or antifermions of triplets by de Broglie wave frequency of these particles, equal to frequency of [C $\rightleftharpoons$ W] pulsations of the 'anchor' Bivacuum fermion ($\mathbf{BVF}_{anc}^{\updownarrow}$)$^i$ of unpaired $\mathbf{F}_{\updownarrow}^{\pm} >^i$. The [C] phase of each sub-elementary fermions of triplets $< [\mathbf{F}_{\uparrow}^+ \bowtie \mathbf{F}_{\downarrow}^-] + \mathbf{F}_{\updownarrow}^{\pm} >^i$ exists as a mass, electric and magnetic asymmetric dipole. The [C $\rightarrow$ W] transition is a result of two stages superposition. The total energy, charge and spin of particle, moving in space with velocity ($\mathbf{v}$) is determined by the unpaired sub-elementary fermion ($\mathbf{F}_{\updownarrow}^{\pm}$)$_z$, as far the energy, charge, spin of paired ones in $[\mathbf{F}_{\uparrow}^+ \bowtie \mathbf{F}_{\downarrow}^-]_{x,y}$ of triplets compensate each other.

*The 1st stage* is a reversible dissociation of [C] phase to Cumulative virtual cloud

$(\mathbf{CVC}^{\pm})_{\mathbf{F}_{\updownarrow}^{\pm}}$ of subquantum particles and the 'anchor' Bivacuum fermion $(\mathbf{BVF}_{anc}^{\updownarrow})$:

$$(\mathbf{I}): \left[ \, (\mathbf{F}_{\updownarrow}^{\pm})_{\mathbf{C}} \, \xrightarrow{\text{Recoil/Antirecoil}} \, \left[ \mathbf{BVF}_{anc}^{\updownarrow} + (\mathbf{CVC}^{\pm})_{\mathbf{F}_{\updownarrow}^{\pm}} \right]_{\mathbf{W}} \, \right]^i \qquad 6.1$$

*The 2nd stage* of $[\mathbf{C} \rightarrow \mathbf{W}]$ transition is a reversible dissociation of the anchor Bivacuum fermion $(\mathbf{BVF}_{anc}^{\updownarrow})^i = [\mathbf{V}^{+} \, \Updownarrow \, \mathbf{V}^{-}]_{anc}^i$ to symmetric $(\mathbf{BVF}^{\updownarrow})^i$ and the anchor cumulative virtual cloud $(\mathbf{CVC}^{\pm})_{\mathbf{BVF}_{anc}^{\updownarrow}}$, with frequency $(\boldsymbol{\omega}_{B}^{ext})_{tr}$, equal to the empirical frequency of de Broglie wave of particle:

$$(\mathbf{II}): \left( \mathbf{BVF}_{anc}^{\updownarrow} \right)_{\mathbf{C}}^{\mathbf{e}_{anc}^{\pm}} \, \xrightarrow{\text{Recoil/Antirecoil}} \, \left[ (\mathbf{BVF}^{\updownarrow})^0 + (\mathbf{CVC}^{\pm})_{\mathbf{BVF}_{anc}^{\updownarrow}}^{\mathbf{e}_{anc}^{\pm}} \right]_{W}^{i} \qquad 6.2$$

The 2nd stage takes a place if $(\mathbf{BVF}_{anc}^{\updownarrow})^i$ is asymmetric only in the case of nonzero external translational - rotational velocity of particle. The beats frequency of $(\mathbf{BVF}_{anc}^{\updownarrow})^{e,p}$ is equal to that of the empirical de Broglie wave frequency: $\boldsymbol{\omega}_B = \hbar/(\mathbf{m}_{\uparrow}^{+}\mathbf{L}_B^2)$. The higher is the external kinetic energy of fermion, the higher is frequency $\boldsymbol{\omega}_B$. The frequency of the stage (II) oscillations modulates the internal frequency of $[\mathbf{C} \rightleftharpoons \mathbf{W}]$ pulsation: $(\boldsymbol{\omega}^{in})^i = \mathbf{R}\,\boldsymbol{\omega}_0^i = \mathbf{R}\,\mathbf{m}_0^i\mathbf{c}^2/\hbar$, related to contribution of the rest mass energy to the total energy of the de Broglie wave (Kaivarainen, 2005; 2005a and http://arxiv.org/abs/physics/0103031).

The $[\mathbf{C} \rightleftharpoons \mathbf{W}]$ pulsations of unpaired sub-elementary fermion $\mathbf{F}_{\updownarrow}^{\pm} >$, of triplets of the electrons or protons $< [\mathbf{F}_{\uparrow}^{+} \bowtie \mathbf{F}_{\downarrow}^{-}] + \mathbf{F}_{\updownarrow}^{\pm} >^{e,p}$ are in counterphase with the in-phase pulsation of paired sub-elementary fermion and antifermion, modulating Bivacuum virtual pressure waves $(\mathbf{VPW}^{\pm})$:

$$\left[ \mathbf{F}_{\uparrow}^{+} \bowtie \mathbf{F}_{\downarrow}^{-} \right]_{W}^{e,p} \, \xrightarrow{\mathbf{CVC}^{+}+\mathbf{CVC}^{-}} \, \left[ \mathbf{F}_{\uparrow}^{+} \bowtie \mathbf{F}_{\downarrow}^{-} \right]_{C}^{e,p} \qquad 6.3$$

For a regular nonrelativistic electron the carrier frequency is $\omega^{in} = R\omega_0^e \sim 10^{21} s^{-1} >> \omega_B^{ext}$. However, for relativistic case at $\mathbf{v} \rightarrow \mathbf{c}$, the situation is opposite: $\omega_B^{ext} >> \omega^{in}$ at $\omega^{in} \rightarrow 0$.

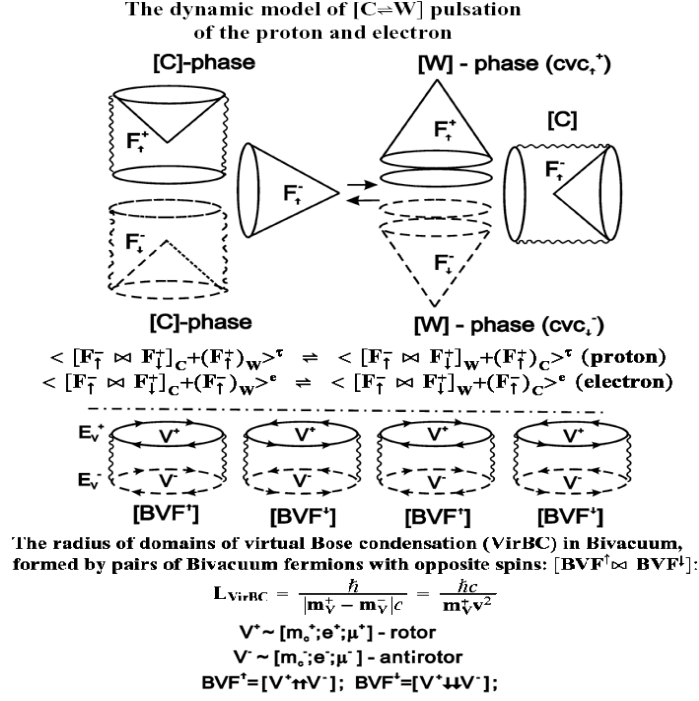

**Fig. 2.** Dynamic model of $[\mathbf{C} \rightleftharpoons \mathbf{W}]$ pulsation of triplets of sub-elementary fermions/antifermions (the reduced by fusion to triplets $\mu$ and $\tau$ electrons) composing, correspondingly, electron and proton $< [\mathbf{F}_\uparrow^+ \bowtie \mathbf{F}_\downarrow^-] + \mathbf{F}_\updownarrow^\pm >^{e,p}$. The pulsation of the pair $[\mathbf{F}_\uparrow^- \bowtie \mathbf{F}_\downarrow^+]$, modulating virtual pressure waves of Bivacuum (VPW$^+$ and VPW$^-$), is counterphase to pulsation of unpaired sub-elementary fermion/antifermion $\mathbf{F}_\updownarrow^\pm >$.

The total energy, charge and spin of triplets - fermions, moving in space with external translational velocity ($\mathbf{v}_{tr}^{ext}$) is determined by the unpaired sub-elementary fermion ($\mathbf{F}_\updownarrow^\pm)_z$, as far the paired ones in $[\mathbf{F}_\uparrow^+ \bowtie \mathbf{F}_\downarrow^-]_{x,y}$ of triplets compensate each other. From (5.1-5.1b) it is easy to get:

$$\mathbf{E}_{tot} = \mathbf{m}_V^+ \mathbf{c}^2 = \hbar\omega_{\mathbf{C}\rightleftharpoons\mathbf{W}} = \mathbf{R}(\hbar\omega_0)_{rot}^{in} + (\hbar\omega_B^{ext})_{tr} = \mathbf{R}(\mathbf{m}_0\mathbf{c}^2)_{rot}^{in} + (\mathbf{m}_V^+\mathbf{v}_{tr}^2)^{ext} \qquad 6.4$$

$$\mathbf{E}_{tot} = \mathbf{m}_V^+\mathbf{c}^2 = -\mathcal{L} + 2\mathbf{T}_k = \mathbf{R}(\mathbf{m}_0\omega_0^2\mathbf{L}_0^2)_{rot}^{in} + \left(\frac{\mathbf{h}^2}{\mathbf{m}_V^+\boldsymbol{\lambda}_B^2}\right) \qquad 6.4a$$

$$\mathbf{E}_{tot} = \mathbf{V} + \mathbf{T_k} = \left[\mathbf{R}(\mathbf{m}_0\mathbf{c}^2)_{rot}^{in} + \frac{1}{2}(\mathbf{m}_V^+\mathbf{v}_{tr}^2)\right] + \frac{1}{2}(\mathbf{m}_V^+\mathbf{v}_{tr}^2) \qquad 6.4b$$

$$or: \quad \mathbf{E}_{tot} = \mathbf{m}_V^+\mathbf{c}^2 = \mathbf{V} + \mathbf{T_k} = \frac{1}{2}(\mathbf{m}_V^+ + \mathbf{m}_V^-)\mathbf{c}^2 + \frac{1}{2}(\mathbf{m}_V^+ - \mathbf{m}_V^-)\mathbf{c}^2 \qquad 6.4c$$

where: $\mathbf{R} = \sqrt{1-(\mathbf{v/c})^2}$ is the relativistic factor; $\mathbf{v} \equiv \mathbf{v}_{tr}^{ext}$ is the external translational group velocity;

$\mathcal{L} = \mathbf{T_k} - \mathbf{V}$ is the Lagrange function; $\boldsymbol{\lambda}_B = h/\mathbf{m}_V^+\mathbf{v} = 2\boldsymbol{\pi}\mathbf{L}_B$ is the external translational de Broglie wave length; the actual inertial mass is $\mathbf{m}_V^+ = \mathbf{m} = \mathbf{m}_0/\mathbf{R}$; $\mathbf{L}_0^i = \hbar/\mathbf{m}_0^i\mathbf{c}$ is a Compton radius of the elementary particle.

It follows from our approach, that the fundamental phenomenon of **corpuscle – wave** duality (Fig.3) is a result of modulation of the primary - carrying frequency of the internal $[\mathbf{C} \rightleftharpoons \mathbf{W}]^{in}$ pulsation of individual sub-elementary fermions (*1st stage*):

$$(\boldsymbol{\omega}^{in})^i = \mathbf{R}\omega_0^i = \mathbf{R} = \sqrt{1-(\mathbf{v/c})^2}\ \ \mathbf{m}_0^i \mathbf{c}^2/\hbar \qquad 6.4d$$

by the frequency of the external empirical de Broglie wave of triplet:
$\omega_B^{ext} = \mathbf{m}_V^+ \mathbf{v}_{ext}^2/\hbar = 2\pi \mathbf{v}_{ext}/\mathbf{L}_B$, equal to angular frequency of $[\mathbf{C} \rightleftharpoons \mathbf{W}]_{anc}$ pulsation of the anchor Bivacuum fermion $(\mathbf{BVF}_{anc}^{\updownarrow})^i$ (*2nd stage*).

The contribution of this external translational dynamics to the total one is determined by asymmetry of the *anchor* $(\mathbf{BVF}_{anc}^{\updownarrow})^i = [\mathbf{V}^+ \Updownarrow \mathbf{V}^-]_{anc}$ of particle, i.e. by second terms in (6.4) and (6.4a):

$$2\mathbf{T}_k = (\hbar\omega_B)_{tr} = \left(\frac{\mathbf{h}^2}{\mathbf{m}_V^+ \boldsymbol{\lambda}_B^2}\right)_{tr} = \left[(\mathbf{m}_V^+ - \mathbf{m}_V^-)\mathbf{c}^2\right]_{tr} \qquad 6.5$$

$$= (\mathbf{m}_V^+ \mathbf{v}^2)_{tr} = (\mathbf{m}_V^+ \boldsymbol{\omega}_B^2 \mathbf{L}_B^2)_{rot} = \frac{\mathbf{p}_B^2}{\mathbf{m}_V^+} \qquad 6.5a$$

This contribution is increasing with particle acceleration and tending to light velocity. At $\mathbf{v} \to \mathbf{c}$, and $\mathbf{R} \to 0$ :

$$2\mathbf{T}_k = (\mathbf{m}_V^+ \mathbf{v}^2)_{tr}^{ext} \to \mathbf{m}_V^+ \mathbf{c}^2 = \mathbf{E}_{tot} = \mathbf{V} + \mathbf{T}_k \qquad 6.5b$$

$$or \quad \mathbf{V} = \mathbf{T}_k = \frac{1}{2}\mathbf{m}_V^+ \mathbf{c}^2 = \frac{1}{2}\hbar\omega_{\mathbf{C} \rightleftharpoons \mathbf{W}} \qquad 6.5c$$

For example, the equality of the averaged potential and kinetic energies of sub-elementary fermions and antifermions should take a place for photon (fig.3).

The properties of the *anchor* Bivacuum fermion $\mathbf{BVF}_{anc}^{\updownarrow}$ can be analyzed (Kaivarainen, 2005), at three conditions:

1. The external translational velocity ($\mathbf{v}$) is zero;
2. The external translational velocity corresponds to Golden mean ($\mathbf{v} = \mathbf{c}\boldsymbol{\phi}^{1/2}$);
3. The relativistic case, when $\mathbf{v} \sim \mathbf{c}$.

Under nonrelativistic conditions ($\mathbf{v} << \mathbf{c}$), the de Broglie wave (modulation) frequency is low: $2\pi(\mathbf{v}_B)_{tr} << (\boldsymbol{\omega}^{in} = \mathbf{R}\omega_0)$. However, in relativistic case ($\mathbf{v} \sim \mathbf{c}$), the modulation frequency of the 'anchor' $(\mathbf{BVF}_{anc}^{\updownarrow})$, equal to that of de Broglie wave, can be higher, than the internal one : $2\pi(\mathbf{v}_B)_{tr} \geq \boldsymbol{\omega}^{in}$.

The paired sub-elementary fermion and antifermion of $[\mathbf{F}_{\uparrow}^- \bowtie \mathbf{F}_{\uparrow}^+]_{S=0}$ also have the 'anchor' Bivacuum fermion and antifermion $(\mathbf{BVF}_{anc}^{\updownarrow})$, similar to that of unpaired. However, the opposite energies of their $[\mathbf{C} \rightleftharpoons \mathbf{W}]$ pulsation compensate each other in accordance with proposed model.

If we proceed from the assumption that the total energy of the corpuscular and wave phase of each sub-elementary fermion, as a sum of their potential and kinetic energies, do not change in the process of $[\mathbf{C} \rightleftharpoons \mathbf{W}]$ pulsation of sub-elementary fermions: $\Delta \mathbf{E}_{tot}^{\mathbf{C} \rightleftharpoons \mathbf{W}} = \Delta \mathbf{V}_{tot} + \Delta \mathbf{T}_{tot} = 0$ in the inertial system ($\mathbf{v} = const$), then we get, that the oscillations of potential and kinetic energy should be opposite and compensating each other:

$$-\mathbf{V}_{tot}\frac{\Delta\mathbf{L}_{\mathbf{V}_{tot}}}{\mathbf{L}_{\mathbf{V}_{tot}}} \overset{\mathbf{C} \rightleftharpoons \mathbf{W}}{\rightleftharpoons} \mathbf{T}_{tot}\frac{\Delta\mathbf{L}_{\mathbf{T}_{tot}}}{\mathbf{L}_{\mathbf{T}_{tot}}} \qquad 6.6$$

The linear dimension of the Wave phase of the electron in nonrelativistic condition $0 < \mathbf{v}_{tr}^{ext} << \mathbf{c}$ $\quad \boldsymbol{\lambda}_B = \mathbf{h}/\mathbf{m}_V^+\mathbf{v}_{tr}^{ext}$ can be much bigger, than that $[\mathbf{C}]$ phase, determined by Compton length of particle: $\boldsymbol{\lambda}_0 = h/\mathbf{m}_0\mathbf{c}$ $(\boldsymbol{\lambda}_B > \boldsymbol{\lambda}_0)$.

The counterphase oscillations of momentum ($\Delta\mathbf{p}$) and dimensions ($\Delta\mathbf{x}$) in the process of $[\mathbf{C} \rightleftharpoons \mathbf{W}]$ pulsation of elementary particles (fig.3) is reflected by the uncertainty principle:

$$\Delta\mathbf{p}\,\Delta\mathbf{x} \geq \hbar/\mathbf{2} \qquad\qquad 6.7$$

The decreasing of momentum uncertainty $\Delta\mathbf{p} \rightarrow \mathbf{0}$ in the Wave [W] phase is accompanied by the increasing of the *effective* de Broglie wave length: $\Delta\mathbf{x} \rightarrow \lambda_B$ and vice verse.

Taking the differential of de Broglie wave length, it is easy to get:

$$\lambda_B = \mathbf{h}/\mathbf{p}_{tr}^{ext} \quad \rightarrow \quad \frac{\Delta\lambda_B}{\lambda_B} = -\frac{\Delta\mathbf{p}}{\mathbf{p}} \qquad\qquad 6.8$$

In conditions, when $\Delta\lambda_B = \lambda_B$ we have $-\Delta\mathbf{p} = \mathbf{p}$. The de Broglie wave length characterize the dimension of cumulative virtual cloud, positive for particles or negative for antiparticles ($\mathbf{CVC}^{\pm}$) in their [W] phase and momentum $\mathbf{p} = \mathbf{m}_V^+\mathbf{v}_{tr}^{ext}$ characterize the corpuscular [C] phase.

The other presentation of uncertainty principle reflects the counterphase oscillation of the kinetic energy and time for free particle in process of $[\mathbf{C} \rightleftharpoons \mathbf{W}]$ pulsation:

$$\Delta\mathbf{T}_k\,\Delta\mathbf{t} \geq \hbar/\mathbf{2} \qquad\qquad 6.9$$

This kind of counterphase energy-time pulsation is in accordance with our theory of time (Kaivarainen, 2005a,b).

The wave function for de Broglie wave of particle with energy $\mathbf{E} = \hbar\boldsymbol{\omega}_B$, moving in direction $\mathbf{x}$ with certain momentum:

$$\mathbf{p} = \mathbf{m}_V^+\mathbf{v}_{tr}^{ext} = \hbar/\mathbf{L}_B = \hbar\mathbf{k} \qquad\qquad 6.10$$

is described by the wave function:

$$\boldsymbol{\Psi}(\mathbf{x},\mathbf{t}) = \mathbf{C}\exp\left[\frac{i}{\hbar}(\mathbf{p}\mathbf{x} - \mathbf{E}\mathbf{t})\right] = \mathbf{C}\exp\left[i\left(\frac{\mathbf{x}}{\mathbf{L}_B} - \boldsymbol{\omega}_B\mathbf{t}\right)\right] \qquad\qquad 6.11$$

where: $\mathbf{C}$ is a permanent complex number. The module of the wave function squared $|\boldsymbol{\Psi}|^2 = \boldsymbol{\Psi}^*\boldsymbol{\Psi} = \mathbf{const}$ is independent on $\mathbf{x}$. This means that the probability to find a particle with permanent $\mathbf{p}$ is equal in any space volume (or it can be localized everywhere). This contradicts the experimental data.

The Quantum Mechanics solve this contradiction assuming the idea of Shrödinger, that particle represents the 'wave packet' with big number of de Broglie waves with different $\mathbf{p} = \hbar\mathbf{k}$, localized in a small interval $\Delta\mathbf{p}$. The amplitude of all this number of de Broglie waves in the packet with spatial dimension $\Delta\mathbf{x} = \lambda_B$ add to each other because of close phase. For the other hand, at the $\Delta\mathbf{x} \gg \lambda_B$ they damper out each other because of phase difference.

The wave packet model can be explained, using eq.6.4 for nonrelativistic particles: $\mathbf{v} \ll \mathbf{c}$ and $\mathbf{R} = \sqrt{1 - (\mathbf{v}/\mathbf{c})^2} \sim 1$. For this case, the carrying internal frequency of $\mathbf{C} \rightleftharpoons \mathbf{W}$ pulsation is much higher, than the external translational de Broglie wave modulation frequency: $\boldsymbol{\omega}_0^{in} \gg \boldsymbol{\omega}_B^{ext}$. The wave packet, consequently, in this case, is formed by the waves, generated by the internal $[\mathbf{C} \rightleftharpoons \mathbf{W}]^{in}$ dynamics, corresponding to *zitterbewegung* (Shrödinger, 1930). However, the wave packet concept itself do not explain the mechanism of $\mathbf{C} \rightleftharpoons \mathbf{W}$ duality.

Our dynamic corpuscle - wave $[\mathbf{C} \rightleftharpoons \mathbf{W}]$ duality theory suggests another possible explanation of the uncertainty principle realization, as a counterphase pulsation of momentum and position, energy and time, described above. The $\mathbf{C} \to \mathbf{W}$ transition is accompanied by conversion of real mass to virtual one, presented by cumulative virtual cloud $\mathbf{CVC}^{\pm}$. As far the energies of both phase $[\mathbf{C}]$ and $[\mathbf{W}]$ are equal, it makes possible to apply the relativistic mechanics to both of them.

### 6.1 The dynamic model of pulsing photon

The model of a photon with integer spin (boson), resulting from fusion (annihilation) of pairs of triplets: electron + positron (see Fig.1), are presented by Fig.3:

$$< [\mathbf{F}^-_\uparrow \bowtie \mathbf{F}^+_\downarrow]_{S=0} + (\mathbf{F}^-_\updownarrow + \mathbf{F}^+_\updownarrow)_{S=\pm 1} + [\mathbf{F}^-_\uparrow \bowtie \mathbf{F}^+_\uparrow]_{S=0} > \qquad 6.12$$

Two side pairs represent a Cooper pairs with zero spin. The central pair $(\mathbf{F}^-_\updownarrow + \mathbf{F}^+_\updownarrow)_{S=\pm 1}$ have uncompensated spin and energy $(\mathbf{E}_{ph} = \mathbf{h}\nu_{ph})$. It determines the properties of photon. Such a structure can originate also, as a result of excitation and fusion of three pairs of asymmetric Bivacuum fermions and antifermions, in the process of transition of the excited state of atom or molecule, i.e. systems: $\sum[electrons + protons]$ to the ground state.

There are *two possible ways* to make the rotation of adjacent sub-elementary fermion and sub-elementary antifermion compatible. One of them is interaction 'side-by-side', like in the 1st and 3d pairs of (6.12). In such a case of Cooper pairs, they are rotating in opposite directions and their angular momenta (spins) compensate each other, turning the resulting spin of such a pair to zero. The resulting energy and charge of such a pair of sub-elementary particle and antiparticle is also zero, because their symmetry shifts with respect to Bivacuum is exactly opposite, compensating each other.

The other way of compatibility is interaction 'head-to-tail', like in a *central pair* of sub-elementary fermions of 7.11a. In this configuration they rotate in the *same direction* and the sum of their spins is: $\mathbf{s} = \pm \mathbf{1}\hbar$. The energy of this pair is a sum of the *absolute values of the energies of sub-elementary fermion and antifermion*, as far their resulting symmetry shift is a sum of the symmetry shifts of each of them.

The energy of photon in Corpuscular phase is a sum of energy of tori of asymmetric sub-elementary fermion and antifermion. Equal to this energy, the energy of the Wave phase $(\mathbf{E}_{ph})_\mathbf{W}$ is determined by the energy of two corresponding cumulative virtual clouds $\mathbf{\varepsilon}_{CVC^+} + \mathbf{\varepsilon}_{CVC^-}$.

The effective mass of photon:

$$\mathbf{m}_{ph} = (\mathbf{m}^+_V + |-\mathbf{m}_{\bar{V}}|) = 2\mathbf{m}^+_\mathbf{V} = \mathbf{2}|-\mathbf{m}_\mathbf{\bar{V}}| = \mathbf{h}\nu_{ph}/\mathbf{c}^2 = \frac{\mathbf{h}}{\mathbf{c}\mathbf{\lambda}_{ph}} \qquad 6.13$$

$\mathbf{\lambda}_{ph} = \mathbf{c}/\mathbf{\nu}_{ph}$ is the photon wave length.

For photon, propagating in space (primordial symmetric Bivacuum) with translational light velocity, the complementary mass is close to zero $|\mathbf{m}_{\bar{V}}| \cong 0$ and potential and kinetic energies are determined by rotational/angle velocity of photon and its central pairs: $\mathbf{v} = \mathbf{L}_{ph}\mathbf{\omega}_{ph}$

$$\mathbf{V} = \mathbf{T}_k = \mathbf{m}^+_{\bar{V}}\mathbf{c}^2 = \frac{\mathbf{m}_0\mathbf{c}^2}{\sqrt{1 - \left(\frac{\mathbf{L}^C_{ph}\mathbf{\omega}_{rot}}{\mathbf{c}}\right)^2}} \qquad 6.14$$

where $\mathbf{L}_{ph} \sim \mathbf{L}^C_0 = \hbar/\mathbf{m}_0\mathbf{c}$ is dimension of photon in corpuscular phase; $\mathbf{\omega}_{rot}$ is the angle frequency of photon in $[C]$ phase rotation around the direction of its propagation.

**Model of photon, as a double
[electron + positron] rotating structure:**
$< 2[\mathbf{F}_\downarrow^- \bowtie \mathbf{F}_\uparrow^+] + (\mathbf{F}_\downarrow^- + \mathbf{F}_\uparrow^+) >_{S=\pm1}$

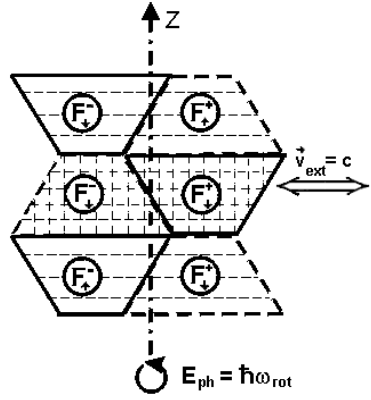

**Fig.3** Model of photon $< 2[\mathbf{F}_\downarrow^- \bowtie \mathbf{F}_\uparrow^+]_{S=0} + (\mathbf{F}_\downarrow^- + \mathbf{F}_\uparrow^+)_{S=\pm1} >$, as result of fusion of electron and positron-like triplets $< [\mathbf{F}_\uparrow^+ \bowtie \mathbf{F}_\downarrow^-] + \mathbf{F}_\uparrow^\pm >$ of sub-elementary fermions , presented on Fig.2. The resulting symmetry shift of such structure is equal to zero, providing the absence or very close to zero rest mass of photon and its propagation in primordial Bivacuum with light velocity or very close to it in the asymmetric secondary Bivacuum.

We may see, that it has axially symmetric configurations in respect to the directions of rotation and propagation, which are normal to each other. These configurations periodically change in the process of sub-elementary fermions and antifermions correlated [*Corpuscle ⇆ Wave*] pulsations in composition of photon (Fig.4). The volume of sextet of sub-elementary fermions in Corpuscular [C] phase is equal to volume, occupied by 6 asymmetric pairs of torus (V$^+$) and antitorus (V$^-$) with geometry of truncated cones and bases: $\mathbf{S}_{V^+} = \pi\mathbf{L}_{V^+}^2$; $\mathbf{S}_{V^-} = \pi\mathbf{L}_{V^-}^2$ (Korn and Korn, 1968):

$$\mathbf{V_C} = 6\mathbf{d}\,\pi(\mathbf{L}_{V^+}^2 + \mathbf{L}_{V^+}\mathbf{L}_{V^-} + \mathbf{L}_{V^-}^2) \qquad 6.15$$

where: $\mathbf{d}$ is the height of truncated cone (eq.2.5a); the radiuses of Compton bases $\mathbf{L}_{V^+}$ and $\mathbf{L}_{V^-}$ and their squares $\mathbf{S}_{V^+}$ and $\mathbf{S}_{V^-}$ of the electron's torus and antitorus can be calculated, using eqs. 3.11.

For the simple case, when the radiuses of torus of sub-elementary fermion and antitorus in paired sub-elementary antifermion in photons are close: $\mathbf{L}_{V^+} \simeq \mathbf{L}_{V^-} \simeq \mathbf{L}_0^6$, then 6.15 turns to: $\mathbf{V_C^0} \simeq 18\,\mathbf{d}\,\pi\mathbf{L}_0^2$.

The volume of Wave phase of photon in general case is much bigger, than that [C] phase. It can be evaluated as a 3D standing wave:

$$\mathbf{V_W} = \frac{3}{8\pi}\lambda_{ph}^3 = \frac{3}{8\pi}\left(\frac{\mathbf{c}}{\mathbf{v}_{ph}}\right)^3 \qquad 6.16$$

The energy density in [C] phase of elementary particles are much higher, than that of [W] phase as far the volume is much less and the energies are equal:

$$\left[\varepsilon_\mathbf{C} = \frac{\mathbf{E_C}}{\mathbf{V_C}} = \frac{\mathbf{m}_V^+\mathbf{v}_{gr}^2}{18\,\mathbf{d}\,\pi\mathbf{L}_0^2}\right] \gg \left[\frac{8\pi\,\mathbf{h}\mathbf{v}_{ph}}{3\,\lambda_{ph}^3} = \frac{\mathbf{E}_W}{\mathbf{V}_W} = \varepsilon_\mathbf{W}\right] \qquad 6.17$$

$$at\ \ \lambda_{ph} > \mathbf{L}_0 \qquad 6.17a$$

The corresponding small charge density of the electrons, protons, deuterons, ions, etc.

in the Wave phase make it easier the overcoming of Coulomb repulsion threshold in catalytic reactions and cold fusion (http://arxiv.org/abs/physics/0207027).

The expanded Wave phase in contrast to compact Corpuscular phase represents a big number ($\mathbf{N}_{BVF}$) of Bivacuum fermions and antifermions in the volume of wave [W] phase $\mathbf{V}_W$ with resulting symmetry shift and uncompensated energy:

$$\mathbf{c}^2 \int_0^{\mathbf{m}_{ph}} [(\mathbf{m}_V^+ - \mathbf{m}_V^-)]_\mathbf{W} \mathbf{d}\Delta\mathbf{m}_V^{\pm} = [\mathbf{m}_{ph}\mathbf{c}_{gr}^2]_\mathbf{C} = \mathbf{hc}^2/(\mathbf{v}_{ph}\lambda_{ph}^2)_W = \mathbf{hv}_{ph} \qquad 6.18$$

For photon in primordial symmetric Bivacuum its group and phase velocities are equal: $\mathbf{v}_{gr} = \mathbf{v}_{ph} = c$. This means that the average kinetic and potential energies are also equal: $\mathbf{T}_k = \mathbf{V}_p$. In the process of $\mathbf{C} \rightleftharpoons \mathbf{W}$ pulsation the rotational-translational local kinetic energy of photon: $\mathbf{m}_0\omega_0\mathbf{L}_0^2 = \mathbf{m}_0\,\mathbf{v}_{gr}\mathbf{v}_{ph}$ in [C] phase turns to non-local mass and charge symmetry shift of Bivacuum dipoles in volume of [W] phase and their corresponding curled rotation around the direction of photon propagation. The *electric component* of electromagnetic field reflects the properties of [C] phase and the *magnetic component* - the properties of [W] phase.

The clockwise and counter clockwise rotation of photons in [C] phase around the z-axis (fig.2) stands for two possible polarizations of photon.

### 6.2 New interpretation of Shrödinger equation and general shape of wave function, describing both the external and internal dynamics of particle

The stationary Shrödinger equation can be easily derived from universal for homogeneous medium wave equation:

$$\nabla^2\Phi(r,t) - \frac{1}{\mathbf{v}^2}\frac{\partial\Phi(r,t)}{\partial t^2} = 0 \qquad 6.19$$

where $\Phi(r,t)$ is the wave amplitude (scalar), depending distance from source (r) and time (t) in the process of its propagation with permanent velocity ($\mathbf{v}$). One of possible form of time and space dependent wave function is like (6.11):

$$\Phi(r,t) = \mathbf{C}\exp\left[i\left(\frac{\mathbf{x}}{\mathbf{L}_B} - \omega_\mathbf{B}\mathbf{t}\right)\right] = \mathbf{C}\exp\left(i\frac{\mathbf{x}}{\mathbf{L}_B}\right)\exp(-i\omega_\mathbf{B}\mathbf{t}) \qquad 6.19a$$

In the case of harmonic dependence of the wave amplitude on time with angle frequency $\omega$, it can be presented as:

$$\Phi(r,t) = \Phi(r)\exp(-i\omega t) \qquad 6.20$$

Putting 6.20 to 6.19, we get the following equation.

$$\nabla^2\Phi^{m,e}(r) + \mathbf{k}^2\Phi^{m,e}(r) = 0 \qquad 6.21$$

where $\mathbf{k}$ is a wave number ($\mathbf{k} = \omega/\mathbf{v} = 2\pi/(\mathbf{vT}) = 2\pi/\lambda = 1/\mathbf{L}$).

The conversion of (6.21) to form describing corpuscle-wave duality can be done using de Broglie relations:

$$\mathbf{k} = \mathbf{p}/\hbar = 2\pi/\mathbf{L}_B; \qquad \mathbf{L}_B = \hbar/\mathbf{p} \qquad 6.22$$

$$\mathbf{k}^2 = \mathbf{p}^2/\hbar^2 = (2\pi/\mathbf{L}_B)^2 = 1/\lambda_B^2 \qquad 6.22a$$

in stationary conditions, when the total energy of de Broglie wave, equal to sum of its external kinetic ($\mathbf{T}_k$) and potential ($\mathbf{V}$) energies, is time-independent, like in standing

waves, for example:

$$\mathbf{E} = \mathbf{T}_k + \mathbf{V} = \frac{\mathbf{p}^2}{2\mathbf{m}} + \mathbf{V} = \mathbf{const} \qquad 6.23$$

$$or : \ \mathbf{p}^2 = 2\mathbf{m}(\mathbf{E} - \mathbf{V}) \qquad 6.23a$$

The de Broglie wave number squared from 6.22a and 6.23a is

$$\mathbf{k}^2 = (2\mathbf{m}/\hbar)(\mathbf{E} - \mathbf{V}) \qquad 6.24$$

Combining 6.24 with 6.21 we get the *stationary* Shrödinger equation:

$$\nabla^2 \Phi(r) + (2\mathbf{m}/\hbar)(\mathbf{E} - \mathbf{V})\Phi(r) = 0 \qquad 6.25$$

It has solutions for continuous wave function, existing as *eigenfunctions* only at certain discreet *eigenvalues* of energy ($\mathbf{E}_n$). It was shown by Shrödinger, that spectra of these energies of the electron in potential electric field ($\mathbf{V}$) describes correctly the absorption spectra of hydrogen atoms.

The time-dependent form of Shrödinger equation includes the time and space dependent wave function, like (6.19a):

$$\Phi(\mathbf{r}, t) = \Phi(\mathbf{r}) \exp(-i\mathbf{E}t/\hbar) = \mathbf{C} \exp\left(i\frac{\mathbf{X}}{\mathbf{L}_B}\right) \exp(-i\boldsymbol{\omega}_{\mathbf{B}}\mathbf{t}) \qquad 6.26$$

The corresponding equation can be presented as:

$$-\frac{\hbar}{i} \frac{\partial \Phi(\mathbf{r}, t)}{\partial t} = \left(-\frac{\hbar}{2\mathbf{m}} \nabla^2 + \mathbf{V}\right) \Phi(\mathbf{r}, t) \qquad 6.27$$

The inertial mass in 6.27, in accordance to our Unified theory, is equal to the actual mass of unpaired/uncompensated sub-elementary fermion of elementary particle: $\mathbf{m} = \mathbf{m}_{\bar{V}}^+$.

The properties of stationary wave function $\Phi(\mathbf{r})$ and time-dependent $\Phi(\mathbf{r}, t)$ should be the same, i.e. they are *continuous, single-valued and finitesimal*. The product of wave function with its *complex conjugate* function, characterize the density of probability of particle location in this point of space at certain time moment:

$$\Phi(\mathbf{r}, t)\Phi^*(\mathbf{r}, t) = |\Phi(\mathbf{r}, t)|^2 \qquad 6.28$$

In solutions of Shrödinger equation the certain eigenvalues of energy ($\mathbf{E}_n$) corresponds to eigenfunctions ($\Phi_n$), describing *anchor sites (primary and secondary)* of elementary particles in their corpuscular [C] phase.

It follows from our theory of wave-corpuscle duality, that de Broglie wave length ($\boldsymbol{\lambda}_B = 2\boldsymbol{\pi}\mathbf{L}_B$) and its frequency ($\boldsymbol{\omega}_{\mathbf{B}}$), as a crucial parameters of wave function (6.26), are determined by properties of the *anchor Bivacuum fermions* of uncompensated sub-elementary fermions of the electron or proton or bosons, like photon.

From eqs.6.4, 6.4a and 6.5 we can see, that the *external* de Broglie wave frequency ($\boldsymbol{\omega}_B^{ext}$) and wave number ($\mathbf{k}_B$) of particle can be expressed via *internal* ($\boldsymbol{\omega}_0^{in}$), *total* ($\boldsymbol{\omega}_{\mathbf{C} \rightleftharpoons \mathbf{W}}$) frequencies and corresponding energies as:

$$\boldsymbol{\omega}_B^{ext} = \frac{1}{\hbar}\left[(\mathbf{m}_{\bar{V}}^+ - \mathbf{m}_{\bar{V}}^-)_{anc}^{ext}\mathbf{c}^2\right]_{tr} = \boldsymbol{\omega}_{\mathbf{C} \rightleftharpoons \mathbf{W}} - \mathbf{R}\boldsymbol{\omega}_0^{in} \qquad 6.29$$

$$or : \ \mathbf{k}_B = \frac{1}{\mathbf{L}_B} = \frac{\mathbf{c}}{\hbar}\left[\mathbf{m}_{\bar{V}}^+(\mathbf{m}_{\bar{V}}^+ - \mathbf{m}_{\bar{V}}^-)\right]^{1/2} = \frac{\mathbf{c}}{\hbar}\left[\mathbf{m}_{\bar{V}}^+(\mathbf{m}_{\bar{V}}^+ - \mathbf{R}\,\mathbf{m}_0)\right]_{tr}^{1/2} \qquad 6.30$$

where relativistic factor: $\mathbf{R} = \sqrt{1 - (\mathbf{v}/\mathbf{c})^2}$ is dependent on the external translational group velocity (v); $\mathbf{m}_V^+ = \mathbf{m}_0/\mathbf{R}$;     $\mathbf{m}_V^- = \mathbf{R}\,\mathbf{m}_0$.

At $\mathbf{v} \to \mathbf{c}$, the $\mathbf{R} \to 0$, the rest mass contribution decreases and $\boldsymbol{\omega}_B^{ext} \to \boldsymbol{\omega}_{\mathbf{C} \rightleftharpoons \mathbf{W}}$ and $\mathbf{k}_B \to (\mathbf{m}_V^+ \mathbf{c}/\hbar)$.

The mass and charge symmetry shifts of asymmetric Bivacuum fermions and antifermions are interrelated (eqs. 4.7- 4.8 in: http://arxiv.org/abs/physics/0103031):

$$\Delta \mathbf{m}_V^+ = (\mathbf{m}_V^+ - \mathbf{m}_V^-) = \mathbf{m}_V^+ \left(\frac{\mathbf{v}}{\mathbf{c}}\right)^2 \qquad 6.31$$

$$\Delta \mathbf{e}_\pm = (\mathbf{e}_+ - \mathbf{e}_-) = \frac{\Delta \mathbf{m}_V^+ \mathbf{e}_+^2}{\mathbf{m}_V^+(\mathbf{e}_+ + \mathbf{e}_-)} = \left(\frac{\mathbf{v}}{\mathbf{c}}\right)^2 \frac{\mathbf{e}_+^2}{\mathbf{e}_+ + \mathbf{e}_-} \qquad 6.31a$$

where the *actual* charge ($\mathbf{e}_+$), in accordance to eq.4.5 from (Kaivarainen, http://arxiv.org/abs/physics/010303) has the following relativistic dependence on the external velocity of Bivacuum dipoles:

$$\mathbf{e}_+ = \frac{\mathbf{e}_0}{[1 - \mathbf{v}^2/\mathbf{c}^2]^{1/4}} \qquad 6.31b$$

The complementary charge ($\mathbf{e}_-$) can be calculated from the earlier obtained relation (eq. 4.18a from http://arxiv.org/abs/physics/010303): $|\mathbf{e}_+\mathbf{e}_-| = \mathbf{e}_0^2$.

Using the relations above, we may present the dimensionless coefficient of wave function (C) in (6.26), as a maximum symmetry shift of the *anchor* Bivacuum fermion, reduced to the rest mass ($\mathbf{m}_0$) and rest charge ($\mathbf{e}_0$):

$$\mathbf{C_m} = \Delta \mathbf{m}_V^+/\sqrt{2}\,\mathbf{m}_0 = (\mathbf{m}_V^+ - \mathbf{m}_V^-)/\sqrt{2}\,\mathbf{m}_0 = \frac{\mathbf{m}_V^+}{\sqrt{2}\,\mathbf{m}_0}\left(\frac{\mathbf{v}}{\mathbf{c}}\right)^2 \qquad 6.32$$

$$\mathbf{C_e} = \Delta \mathbf{e}_\pm/\sqrt{2}\,\mathbf{e}_0 = (\mathbf{e}_+ - \mathbf{e}_-)/\sqrt{2}\,\mathbf{e}_0 = \left(\frac{\mathbf{v}}{\mathbf{c}}\right)^2 \frac{\mathbf{e}_+^2/\sqrt{2}\,\mathbf{e}_0}{\mathbf{e}_+ + \mathbf{e}_-} \qquad 6.32a$$

We have to keep in mind, that the complementary mass and charge are undetectable directly and we may consider them as imaginary ones: $i\mathbf{m}_V^-$ and $i\mathbf{e}_-$. Consequently, using 6.30; 6.30a and 6.32, we may present the wave function (6.26) and its complex conjugate in terms of Bivacuum dipoles symmetry shifts for understanding the mechanism of particle internal dynamics and its propagation in space:

$$\Phi(\mathbf{r},t) = \mathbf{C}\exp\left(i\frac{\mathbf{X}}{\mathbf{L}_B}\right)\exp(-i\boldsymbol{\omega}_\mathbf{B}\mathbf{t}); \qquad \Phi^*(\mathbf{r},t) = \mathbf{C}\exp\left(-i\frac{\mathbf{X}}{\mathbf{L}_B}\right)\exp(i\boldsymbol{\omega}_\mathbf{B}\mathbf{t}) \qquad 6.33$$

$$\Phi(\mathbf{r},t) = \frac{\mathbf{m}_V^+ - i\mathbf{m}_V^-}{\sqrt{2}\,\mathbf{m}_0}\exp\left[i\frac{\mathbf{X}}{\hbar}\mathbf{c}\,[\mathbf{m}_V^+(\mathbf{m}_V^+ - i\mathbf{m}_V^-)]^{1/2}\right]\exp\left\{-i\frac{1}{\hbar}[(\mathbf{m}_V^+ - i\mathbf{m}_V^-)\mathbf{c}^2]_{tr}\mathbf{t}\right\} = \qquad 6.33a$$

$$\Phi(\mathbf{r},t) = \frac{\mathbf{m}_V^+ - i\mathbf{R}\,\mathbf{m}_0}{\sqrt{2}\,\mathbf{m}_0}\exp\left[i\frac{\mathbf{X}}{\hbar}\mathbf{c}\,[\mathbf{m}_V^+(\mathbf{m}_V^+ - i\mathbf{R}\,\mathbf{m}_0)]^{1/2}\right]\exp\left\{-i\frac{1}{\hbar}[(\mathbf{m}_V^+ - i\mathbf{R}\,\mathbf{m}_0)\mathbf{c}^2]_{tr}\mathbf{t}\right\} \qquad 6.33b$$

$$\Phi^*(\mathbf{r},t) = \frac{\mathbf{m}_V^+ + i\mathbf{m}_V^-}{\sqrt{2}\,\mathbf{m}_0}\exp\left[i\frac{\mathbf{X}}{\hbar}\mathbf{c}\,[\mathbf{m}_V^+(\mathbf{m}_V^+ + i\mathbf{m}_V^-)]^{1/2}\right]\exp\left\{-i\frac{1}{\hbar}[(\mathbf{m}_V^+ + i\mathbf{m}_V^-)\mathbf{c}^2]_{tr}\mathbf{t}\right\} = \qquad 6.34$$

$$\Phi^*(\mathbf{r},t) = \frac{\mathbf{m}_V^+ + i\mathbf{R}\,\mathbf{m}_0}{\sqrt{2}\,\mathbf{m}_0}\exp\left[i\frac{\mathbf{X}}{\hbar}\mathbf{c}\,[\mathbf{m}_V^+(\mathbf{m}_V^+ + i\mathbf{R}\,\mathbf{m}_0)]^{1/2}\right]\exp\left\{-i\frac{1}{\hbar}[(\mathbf{m}_V^+ + i\mathbf{R}\,\mathbf{m}_0)\mathbf{c}^2]_{tr}\mathbf{t}\right\} \qquad 6.34a$$

From 6.33b and 6.34a it follows, that at $\mathbf{v} = \mathbf{c}$ and $\mathbf{R} = \mathbf{0}$ these wave functions turn to that, describing the photons with effective mass $\mathbf{m}_V^+ = \hbar\omega/\mathbf{c}^2$; and frequency $\boldsymbol{\omega} = \frac{1}{\hbar}[\mathbf{m}_V^+\mathbf{c}^2]_{tr}$.

$$\Phi(\mathbf{r},t) = \Phi^*(\mathbf{r},t) = \frac{\mathbf{m}_{\bar{V}}^+}{\sqrt{2}\,\mathbf{m}_0}\exp\left[i\frac{\mathbf{x}}{\hbar}\mathbf{m}_{\bar{V}}^+\mathbf{c}\right]\exp\left\{-i\frac{1}{\hbar}[\mathbf{m}_{\bar{V}}^+\mathbf{c}^2]_{tr}\mathbf{t}\right\} \qquad 6.35$$

The product of the conventional forms of complex conjugate wave functions (6.33) gives the space and time independent pre-exponential coefficient squared: $|\Phi(\mathbf{r},t)|^2 = \mathbf{C}^2 = const.$

From product of 6.33b and 6.34a we get the new general formula for density of probability of particle, dependent on space and time $|\Phi(\mathbf{r},t)|^2$:

$$|\Phi(\mathbf{r},t)|^2 = \Phi(\mathbf{r},t)\Phi^*(\mathbf{r},t) = \qquad\qquad 6.36$$

$$= \frac{(\mathbf{m}_{\bar{V}}^+)^2 + (\mathbf{m}_{\bar{V}}^-)^2}{2\mathbf{m}_0^2}\exp\left[i\frac{\sqrt{2}\,\mathbf{x}}{\hbar}\mathbf{m}_{\bar{V}}^+\mathbf{c}\right]\exp\left\{-i\frac{(2\mathbf{m}_{\bar{V}}^+)\mathbf{c}^2}{\hbar}\mathbf{t}\right\}$$

The resulting energy of this state is characterized by the length of hypotenuse of triangle with adjacent cathetus squared:

$$\mathbf{E}_{\mathbf{V}^+ \mathbf{0}\mathbf{V}^-}^{Res} = \mathbf{m}_{\mathbf{V}^+ \mathbf{0}\mathbf{V}^-}^{\pm}\mathbf{c}^2 = \sqrt{(\mathbf{m}_{\bar{V}}^+)^2 + (\mathbf{m}_{\bar{V}}^-)^2}\,\mathbf{c}^2 \qquad 6.37$$

It is important to point out, that in state of rest, when the external translational velocity of elementary particle is zero ($\mathbf{v} = \mathbf{0}$), the real and complementary mass are equal to the rest mass: $\mathbf{m}_{\bar{V}}^+ = \mathbf{m}_{\bar{V}}^- = \mathbf{m}_0$, the external de Broglie wave length tends to infinity ($\lambda_B = 2\pi L_B = \infty$) and its frequency to zero ($\omega_B = 0$), the wave function, described by conventional expression (6.19a) becomes equal to coefficient $\mathbf{C}$. This coefficient itself, as a square root of pre-exponential factor $\mathbf{C} = \sqrt{\frac{(\mathbf{m}_{\bar{V}}^+)^2 + (\mathbf{m}_{\bar{V}}^-)^2}{2\mathbf{m}_0^2}}$ at these conditions is equal to $\mathbf{C} = \mathbf{1}$. The corresponding density of probability describing only the external properties of particle $\mathbf{C}^2 = \mathbf{1}$ is a permanent value, independent on space and time.

However, the general expression of density of probability (6.36) of particle location in selected point of space-time, when its external translational velocity is equal to zero ($\mathbf{v}^{ext} = \mathbf{0}$), following from our theory, turns to:

$$|\Phi(\mathbf{r},t)|^2 = \exp\left(i\sqrt{2}\,\frac{\mathbf{x}}{\mathbf{L}_0}\right)\exp\left(-\frac{i}{\pi}\boldsymbol{\omega}_0\mathbf{t}\right) \qquad 6.38$$

where the Compton wave length and frequency of particle are equal, correspondingly, to:

$$\mathbf{L}_0 = \frac{\mathbf{c}}{\boldsymbol{\omega}_0} = \frac{\hbar}{\mathbf{m}_0\mathbf{c}} \quad \text{and} \quad \boldsymbol{\omega}_0 = \frac{\mathbf{m}_0\mathbf{c}^2}{\hbar} \qquad 6.38a$$

We can see, that the general expression of density probability of particle in [C] phase location (6.38), in contrast to conventional, the permanent one, is oscillating due to internal $[\mathbf{C} \rightleftharpoons \mathbf{W}]_{in}$ pulsation of sub-elementary fermions, rotating around common axes, as presented in Fig.1 and Fig.3. At fixed coordinate ($\mathbf{x}$), the probability of particle in [C] phase location is dependent on time, i.e. phase of pulsation. At fixed time ($\mathbf{t}$) this probability is dependent on coordinate of particle in [C] phase location.

### 6.3 The mechanism of free particle propagation in space

The propagation of elementary particles (triplets-fermions $< [\mathbf{F}_{\uparrow}^+ \bowtie \mathbf{F}_{\downarrow}^-] + \mathbf{F}_{\updownarrow}^{\pm} >^{e,p}$ or sextets-bosons $< \mathbf{2}[\mathbf{F}_{\uparrow}^- \bowtie \mathbf{F}_{\downarrow}^+]_{S=0} + (\mathbf{F}_{\updownarrow}^- + \mathbf{F}_{\updownarrow}^+)_{S=\pm 1} >^{ph}$) in pure Bivacuum or in Bivacuum, perturbed by matter, transparent for these particles, can be considered as a two stage

process:

**Stage I**: It corresponds to elementary particle state, when the unpaired/uncompensated sub-elementary fermions $\mathbf{F}_{\updownarrow}^{\pm} >^{e,p}$ or $(\mathbf{F}_{\updownarrow}^{-} + \mathbf{F}_{\updownarrow}^{+})_{S=\pm 1} >^{ph}$ are in [C] phase and compensated each other in pairs $[\mathbf{F}_{\uparrow}^{+} \bowtie \mathbf{F}_{\downarrow}^{-}]$ are in [W] phase. On this stage the elastic deformational waves in Bivacuum matrix and corresponding reversible Bivacuum dipoles symmetry shifts, provided by uncompensated sub-elementary fermions in [C] phase, stand for *kinetic* energy and momentum contributions to big number of secondary *anchor sites* of elementary particle in Bivacuum matrix, using its *nonlocal* properties. The wave [W] phase of symmetric pairs $[\mathbf{F}_{\uparrow}^{-} \bowtie \mathbf{F}_{\downarrow}^{+}]_{S=0}$ transfer the *potential* energy to the same secondary *anchor sites.* The anchor sites corresponds to particle's eigenfunctions and corpuscular eigen states.

These eigenfunctions are alternative, i.e. incompatible with each other - *orthogonal*.

The energy conservation law demands, that the resulting energy of all activated anchor sites should be zero. It is possible, if we assume that all *anchor sites* (**AS**) are composed from two or three pairs of conjugated and correlated Cooper pairs of asymmetric Bivacuum fermions:

$$\mathbf{AS} = \sum^{N} 3[\mathbf{BVF}_{\downarrow}^{-} \bowtie \mathbf{BVF}_{\uparrow}^{+}]_n \longrightarrow \sum^{N} 3[\mathbf{F}_{\downarrow}^{-} \bowtie \mathbf{F}_{\uparrow}^{+}]_n \qquad 6.39$$

The opposite asymmetry of Bivacuum fermions and antifermions, forming virtual Cooper pairs, is provided by their rotation around common basic axis. Such anchor sites are suitable for accommodation of the electrons, positrons and photons.

**Stage II**: Corresponds to particle state, when the unpaired/uncompensated sub-elementary fermions $\mathbf{F}_{\updownarrow}^{\pm} >^{e,p}$ or $(\mathbf{F}_{\updownarrow}^{-} + \mathbf{F}_{\updownarrow}^{+})_{S=\pm 1} >^{ph}$ are in expanded [W] phase, representing cumulative virtual cloud ($\mathbf{CVC}^{\pm}$), modulated by de Broglie wave of particles. The symmetric pairs $[\mathbf{F}_{\uparrow}^{+} \bowtie \mathbf{F}_{\downarrow}^{-}]$ on this stage are in the compact [C] phase.

The jumps of the triplets (fermions) or sextet (photons) with group velocity of wave packet to one of prepared by previous stage *anchor sites* occur on this stage. The properties of the anchor site can change after complex formation with particle, however without violation of energy conservation and energy dissipation.

The most probable distance of such 'jump' is determined by de Broglie wave length of particle ($\lambda_B = h/\mathbf{p}$), equal to that of cumulative virtual cloud of uncompensated sub-elementary fermions and the most probable direction of jump coincide with particle momentum in its [C] phase. However the new location of particle, as only one of many possible, is not predetermined and 'jumps' can be considered as the stochastic process.

*The principle of superposition in quantum mechanics has the same formal shape as in classical mechanics:*

$$\Phi(\mathbf{r},t) = c_1 \Phi(\mathbf{r},t)_1 + c_2 \Phi(\mathbf{r},t)_2 + \dots c_n \Phi(\mathbf{r},t)_n \qquad 6.40$$

where: $c_n$ are arbitrary complex numbers; $\Phi(\mathbf{r},t)_n$ is wave function, describing different ($n$) states of quantum system. In accordance to our theory these states correspond to possible *anchor sites* of moving in space particle.

However, in contrast to state/wave superposition of classical systems, in quantum system any state is not the result of 'mixing' of other states, but always the alternative or *orthogonal*, i.e. only one state of many possible.

## 7  The Problem of Time, 'Free Energy' Source and Fields origination in the Framework of Unified Theory

The dimensionless **pace of time** for any closed system is determined by the pace of its kinetic energy change (anisotropic in general case), related to change of Bivacuum Tuning Energy (Kaivarainen, http://arxiv.org/abs/physics/0103031):
$\mathbf{TE}^i = |\mathbf{m}_V^+ \mathbf{c}^2 - \mathbf{q}\,\mathbf{m}_0 \mathbf{c}^2|^i = \hbar|\boldsymbol{\omega}_{C \rightleftharpoons W} - \mathbf{q}\,\boldsymbol{\omega}_0^i|$, introduced, as a difference between the total energy of particle, related to frequency of its $[\mathbf{C} \rightleftharpoons \mathbf{W}]$ pulsation and fundamental energy of Bivacuum $(\mathbf{n}\,\mathbf{m}_0\mathbf{c}^2)$:

$$[\mathbf{dt}/\mathbf{t} = \mathbf{d}\ln \mathbf{t} = -\mathbf{d}\ln \mathbf{T}_k]_{x,y,z} = -\mathbf{d}\ln[(1 + \mathbf{R})\,\mathbf{TE}^i]_{x,y,z} \qquad 7.1$$

$$= -\mathbf{d}\ln[(1 + \mathbf{R})|\boldsymbol{\omega}_{C \rightleftharpoons W} - \mathbf{q}\,\boldsymbol{\omega}_0^i|]_{x,y,z} \qquad 7.1a$$

where: $\mathbf{R} = [1 - (\mathbf{v}/\mathbf{c})^2]^{1/2}$ is relativistic factor; $\mathbf{q} = 1, 2, 3\ldots$ is quantization number of $\mathbf{VPW}^\pm$ energy $\mathbf{E}_{VPW} = \mathbf{q}\,\mathbf{m}_0^i \mathbf{c}^2 = \mathbf{q}\,\hbar\,\boldsymbol{\omega}_0^i$ and fundamental Compton frequency: $\boldsymbol{\omega}_0^i = \mathbf{m}_0^i \mathbf{c}^2/\hbar$; $\mathbf{T}_k^\pm = \mathbf{m}_V^\pm \mathbf{v}^2/2$ is kinetic energy of particle.

Using relation (7.1), the pace of the internal time and time itself for closed system of particles can be presented via their acceleration and velocity:

$$\left[ \frac{\mathbf{dt}}{\mathbf{t}} = \mathbf{d}\ln \mathbf{t} = -\frac{\mathbf{d}\overrightarrow{\mathbf{v}}}{\overrightarrow{\mathbf{v}}} \frac{2 - (\mathbf{v}/\mathbf{c})^2}{1 - (\mathbf{v}/\mathbf{c})^2} \right]_{x,y,z} \qquad 7.1b$$

$$\left[ \mathbf{t} = -\frac{\overrightarrow{\mathbf{v}}}{\mathbf{d}\overrightarrow{\mathbf{v}}/\mathbf{dt}} \frac{1 - (\mathbf{v}/\mathbf{c})^2}{2 - (\mathbf{v}/\mathbf{c})^2} = -\frac{\mathbf{dt}}{\mathbf{dT}_k} \mathbf{T}_k \right]_{x,y,z} \qquad 7.1c$$

The time for each selected closed system of particles is a parameter, characterizing the average velocity and acceleration of these particles, i.e. this system dynamics. The pace of time is zero and time - infinitive, if kinetic energy of closed system is permanent. This means the infinitive life-time of torus and antitorus of our Bivacuum dipoles, following from 1st postulate of Unified theory (see eq. 3.1). The pace of time and time itself are uncertain ($\mathbf{dt}/\mathbf{t} = \mathbf{0}/\mathbf{0}$; $\mathbf{t} = \mathbf{0}/\mathbf{0}$), if the translational velocity is permanent and equal to zero ($\mathbf{v} = \mathbf{0} = \mathbf{const}$).

A lot of experimental results, like Kozyrev's ones, incompatible with existing paradigm, find the explanations in terms of our theory, confirming the existence of *new kind of Bivacuum mediated nonlocal interaction between distant systems, realized via Bivacuum mediated interaction (BMI), including nonlocal Virtual Guides* ($\mathbf{VirG}_{SME}^{ext}$) of spin, momentum and energy between Sender (nonequilibrium system) and Receiver. The worm hole like structures $\mathbf{VirG}_{SME}^{ext}$ can be formed by Bivacuum bosons or Cooper pairs of Bivacuum fermions. Synchronization of $[\mathbf{C} \rightleftharpoons \mathbf{W}]$ pulsation between remote elementary particles under the action of all-pervading Bivacuum virtual pressure waves ($\mathbf{VPW}^\pm$) is important factor in quantum entanglement.

*The main source of 'free' energy of Bivacuum, used by overunity devices,* is forced combinational resonance of de Broglie waves of real elementary particles with fundamental Bivacuum virtual pressure waves ($\mathbf{VPW}^\pm$) of basic quantized frequency $\boldsymbol{\omega}_0^i = (\mathbf{m}_0^i \mathbf{c}^2)/\hbar = \boldsymbol{\omega}_{rot}^i$ . The condition of combinational resonance is:

$$\boldsymbol{\omega}_{\mathbf{VPW}} = \mathbf{q}\,\boldsymbol{\omega}_0^i = \mathbf{k}\,\boldsymbol{\omega}_{C \rightleftharpoons W} \qquad 7.2$$

$$or: \quad \mathbf{E}_{VPW} = \mathbf{n}\,\mathbf{m}_0^i \mathbf{c}^2 = \mathbf{k}\,\mathbf{m}_V^+ \mathbf{c}^2 \qquad 7.2a$$

where: $\mathbf{q}$ and $\mathbf{k}$ are the integer numbers.

The energy exchange between $\mathbf{VPW}^+ + \mathbf{VPW}^-$ and real particles in the process of $[\mathbf{C} \rightleftharpoons \mathbf{W}]$ pulsation of pairs $[\mathbf{F}_\uparrow^+ \bowtie \mathbf{F}_\downarrow^-]_{x,y}$ of triplets $< [\mathbf{F}_\uparrow^+ \bowtie \mathbf{F}_\downarrow^-]_{x,y} + \mathbf{F}_\updownarrow^\pm >_z^i$ at *pull-in*

-*range* state accelerate them, driving to resonant conditions (7.2 and 7.2a).

In accordance to rules of combinational resonance of Bivacuum virtual pressure waves with elementary particles, we have the following relation between quantized energy and frequency of $\mathbf{VPW}^\pm$ and energy/frequency of triplets $\mathbf{C} \rightleftharpoons \mathbf{W}$ pulsation in resonance conditions:

$$\mathbf{E_{VPW^\pm}} = \hbar\boldsymbol{\omega}^i_{\mathbf{VPW^\pm}} = q\hbar\boldsymbol{\omega}^i_0 = \hbar\boldsymbol{\omega}^i_{\mathbf{C \rightleftharpoons W}} = \mathbf{R}\,\hbar\boldsymbol{\omega}_0 + \hbar\boldsymbol{\omega}_B = \mathbf{R}\,\hbar\boldsymbol{\omega}_0 + \mathbf{h}^2/2\mathbf{m}_V^+\boldsymbol{\lambda}_B^2 \qquad 7.3$$

$$or: \quad \mathbf{q}\,\mathbf{m}_0^i\mathbf{c}^2 = \mathbf{R}\,\mathbf{m}_0^i\mathbf{c}^2 + \mathbf{m}_V^+\mathbf{v}^2 = \mathbf{R}\,\mathbf{m}_0^i\mathbf{c}^2 + \frac{\mathbf{m}_0^i\mathbf{c}^2(\mathbf{v}/\mathbf{c})^2}{\mathbf{R}} \qquad 7.3a$$

$$\mathbf{R} = \sqrt{1 - (\mathbf{v}/\mathbf{c})^2}; \quad \mathbf{q} = 1, 2, 3\dots \text{(integer numbers)}$$

where the angle frequency of de Broglie waves is: $(\boldsymbol{\omega}_B)_{1,2,3}$ their length is: $\boldsymbol{\lambda}_B = 2\boldsymbol{\pi}\mathbf{L}_B^2 = \mathbf{h}/\mathbf{m}_V^+\mathbf{v}$ and:

$$\boldsymbol{\omega}_B = \hbar/2\mathbf{m}_V^+\mathbf{L}_B^2 = \mathbf{m}_V^+\mathbf{v}^2/2\hbar \qquad 7.4$$

The relation between translational most probable velocity of particle and quantization number ($\mathbf{q}$), corresponding to resonant interaction of Bivacuum $\mathbf{VPW}^\pm$ with pulsing particles, is:

$$\mathbf{v} = \mathbf{c}\left(\frac{\mathbf{q}^2 - 1}{\mathbf{q}^2}\right)^{1/2} \qquad 7.5$$

At the conditions of triplets fusion, when $\mathbf{q} = 1$, the *translational* velocity of particle is zero: $\mathbf{v}_{q=1} = \mathbf{0}$. When the quantized energy of $\mathbf{VPW}^\pm_q$, corresponds to $\mathbf{q} = 2$, the resonant translational velocity of particle should be: $\mathbf{v}_{q=2} = \mathbf{c} \cdot \mathbf{0.866} = \mathbf{2,6} \cdot \mathbf{10}^{10}$ cm/s. At $\mathbf{q} = 3$, we have from (7.5): $\mathbf{v}_{q=3} = \mathbf{c} \cdot \mathbf{0.942} = \mathbf{2,83} \cdot \mathbf{10}^{10}$ cm/s.

It is anticipated, that at the velocity, corresponding to $\mathbf{n} < \mathbf{1,5}$, the interaction of pulsing particles with basic $\mathbf{VPW}^\pm_{n=1}$ after forcing of oscillation should slow down the their velocity, driving translational mobility to resonant conditions: $\mathbf{q} = 1, \ \mathbf{v} \rightarrow \mathbf{0}$. The 2nd and 3d laws of thermodynamics, reflecting the 'spontaneous' cooling of matter can be a consequence of just this conditions.

For the other hand, if velocity of particles corresponds to $\mathbf{q} > \mathbf{1,5}$, their pull-in range interaction with $\mathbf{VPW}^\pm_{n=2}$ can accelerate them up to conditions: $\mathbf{q} = 2, \ \mathbf{v} \rightarrow \mathbf{2,6} \cdot \mathbf{10}^{10}$ cm/s.

Virtual particles and antiparticles have mass $\left|\widetilde{\mathbf{m}_V^+} - \widetilde{\mathbf{m}_V^-}\right| < \mathbf{m}_0$ and charge $|\widetilde{\mathbf{e}}_+ - \widetilde{\mathbf{e}}_-| < \mathbf{e}_0$ symmetry shift smaller, than the rest mass and charge of real sub-elementary fermions and antifermions. The frequency of their $[\widetilde{\mathbf{C}} \rightleftharpoons \widetilde{\mathbf{W}}]$ pulsation is lower than fundamental Compton's one: $\boldsymbol{\omega}^i_{\widetilde{\mathbf{C}} \rightleftharpoons \widetilde{\mathbf{W}}} < \boldsymbol{\omega}^i_0 = \mathbf{m}_0^i\mathbf{c}^2/\hbar$. These reasons prevent the fusion of virtual particles and antiparticles to triplets. The action of fundamental virtual pressure waves ($\mathbf{VPW}^\pm$) of Bivacuum with Compton frequency ($\boldsymbol{\omega}^i_0$) on groups of Cooper pairs of *virtual* particles: $\mathbf{N}[\mathbf{BVF}^\uparrow \bowtie \mathbf{BVF}^\downarrow]^{as}$ ($\mathbf{N} \gg \mathbf{1}$), forming big vortices with radiuses: $\left[\mathbf{R} = \hbar/\left|\widetilde{\mathbf{m}_V^+} - \widetilde{\mathbf{m}_V^-}\right|^i\mathbf{c}\right] \gg \left[\mathbf{L}_0^i = (\mathbf{L}^+\mathbf{L}^-)^{1/2} = \hbar/(\mathbf{m}_0\mathbf{c})\right]$, can *accelerate their rotation around common axis*, driving to Golden mean conditions from the lower limit: $\mathbf{v}^{ext}_{rot} \rightarrow \phi^{1/2}\mathbf{c}$. It is another - *virtual source* of superfluous energy of Bivacuum, standing for mechanism of overunity devices. This acceleration finally may transform the virtual particles to the real elementary particles, accompanied by release of huge amount of energy. It may happen in strong enough fields, like in hot dense plasma, like in stars, providing the necessary symmetry shift of Bivacuum fermions and frequency of $[\widetilde{\mathbf{C}} \rightleftharpoons \widetilde{\mathbf{W}}]$ pulsation, sufficient for

their *forced resonance* with basic $\mathbf{VPW}^{\pm}_{q=1}$.

### 7.1 The mechanism of electrostatic, magnetic and gravitational fields origination

*The electric field* of triplets of elementary particles is represented by elastic spherical waves in matrix of Bivacuum, excited by reversible recoil-antirecoil effects, accompanied $[\mathbf{C} \rightleftharpoons \mathbf{W}]$ pulsation of unpaired sub-elementary fermion of triplets. Superposition of such virtual elastic waves, excited by the opposite charges, is followed by linear alignment of Bivacuum dipoles between them and their mass and charge symmetry shift. This effects provide the Coulomb attraction between opposite charges due to decreasing of resulting symmetry shift between charges. The Coulomb repulsion is also result of system tendency to reduce the resulting symmetry shift of Bivacuum in space between similar by sign charges.

*The gravitational field* of triplets is a result of $\mathbf{VPW}^{+}_{q}$ and $\mathbf{VPW}^{-}_{q}$ excitation in Bivacuum by $[\mathbf{C} \rightleftharpoons \mathbf{W}]$ pulsation of pairs $[\mathbf{F}^{+}_{\uparrow} \bowtie \mathbf{F}^{-}_{\downarrow}]_{x,y}$ of triplets $< [\mathbf{F}^{+}_{\uparrow} \bowtie \mathbf{F}^{-}_{\downarrow}]_{x,y} + \mathbf{F}^{\pm}_{\updownarrow} >^{i}_{z}$ and their superposition, increasing or decreasing the virtual pressure waves between two or more pulsing triplets, independently of their electric charge. In accordance to our conjecture, the gravitation and antigravitation has some similarity with hydrodynamic Bjerknes attraction and repulsion between pulsing objects.

*The magnetic field* is a result of triplets fast rotation, inducing their symmetry shift and circular alignment of Bivacuum dipoles dependent on direction of triplets propagation.

The detailed description of the above phenomena is presented in our works: http://arxiv.org/abs/physics/0103031 and http://arxiv.org/abs/physics/0207027.

## 8. The Virtual Replica (VR) of Material Objects and its Multiplication (VRM)

Theory of Virtual Replica (**VR**) of material objects in Bivacuum and **VR** Multiplication in space and time: $\mathbf{VRM(r,t)}$ is proposed. The *primary* $\mathbf{VR}_0$ in initial time represents a three-dimensional (3D) superposition of Bivacuum virtual standing waves $\mathbf{VPW}^{\pm}_{m}$ and $\mathbf{VirSW}^{\pm 1/2}_{m}$, modulated by $[\mathbf{C} \rightleftharpoons \mathbf{W}]$ pulsation of elementary particles and translational and librational de Broglie waves of molecules of macroscopic object (http://arxiv.org/abs/physics/0207027).

*For the end of energy, charge and spin conservation* in Bivacuum, we have to assume, that symmetry shifts of Bivacuum dipoles, involved in **VR** formation, should compensate each other. This condition is satisfied, if we assume, that the primary and secondary **VR** is formed by certain number (N) of virtual Cooper pairs of Bivacuum fermions and antifermions of opposite spins and symmetry shifts:

$$\mathbf{VR} = \sum_{n}^{N} [\mathbf{BVF}^{\uparrow} \bowtie \mathbf{BVF}^{\downarrow}]_{n}$$

The *isotropic* infinitive multiplication of primary $\mathbf{VR}_0$ in space and time in form of 3D packets of virtual standing waves, representing huge number (M) of *secondary* $\mathbf{VR}_m$, is a result of interference of all pervading external coherent basic *reference waves* - Bivacuum Virtual Pressure Waves ($\mathbf{VPW}^{\pm}_{q=1}$) and Virtual Spin Waves ($\mathbf{VirSW}^{\pm 1/2}_{q=1}$) with similar kinds of modulated standing waves, like that, forming the primary **VR** and changing of the object itself with time. The latter has a properties of the *object waves*. Consequently, the $\mathbf{VRM(r,t)}$, as a result of mixing of the *object waves* with *reference waves* can be named **Holoiteration** by analogy with hologram (in Greece *'holo'* means the 'whole' or 'total'). The $\mathbf{VRM(r,t)}$ can be considered as a result of linear superposition of primary $\mathbf{VR}_0$ of different states with corresponding amplitude of probability ($c_m$):

$$\mathbf{VRM(r,t)} = \sum_{m}^{M} c_m [\mathbf{VR}_m >_m$$

The frequencies of basic reference virtual pressure waves ($\mathbf{VPW}_{q=1}^{\pm} \equiv \mathbf{VPW}_0^{\pm}$) and virtual spin waves ($\mathbf{VirSW}_{q=1}^{\pm 1/2} \equiv \mathbf{VirSW}_0^{\pm 1/2}$) of Bivacuum are equal to Compton frequencies of three electron generation ($i = e, \mu, \tau$):

$$[\boldsymbol{\omega}_{VPW_0} = \boldsymbol{\omega}_{VirSW_0} = \boldsymbol{\omega}_0 = \boldsymbol{\omega}_{\mathbf{C \rightleftharpoons W}}^{\mathbf{v=0}} = \mathbf{m}_0\mathbf{c}^2/\hbar\,]^i \qquad 8.1$$

The *Bivacuum virtual pressure waves modulation (*$\mathbf{VPW}_{\mathbf{m}}^{\pm}$*)* can be realized by pairs of positive and negative cumulative virtual clouds ($\mathbf{CVC}^+ \bowtie \mathbf{CVC}^-$), emitted/absorbed in the process of [$\mathbf{C} \rightleftharpoons \mathbf{W}$] pulsation of *pairs:* [$\mathbf{F}_\uparrow^+ \bowtie \mathbf{F}_\downarrow^-]_C \rightleftharpoons [\mathbf{F}_\uparrow^+ \bowtie \mathbf{F}_\downarrow^-]_W$ of elementary triplets (electrons, protons, neutrons) $< [\mathbf{F}_\uparrow^+ \bowtie \mathbf{F}_\downarrow^-] + \mathbf{F}_\updownarrow^\pm >^i$ of the object. These kinds of waves superposition are responsible for gravitational attraction or repulsion between two or more objects and do not depend on the charge of triplets (http://arxiv.org/abs/physics/0207027).

The *Bivacuum virtual spin waves modulation* ($\mathbf{VirSW}^{\pm 1/2}$) can be a consequence of *recoil angular momentum oscillation,* accompanied the $\mathbf{CVC}^\pm$ emission $\rightleftharpoons$absorption in the process of [$\mathbf{C} \rightleftharpoons \mathbf{W}$] pulsation of *unpaired* sub-elementary fermion or antifermion $\mathbf{F}_\updownarrow^\pm >^i$ of triplets:

$$[(\mathbf{F}_\uparrow^+ \bowtie \mathbf{F}_\downarrow^-)_C + (\mathbf{F}_\updownarrow^\pm)_W] \overset{+\mathbf{CVC}^\pm - \mathbf{Recoil}}{\underset{-\mathbf{CVC}^\pm + \mathbf{Antirecoil}}{<=\!=\!=\!=\!=\!=>}} [(\mathbf{F}_\uparrow^+ \bowtie \mathbf{F}_\downarrow^-)_W + (\mathbf{F}_\updownarrow^\pm)_C] \qquad 8.2$$

The recoil energy of the in-phase [$\mathbf{C} \rightleftharpoons \mathbf{W}$] pulsation of a sub-elementary fermion $\mathbf{F}_\downarrow^+$ and antifermion $\mathbf{F}_\uparrow^-$ of pair [$\mathbf{F}_\uparrow^- \bowtie \mathbf{F}_\downarrow^+$] and the angular momenta of CVC$^+$ and CVC$^-$ of $\mathbf{F}_\uparrow^-$ and $\mathbf{F}_\downarrow^+$ in pairs compensate each other and the resulting recoil momentum and energy of [$\mathbf{F}_\downarrow^+ \bowtie \mathbf{F}_\uparrow^+$] is zero.

The stability of VR of object, as a *hierarchical system of quantized metastable torus-like and vortex filaments structures formed by* $\mathbf{VPW}_{\mathbf{m}}^{\pm}$ *and by* $\mathbf{VirSW}_{\mathbf{m}}^{\pm 1/2}$ *excited by paired and unpaired sub-elementary fermions, correspondingly,* in superfluid Bivacuum, could be responsible for so-called "**phantom effect**" of object after its destroyment or removing to remote place.

For free elementary particles the notion of secondary virtual replica, as one of multiplicated primary VR$_0$ coincide with notion of one of possible 'anchor sites' (eq.6.36), as a conjugated dynamic complex of three Cooper pair of asymmetric fermions.

### *8.1 Modulation of the basic Virtual Pressure Waves (*$VPW_{q=1}^{\pm}$*)$^i$ and Virtual Spin Waves (*$VirSW_{q=1}^{\pm 1/2}$*)$^i$ of Bivacuum by molecular translations and librations, as a condition of VR formation*

The external translational and librational kinetic energy of particle ($\mathbf{T}_k)_{tr,lb}$ is directly related to corresponding de Broglie wave length ($\boldsymbol{\lambda}_B$), the group ($\mathbf{v}$), phase velocity ($\mathbf{v}_{ph}$) and frequency ($\boldsymbol{\nu}_B = \boldsymbol{\omega}_B/2\pi$):

$$\left(\boldsymbol{\lambda}_B = \frac{\mathbf{h}}{\mathbf{m}_V^+\mathbf{v}} = \frac{\mathbf{h}}{2\mathbf{m}_V^+\mathbf{T}_k} = \frac{\mathbf{v}_{ph}}{\mathbf{v}_B} = 2\pi\frac{\mathbf{v}_{ph}}{\boldsymbol{\omega}_B}\right)_{tr,lb} \qquad 8.3$$

where the de Broglie wave frequency is related to its length and kinetic energy of particle as:

$$\left[ \mathbf{\nu}_B = \frac{\boldsymbol{\omega}_B}{2\pi} = \frac{h}{\mathbf{m}_V^+ \boldsymbol{\lambda}_B^2} = \frac{\mathbf{m}_V^+ \mathbf{v}^2}{h} \right]_{tr,lb} \qquad 8.4$$

The total energy/frequency of de Broglie wave and *resulting* frequency of pulsation $(\boldsymbol{\omega}_{\mathbf{C} \rightleftharpoons \mathbf{W}})_{tr,lb}$ (see eq. 6.4) is a result of modulation/superposition of the internal frequency, related to the rest mass of particle, by the external most probable frequency of de Broglie wave of the whole particle $(\boldsymbol{\omega}_B)_{tr,lb}$, determined by its most probable external momentum: $(\mathbf{p} = \mathbf{m}_V^+ \mathbf{v})_{tr,lb}$, related to translations or librations:

$$\left[ \mathbf{E}_{tot} = \mathbf{m}_V^+ \mathbf{c}^2 = \hbar \boldsymbol{\omega}_{\mathbf{C} \rightleftharpoons \mathbf{W}} \right]_{tr,lb} = \mathbf{R}(\hbar \boldsymbol{\omega}_0)_{rot}^{in} + (\hbar \boldsymbol{\omega}_B^{ext})_{tr,lb} = \mathbf{R}(\mathbf{m}_0 \boldsymbol{\omega}_0^2 \mathbf{L}_0^2)_{\mathbf{rot}}^{in} + \left( \frac{h^2}{\mathbf{m}_V^+ \boldsymbol{\lambda}_B^2} \right)_{tr,lb}^{ext} \quad 8.5$$

where relativistic factor: $\mathbf{R} = \sqrt{1 - (\mathbf{v}/\mathbf{c})^2}$ is tending to zero at $\mathbf{v} \to \mathbf{c}$.

In composition of condensed matter the value of $(\boldsymbol{\lambda}_B)_{tr,lb}$ is bigger for librations than for translation of molecules. The corresponding most probable modulation frequencies of translational and librational de Broglie waves is possible to calculate, using our Hierarchic theory of condensed matter and based on this theory computer program (Kaivarainen, 2001; 2003; 2004; 2005).

The *frequencies* of Bivacuum virtual pressure waves $(\mathbf{VPW_m^\pm})$ and virtual spin waves $(\mathbf{VirSW_m^{\pm 1/2}})$ are modulated by the *resulting* frequencies of de Broglie waves of the object molecules, related to librations $(\boldsymbol{\omega}_{lb})$ and translations $(\boldsymbol{\omega}_{tr})$, correspondingly.

The combinational resonance between the basic Bivacuum virtual waves $(q = 1)$ and resulting frequency of $[\mathbf{C} \rightleftharpoons \mathbf{W}]$ pulsation of electrons, protons and neutrons, composing atoms and molecules of the object, is possible at conditions:

$$\boldsymbol{\omega}_{\mathbf{VPW}_{\mathbf{q=1}}^\pm}^i = \mathbf{z} \, \mathbf{R} \boldsymbol{\omega}_0^i + \mathbf{g} \boldsymbol{\omega}_B^{tr} + \mathbf{r} \boldsymbol{\omega}_B^{lb} \cong \mathbf{z} \, \mathbf{R} \boldsymbol{\omega}_0^i + \mathbf{g} \boldsymbol{\omega}_B^{tr} \qquad 8.6$$

$$\boldsymbol{\omega}_{\mathbf{VirSW}_{\mathbf{q=1}}^{\pm 1/2}}^i = \mathbf{z} \, \mathbf{R} \boldsymbol{\omega}_0^i + \mathbf{g} \boldsymbol{\omega}_B^{tr} + \mathbf{r} \boldsymbol{\omega}_B^{lb} \cong \mathbf{z} \, \mathbf{R} \boldsymbol{\omega}_0^i + \mathbf{r} \boldsymbol{\omega}_B^{lb} \qquad 8.6a$$

$$\mathbf{R} = \sqrt{1 - (\mathbf{v}/\mathbf{c})^2}; \quad \mathbf{z}, \mathbf{g}, \mathbf{r} = 1, 2, 3 \ldots \text{(integer numbers)}$$

Each of 24 collective excitations of condensed matter, introduced in our Hierarchic theory (Kaivarainen, 1995; 2001; 2004), has the own characteristic frequency and can contribute to Virtual Replica of the object.

In contrast to regular hologram, VR contains information not only about surface and shape properties of the object, but also about its internal properties.

Three kind of modulations: *frequency, amplitude and phase* of Bivacuum virtual waves $(\mathbf{VPW_m^\pm})$ and $(\mathbf{VirSW_m^{\pm 1/2}})$ by $[\mathbf{C} \rightleftharpoons \mathbf{W}]$ pulsation of elementary particles of molecules and their de Broglie waves may be described by known relations (Prochorov, 1999):

1. *The frequencies* of virtual pressure waves $(\omega_{_{VPW^\pm}}^M)$ and spin waves $(\omega_{_{VirSW^\pm}}^M)$, *modulated* by translational and librational de Broglie waves of the object's molecules, correspondingly, can be presented as:

$$\boldsymbol{\omega}_{_{VPW_m^\pm}}^M = \mathbf{z} \mathbf{R} \boldsymbol{\omega}_0^i + \Delta \boldsymbol{\omega}_B^{tr} \cos \boldsymbol{\omega}_B^{tr} t \qquad 8.7$$

$$\boldsymbol{\omega}_{_{VirSW_m^{\pm 1/2}}}^M = \mathbf{z} \mathbf{R} \boldsymbol{\omega}_0^i + \Delta \boldsymbol{\omega}_B^{lb} \cos \boldsymbol{\omega}_B^{lb} t \qquad 8.7a$$

The Compton pulsation frequencies of the electrons, protons and neutrons are close to basic frequency of Bivacuum virtual waves, corresponding to $\mathbf{q} = \mathbf{j} - \mathbf{k} = \mathbf{1}$:

$$\omega_0^i = \mathbf{m}_0^i \mathbf{c}^2/\hbar \sim \omega_{VPW_{q=1}^\pm, ViSW_{q=1}}^i \qquad 8.8$$

Such kind of modulation is accompanied by two satellites with frequencies: $(\omega_0^i + \omega_B^{tr,lb})$ and $(\omega_0^i - \omega_B^{tr,lb}) = \Delta\omega_B^{tr,lb}$. The latter is named frequency deviation. In our case: $\omega_0^e (\sim 10^{21} s^{-1}) >> \omega_B^{tr,lb} (\sim 10^{12} s^{-1})$ and $\Delta\omega_{tr,lb} >> \omega_B^{tr,lb}$.

The temperature of condensed matter and phase transitions may influence the modulation frequencies of de Broglie waves of its molecules.

2. *The amplitudes of virtual pressure waves ($VPW_m^\pm$) and virtual spin waves $VirSW_m^{\pm 1/2}$ (informational waves), modulated by the object elementary particles and molecules dynamics, are dependent on translational and librational de Broglie waves frequencies as:*

$$\mathbf{A}_{VPW_m^\pm} \approx \mathbf{A}_0(\sin\mathbf{R}\omega_0^i\mathbf{t} + \gamma\omega_B^{tr}\sin\mathbf{t}\cdot\cos\omega_B^{tr}t) \qquad 8.9$$

$$\mathbf{I}_{VirSW_m^{\pm 1/2}} \approx \mathbf{I}_0(\sin\mathbf{R}\omega_0^i\mathbf{t} + \gamma\omega_B^{lb}\sin\mathbf{t}\cdot\cos\omega_B^{lb}\mathbf{t}) \qquad 8.9a$$

where: the informational/spin field amplitude is determined by the amplitude of Bivacuum fermions $[\mathbf{BVF}^\uparrow \rightleftarrows \mathbf{BVF}^\downarrow]$ equilibrium constant oscillation:
$\mathbf{I}_S = \mathbf{I}_{\mathbf{VirSW}^{\pm 1/2}} \sim \Delta\mathbf{K}_{BVF^\uparrow \rightleftarrows BVF^\downarrow}(\mathbf{t})$

The index of frequency modulation is defined as: $\gamma = (\Delta\omega_{tr,lb}/\omega_B^{tr,lb})$. The carrying zero-point pulsation frequency of particles is equal to the basic frequency of Bivacuum virtual waves: $\omega_{VPW_0^\pm, ViSW_0}^i = \omega_0^i$. Such kind of modulation is accompanied by two satellites with frequencies: $(\omega_0^i + \omega_B^{tr,lb})$ and $(\omega_0^i - \omega_B^{tr,lb}) = \Delta\omega_{tr,lb}$. In our case: $\omega_0^e (\sim 10^{21} s^{-1}) >> \omega_B^{tr,lb} (\sim 10^{12} s^{-1})$ and $\gamma >> 1$.

The fraction of molecules in state of mesoscopic molecular Bose condensation (mBC), representing, coherent clusters (Kaivarainen, 2001a,b; 2004) is a factor, influencing the amplitude ($A_0$) and such kind of modulation of Virtual replica of the object;

3. *The phase modulated $VPW_m^\pm$ and $VirSW_m^{\pm 1/2}$* by de Broglie waves of molecules, related to their translations and librations, can be described like:

$$\mathbf{A}_{VPW_m^\pm}^M = \mathbf{A}_0\sin(\mathbf{R}\omega_0\mathbf{t} + \Delta\varphi_{tr}\sin\omega_B^{tr}\mathbf{t}) \qquad 8.10$$

$$\mathbf{I}_{VirSW_m^{\pm 1/2}}^M = \mathbf{I}_0\sin(\mathbf{R}\omega_0\mathbf{t} + \Delta\varphi_{lb}\sin\omega_B^{lb}\mathbf{t}) \qquad 8.10a$$

The value of phase increment $\Delta\varphi_{tr,lb}$ of modulated virtual waves of Bivacuum ($\mathbf{VPW}_m^\pm$ and $\mathbf{VirSW}_m^{\pm 1/2}$), contains the information about external and internal geometrical properties of the object.

The phase modulation takes place, if the phase increment $\Delta\varphi_{tr,lb}$ is independent on modulation frequency $\omega_B^{tr,lb}$.

### 8.2 The superposition of internal and surface Virtual Replicas of the object, as the "Ether Body"

The superposition of individual microscopic $\mathbf{VR}_{mic}$ of the electrons, protons, neutrons and atoms/molecules of the object (internal and surface ones), formed by interference of de Broglie waves of these particles with basic virtual waves of Bivacuum ($\mathbf{VPW}_{q=1}^\pm$ and $\mathbf{VirSW}_{q=1}^{\pm 1/2}$), stands for *internal macroscopic virtual replica* of the object ($\mathbf{VR}^{in}$), describing its *internal* bulk properties. The overall shape of ($\mathbf{VR}^{in}$) should be close to shape of the object itself, for example, such as the human's body and it organs shape.

*Spatial stability of condensed systems* means that the macroscopic internal virtual replica: $\mathbf{VR}^{in} = \sum \mathbf{VR}_{mic}^{in}$, as a result of 3D standing waves superposition of microscopic $\mathbf{VR}_{mic}^{in}$ in superfluid Bivacuum, should have location of nodes, coinciding with the most probable positions of the atoms and molecules in condensed matter.

The superposition of coherent de Broglie waves of atoms and molecules in clusters, forming 3D standing waves B, determined by their librations and translations, represents the *mesoscopic Bose condensate:* [**mBC**] (Kaivarainen, 2001 b,c). In accordance to our theory, this means also the coherent [$\mathbf{C} \rightleftharpoons \mathbf{W}$] pulsations of elementary particles of these molecules and atoms. The violation of this coherency is accompanied by density fluctuation and defects origination or cavitational fluctuations in solids and liquids.

The *surface macroscopic virtual replica of the object:* $\mathbf{VR}^{sur} = \sum \mathbf{VR}_{mic}^{sur}$ is a part of the **Ether body**. The mechanism of its origination is similar to *internal macroscopic virtual replica* of the object $\mathbf{VR}^{in} = \sum \mathbf{VR}_{mic}^{in}$. It is a result of modulation of Bivacuum virtual waves by de Broglie waves of elementary particles of the atoms and molecules on the surface of the object. Its dimension can exceed the dimensions of the object.

The superposition of the internal and surface virtual replicas corresponds to notion of the "*ether body*" in Eastern philosophy:

$$\textbf{Ether Body} \equiv \mathbf{VR} = \mathbf{VR}^{in} + \mathbf{VR}^{sur} = \sum (\mathbf{VR}_{mic}^{in} + \mathbf{VR}_{mic}^{sur}) \qquad 8.11$$

Stability of hierarchic system of whirls, forming Ether Body, as a hierarchical system of virtual standing waves and *curls* in superfluid Bivacuum (like permanent circular currents and whirls in superfluid $^4\mathbf{He}$), could be responsible for so-called "*phantom effect*" of this object.

### 8.3 The infinitive spatial Virtual Replica Multiplication VPM(r). The "Astral" and "Mental" bodies, as a distant and nonlocal components of VRM(r)

The mechanism of primary *Virtual Replica Multiplication* (**VPM***)* have general features with hologram origination, however without photomaterials or screens, fixing **VR**. The role of coherent *reference waves* play unperturbed by the object basic Bivacuum virtual waves (**VPW**$_{q=1}^{\pm}$ and nonlocal **VirSW**$_{q=1}^{\pm 1/2}$). The role of *subject waves* is represented by the primary Virtual Replica of the object, containing information not only on shape/surface, but also about internal properties of the subject.

The VRM is a spatially *isotropic* process, like excitation of spherical waves. It can be subdivided on two components - *distant (translational)* and *nonlocal (rotational or librational)*:

1) the *distant component* of **VRM(r)**$^{dis}$ is a result of replication of the translational component of primary **VR** outside the volume of the object, by means of Virtual Pressure Waves (**VPW**$_{q=1}^{\pm}$). The front of 3D VRM$^{dis}$ in form of huge number of *secondary* VR isotropicaly expand in space like gravitation waves with light velocity (http://arxiv.org/abs/physics/0207027).

The virtual replica of the object can be reproducible in form of distant **VRM**$^{dis}$ (like the hologram) in any remote space regions, where the interference pattern of the *reference wave* **VPW**$_{q=1}^{\pm}$ with *primary* **VR** as the object wave, is existing. The volume of space, occupied by distant **VRM**$^{dis}$, is expanding with light velocity (**c**) during the life-time of primary **VR** and atoms, composing the object.

The expanding with light velocity population of **VR(ct)**, spatially separated from the body/object, may correspond to Eastern ancient notion of the "*astral body*":

$$\textbf{Astral Body} = \sum^{t} \mathbf{VR}_{tr}(\mathbf{ct}) = \mathbf{VRM}^{dis} \qquad 8.12$$

As far each individual secondary **VR** in population $\sum^{t}$ **VR(ct)** in the absence of dissipation in superfluid Bivacuum is the exact copy of the primary **VR**, they should be spatially also indistinguishable, like particles in state of Bose condensate. The detected by psychic or by special device secondary replica displays its properties

The dielectric permittivity ($\varepsilon_0$) and permeability ($\mu_0$) in the volume of the Astral bodies may differ from their averaged values in Bivacuum because of small charge symmetry shift in Bivacuum fermions (**BVF$^{\updownarrow}$**): $\Delta e = |e_+ - e_-| > 0$, induced by *recoil $\rightleftharpoons$ antirecoil* effects, accompanied [**C $\rightleftharpoons$ W**] pulsation of elementary particles. Consequently, the probability of atoms and molecules excitation and ionization (dependent on Coulomb interaction between electrons and nuclears), as a result of their thermal collisions with excessive kinetic energy, may be higher in volumes of the Astral bodies, than outside of them. This may explain their special optical effects - a shining of some objects phantoms (ghosts) in darkness, or their specific spectrogram, representing astral bodies. Their spatial instability of phantoms can be explained by spatial similarity of astral bodies, composing **VRM$^{dis}$**. The possibility of phenomena like *remote vision and remote healing* also follow from our holomovement like mechanism of **VRM(r)**$_S$ $\bowtie$ **VRM(r)**$_R$ superposition of Sender and Receiver and their 'tuning'.

The sensitivity of Kirlian effect or Gas Discharge Visualization (GDV) to internal process of macroscopic object, like human body, also can be explained by specific properties of the surface Ether and Astral bodies, changing the probability of the air molecules excitation/ionization after thermal collision;

2) the *nonlocal component* of **VRM$^{nl}$** is a result of 3D replication of the rotational/librational component of *primary* virtual replica (**VR**$_{lb}$) outside the volume of the object, by means of nonlocal (informational) Virtual Spin Waves (**VirSW$^{\pm}_{q=1}$**), propagating in symmetric Bivacuum instantly, i.e. without light velocity limitation.

The *nonlocal macroscopic virtual replica multiplication* (**VRM$^{nl}$**) or **VR** *iteration*, is a result of interference of modulated by librational de Broglie waves the *recoil $\rightleftharpoons$ antirecoil* effects the Bivacuum virtual spin waves: **VirSW$^{\pm 1/2}_{\mathbf{m}}$** – *object spin waves* with corresponding *reference spin waves* of Bivacuum (**VirSW$^{\pm 1/2}_{\mathbf{q}}$**).

The Eastern notion of *mental body* may correspond to **VRM$^{nl}$**, as a multiplication (holoiteration or holomovement after Bohm) of informational Virtual Replicas [**VR**$_{lb}$]:

$$\mathbf{Mental\,Body} \;=\; \sum^{t} \mathbf{VR}_{lb} \;=\; \mathbf{VRM}^{nl} \qquad\qquad 8.13$$

Hierarchical superposition of huge number of Astral and Mental Bodies of all human population on the Earth can be responsible for Global Informational Field origination, like Noosphere, proposed by Russian scientist Vernadsky in the beginning of 20th century. The Astral and Mental bodies may partly be overlapped with Ether body. This provide the possibility of dynamic exchange interaction and feedback reaction between all three virtual bodies of the object: Ether, Astral and Mental.

*One important conjecture,* following from our approach to distant **VRM$^{dis}$** can be discussed. We proceed from the consequence of our theory, that the volume of space, occupied by distant **VRM$^{dis}$** is expanding isotropicaly with light velocity (**c**) in 3D space during the life-time of **VR** and atoms, composing the object.

The life span of the individual stable atoms, including hydrogen, carbons, oxygen, composing biological objects is comparable with life-time of the Universe, i.e. over ten billions of years. This means, that not only nonlocal **VRM$^{nl}$**, but as well the distant **VRM$^{dis}$** of these atoms may involve all the Universe. It is a conditions of Virtual Guides of spin, momentum and energy (**VirG**$_{SME}$) 3D net formation in the Universe, connecting virtually all similar and coherent elementary particles and atoms of the equal de Broglie wave length. We suppose, however, that only in composition of biosystems these atoms may become enough coherent and orchestrated to provide a strong enough *cumulative interaction* between Sender and Receiver, for example, between *psychic* and very remote objects (inorganic or biological) via 3D net of $\mathbf{VirG}_{SME}$ and $\mathbf{VRM}$ as a factors of Bivacuum mediated interaction (BMI). The construction of $\mathbf{VirG}_{SME}$ and mechanism of their action will be described in the next section.

*A complex Hierarchical system $\sum VRM(r,t)$ of Solar system, galactics, including Noosphere, may be considered as Hierarchical quantum supercomputer or Superconsciousness, able to simulate all probable situations of virtual future and past. It is possible in conditions of time uncertainty: $t = 0/0$ when the translational velocity $v = 0$ and accelerations $(dv/dt) = 0$ in the volume of $\sum VRM(r,t)$ are zero (Kaivarainen, 2005: http://arXiv.org/abs/physics/0103031 ).*

Our theory admit a possibility of feedback reaction between the iterated VR and primary one and between primary VR and the object physical properties. Consequently, the phenomena of most probable event anticipation by enough sensitive physical detectors and human beings (psychics) is possible in principle. This may explain the reproducible results of unconsciousness response (by changes of human skin conductance) of future events (presponse), obtained by Dick Bierman and Dean Radin (2002). However, in these experiments the possibility of influence of intention of participant on random events generator (REG), choosing next photo (calm or emotional), like in Bierman's experiments, also should be taken into account. Such kind of weak influence of humans intention on REG was demonstrated in long term studies of Danne and Jahn (2003).

*In contrast to virtual time, the reversibility of real time looks impossible,* as far it needs the reversibility of all dynamic process in Universe due to interrelations between closed systems of different levels of hierarchy. It is evident that such 'play back' of the Universe history needs the immense amount of energy redistribution in the Universe.

*All three described Virtual Replicas: Ether, Astral and Mental bodies are interrelated with each other and physical body.* The experimental evidences are existing, that between properties of the *Ether* bodies and corresponding physical bodies of living organisms or inorganic matter, the correlation takes a place. It is confirmed by the Kirlian effect, reflecting the ionization/excitation threshold of the air molecules in volume of Ether and astral bodies.

The perturbation of the Ether body of one object (Receptor) by the astral or mental body of the other object (Sender) can be imprinted in properties of physical body (condensed matter) of Receptor for a long time in form of subtle, but stable structural perturbations. The stability of such kind of informational 'taping' is determined by specific properties of material, as a matrix for imprinting. For example, water and aqueous systems, like biological ones, are very good for stable imprinting of virtual information and energy via introduced VRM and Virtual Guides (see next chapter). However, some 'sensitive' stones or other rigid materials have a much longer life span.

*The Ghost phenomena can be explained by reproducing of such imprinted in walls, cells and floor information, mediated by distant virtual replica multiplication ($VRM^{dis}$). The reproduction of VR from imprinted in condensed matter $VRM^{dis}$ is a process, similar to treatment of regular hologram by the reference waves. In the case of 'Ghost' the reference waves can be presented by basic $VPW_m^{\pm}$ and $VirSW_m^{\pm 1/2}$, modulated by special superposition of Virtual replicas of other objects, for example, Earth, Moon and Sun.*

The *nonlocal* Mental - Informational body formation in living organisms and humans, in accordance to our theory (Kaivarainen, 2001; 2003), is related to equilibrium shift of dynamic equilibrium of [assembly ⇌ disassembly] of coherent water clusters in microtubules of the neurons (librational effectons), accompanied series of elementary acts of consciousness in *nonequilibrium processes* of meditation, intention and braining. Corresponding variations of kinetic energy and momentum of water molecules can be transmitted from Sender to remote Receiver via nonlocal virtual spin-momentum-energy guides (**VirG**$_{SME}$), described in next chapter.

In complex process of Psi phenomena, the first stage is a 'target searching' by nonlocal [mental body] of psychic, then formation of **VirG**$_{SME}$, then activation of psychic's [astral body] by its [ether body]. The latter can be interrelated with specific processes of physical body of psychic, like dynamics of water in microtubules of neurons ensembles, realizing elementary acts of perception and consciousness, in accordance to our model (Kaivarainen, 2000; 2005).

## 9  Possible Mechanism of Quantum entanglement between distant elementary particles via Virtual Guides of spin, momentum and energy (VirG$_{S,M,E}$)

In accordance to our theory, the instant nonlocal quantum entanglement between two or more distant *similar* elementary particles (electrons, protons, neutrons, photons), named [Sender (S)] and [Receiver (R)], revealed in a lot of experiments, started by Aspect and Grangier (1983), involves a few stages:

**1**. Superposition of the nonlocal components of Virtual replicas multiplication (**VRM**) or *holoiteration* pattern, formed by interference of modulated by de Broglie waves of [S] and [R] the counterphase virtual spin waves: [**VirSW**$_{\mathbf{m}}^{\pm 1/2}(S)$ <==> **VirSW**$_{\mathbf{m}}^{\pm 1/2}(R)$], with corresponding *reference waves* of Bivacuum **VirSW**$_{\mathbf{q}}^{\pm 1/2}$;

**2**. Tuning (frequency and phase synchronization) of de Broglie waves of remote interacting identical particles (like electrons, protons) and their complexes in form of atoms and molecules, as a condition of Virtual Guides of spin-momentum-energy (**VirG**$_{\mathbf{SME}}$) between [S] and [R] formation (see Fig.4).

The formation of two spatial configurations of Virtual Guides, representing quasi 1D virtual Bose condensate (BC), is possible:

a) single *nonlocal virtual guides* **VirG**$_{SME}^{(\mathbf{BVB}^{\pm})^i}$ - virtual microtubules from Bivacuum bosons (**BVB**$^{\pm}$)$^i$. In this case the **VirG**$_{SME}^{(\mathbf{BVB}^{\pm})^i}$ is not rotating as a whole around its axis and the resulting spin is zero. The *longitudinal* momentum of BVB$^{\pm}$ is also zero, providing conditions for virtual BC;

b) twin *nonlocal virtual guides* **VirG**$_{SME}^{[\mathbf{BVF}^{\uparrow} \bowtie \mathbf{BVF}^{\downarrow}]^i}$ from Cooper pairs of Bivacuum fermions [**BVF**$^{\uparrow}$⋈ **BVF**$^{\downarrow}$]$^i$. In this case each of two adjacent microtubules rotate as respect to each other and around their own axes in opposite directions. The resulting angular moment (spin) of such pair, as well as *longitudinal* momentum is also zero.

Two remote coherent triplets - elementary fermions, like (electron - electron) or (proton - proton) with similar frequency of [**C** ⇌ **W**]$_{e,p}$ pulsation and opposite spins (phase) can be connected by nonlocal Virtual guides (**VirG**$_{SME}^{e,p}$) of spin (S), momentum (M) and energy (E) from Sender to Receiver. The spin - information (qubits), momentum and kinetic energy instant transmission via such **VirG**$_{SME}^{e,p}$ from [S] to [R] is possible, as described (Fig.4). The similar mechanism can work between two synchronized photons (bosons) of the opposite spins.

The double **VirG**$_{SME}^{[\mathbf{BVF}^{\uparrow} \bowtie \mathbf{BVF}^{\downarrow}]^i}$ can be transformed to single **VirG**$_{SME}^{(\mathbf{BVB}^{\pm})^i}$ by conversion of opposite Bivacuum fermions: **BVF**$^{\uparrow}$ = [**V**$^+$ ⇈ **V**$^-$] and **BVF**$^{\downarrow}$ = [**V**$^+$ ⇊ **V**$^-$] to the pair of Bivacuum bosons of two possible alternatives of polarization:

$$\mathbf{BVB}^+ = [\mathbf{V}^+ \uparrow\downarrow \mathbf{V}^-] \quad or \quad \mathbf{BVB}^- = [\mathbf{V}^+ \downarrow\uparrow \mathbf{V}^-]$$

Superposition of two nonlocal virtual spin waves excited by similar elementary particles (electrons or protons) of Sender $(\mathbf{VirSW_m^{S=+1/2}})_S$ and Receiver $(\mathbf{VirSW_m^{S=-1/2}})_R$ of the same pulsation frequency and opposite spins, i.e. opposite phase of $[\mathbf{C} \rightleftharpoons \mathbf{W}]$ pulsation, is necessary for *Virtual Guide* $(\mathbf{VirG}_{SME})^i$ formation (Fig.4).:

$$\left[ < [\mathbf{F_\downarrow^+} \bowtie \mathbf{F_\uparrow^-}]_C + (\mathbf{F_\downarrow^-})_W >_S \overset{\mathbf{VirSW_S}}{\rightarrow} \begin{matrix} \mathbf{BVB^+} \\ \circleddash = \circleddash \\ \mathbf{BVB^-} \end{matrix} \overset{\mathbf{VirSW_R}}{\leftarrow} < (\mathbf{F_\uparrow^-})_C + [\mathbf{F_\downarrow^-} \bowtie \mathbf{F_\uparrow^+}]_W >_R \right]^i \qquad 9.1$$

$$= [\mathbf{n_+BVB^+}(\mathbf{V^+} \uparrow\downarrow \mathbf{V^-}) + \mathbf{n_-BVB^-}(\mathbf{V^+} \downarrow\uparrow \mathbf{V^-})]^i = (\mathbf{VirG}_{SME}^{ext})^i \qquad 9.1a$$

The radius of virtual microtubules of $\mathbf{VirG}_{SME}^i$ is determined by Compton radius of three generation of torus and antitorus ($i = e, \mu, \tau$), forming them:

$$\mathbf{L}_V^e = \hbar/\mathbf{m_0^e c} >> \mathbf{L}_V^\mu = \hbar/\mathbf{m_0^\mu c} > \mathbf{L}_V^\tau = \hbar/\mathbf{m_0^\mu c} \qquad 9.2$$

The radius of $\mathbf{VirG}_{SME}^e$, connecting two remote electrons, is the biggest one ($\mathbf{L}^e$). The radius of $\mathbf{VirG}_{SME}^\tau$, connecting two protons or neutrons ($\mathbf{L}^\tau$) is about $3.5 \times 10^3$ times smaller. The entanglement between similar and coherent atoms in pairs of [S] and [R], like hydrogen, oxygen, carbon or nitrogen atoms, can be realized via complex virtual guides of atomic ($\mathbf{VirG}_{SME}^{at}$), representing *multi-shell coaxial constructions*.

### 9.1. The mechanism of momentum and energy transmission between similar elementary particles of Sender and Receiver via VirG$_{\mathbf{S,M,E}}$

The increments or decrements of momentum $\pm\Delta\mathbf{p} = \Delta(\mathbf{m}_V^+\mathbf{v})_{tr,lb}$ and kinetic $(\pm\Delta\mathbf{T}_k)_{tr,lb}$ energy transmission from [S] to [R] of *selected generation of elementary particles* is determined by the translational and librational velocity variation ($\Delta\mathbf{v}$) of nuclei of (S). This means, that energy/momentum transition from [S] to [R] is possible, if they are in nonequilibrium state.

The variation of kinetic energy of atomic nuclei under external force application, induces nonequilibrium in a system ($\mathbf{S + R}$) and decoherence of $[\mathbf{C} \rightleftharpoons \mathbf{W}]$ pulsation of protons and neutrons of [S] and [R]. The nonlocal energy transmission from [S] to [R] is possible, if the decoherence is not big enough for disassembly of the virtual microtubules and their systems in the case of atoms. The electronic $\mathbf{VirG}_{SME}^e$, as more coherent (not so dependent on thermal vibrations), can be responsible for stabilization of the complex atomic Virtual Guides $\sum \mathbf{VirG}_{SME}^{e,p,n}$.

The values of the energy and velocity increments or decrements of free elementary particles are interrelated.

The instantaneous energy flux via $(\mathbf{VirG}_{SME})^i$, is mediated by pulsation of energy and radii of torus ($\mathbf{V^+}$) and antitorus ($\mathbf{V^-}$) of Bivacuum bosons: $\mathbf{BVB^+} = [\mathbf{V^+}\uparrow\downarrow \mathbf{V^-}]$. Corresponding energy increments of the actual torus and complementary antitorus of $\mathbf{BVB^\pm}$, forming $(\mathbf{VirG}_{SME})^i$, are directly related to increments of Sender particle external velocity ($\Delta\mathbf{v}$):

$$\Delta\mathbf{E}_{V^+} = +\Delta\mathbf{m}_V^+c^2 = \left( +\frac{\mathbf{p}^+}{\mathbf{R}^2}(\Delta\mathbf{v})_{\mathbf{F_\uparrow^-}}^{[\mathbf{F_\uparrow^+} \bowtie \mathbf{F_\downarrow^-}]} = \mathbf{m}_V^+\mathbf{c}^2 \frac{\Delta\mathbf{L}_{V^+}}{\mathbf{L}_{V^+}} \right)_{N,S} \quad \text{actual} \qquad 9.3$$

$$\Delta\mathbf{E}_{V^-} = -\Delta\mathbf{m}_V^-c^2 = \left( -\frac{\mathbf{p}^-}{\mathbf{R}^2}(\Delta\mathbf{v})_{\mathbf{F_\uparrow^-}}^{[\mathbf{F_\uparrow^+} \bowtie \mathbf{F_\downarrow^-}]} = -\mathbf{m}_V^-\mathbf{c}^2 \frac{\Delta\mathbf{L}_{V^-}}{\mathbf{L}_{V^-}} \right)_{N,S} \quad \text{complementary} \qquad 9.4$$

where: $\mathbf{p}^+ = \mathbf{m}_V^+\mathbf{v}$; $\mathbf{p}^- = \mathbf{m}_V^-\mathbf{v}$ are the actual and complementary momenta; $\mathbf{L}_{V^+} = \hbar/\mathbf{m}_V^+\mathbf{c}$ and $\mathbf{L}_{V^-} = \hbar/\mathbf{m}_V^-\mathbf{c}$ are the radii of torus and antitorus of $\mathbf{BVB^\pm} = [\mathbf{V^+} \Updownarrow \mathbf{V^-}]$, forming $\mathbf{VirG}_{S,M,E}$.

The nonlocal energy exchange between [S] and [R] is accompanied by the instant pulsation of radii of tori ($V^+$) and antitori ($V^-$) of $BVF^{\updownarrow}$ and $BVB^\pm$, accompanied by corresponding pulsation $|\Delta L_{V^\pm}/L_{V^\pm}|$ of the whole virtual microtubule $\mathbf{VirG}_{SME}$ (Fig.4).

The nonequilibrium state of elementary particles of [S] and [R], connected by $\mathbf{VirG}_{S,M,E}$, means difference in their kinetic and total energies and frequency of de Broglie waves and that of $[\mathbf{C} \rightleftharpoons \mathbf{W}]$ pulsation. The consequence of this difference are beats between states of [S] and [R], equal to frequency of $\mathbf{VirG}_{SME}$ radius pulsation (see eq.7.3):

$$\Delta v_{\mathbf{VirG}}^{S,R} = v_{\mathbf{C} \rightleftharpoons \mathbf{W}}^{S} - v_{\mathbf{C} \rightleftharpoons \mathbf{W}}^{R} = \frac{\mathbf{c}^2}{h}\Big[(\mathbf{m}_V^+)^S - (\mathbf{m}_V^+)^R\Big] = \qquad 9.4a$$

$$= \frac{1}{h}\left[\mathbf{m}_0\mathbf{c}^2(\mathbf{R}^S - \mathbf{R}^R) + \left(\frac{h^2}{(\mathbf{m}_V^+\boldsymbol{\lambda}_B^2)^S} - \frac{h^2}{(\mathbf{m}_V^+\boldsymbol{\lambda}_B^2)^R}\right)\right]$$

The beats between the total frequencies of [S] and [R] states (electrons, protons or neutrons), connected by $\mathbf{VirG}_{S,M,E}$ and different excitation states ($j-k$) of $[\mathbf{BVF}^\uparrow \bowtie \mathbf{BVF}^\downarrow]_{j-k}$ are accompanied by *emission* $\rightleftharpoons$ *absorption* of positive and negative virtual pressure waves: $\mathbf{VPW}^+$ and $\mathbf{VPW}^-$, generating positive and negative virtual pressure: $\mathbf{VirP}^+$ and $\mathbf{VirP}^-$.

The difference between total energies of elementary particles of Sender and Receiver can be expressed via these virtual pressures as:

$$\mathbf{E}_{tot}^S - \mathbf{E}_{tot}^S = h\Delta v_{\mathbf{VirG}}^{S,R} = \Delta(\mathbf{m}_V^+\mathbf{c}^2)^{S,R} = \Delta\mathbf{V} + \Delta\mathbf{T_k} = \qquad 9.4b$$

$$= \frac{1}{2}\Delta(\mathbf{m}_V^+ + \mathbf{m}_V^-)^{S,R}\mathbf{c}^2 + \frac{1}{2}\Delta(\mathbf{m}_V^+ - \mathbf{m}_V^-)^{S,R}\mathbf{c}^2 \sim$$

$$\sim \Delta(\mathbf{VirP}^+ + \mathbf{VirP}^-)^{S,R} + \Delta(\mathbf{VirP}^+ - \mathbf{VirP}^-)^{S,R} \qquad 9.4c$$

If the temperature or kinetic energy of [S] is higher, than that of [R]: $\mathbf{T}_S > \mathbf{T}_R$, then $\Delta v_{\mathbf{VirG}}^{S,R} > 0$ and the *direction* of momentum and energy flux, mediated by positive and negative virtual pressure of subquantum particles and antiparticles: $\Delta\mathbf{VirP}^+$ and $\Delta\mathbf{VirP}^-$, is from [S] to [R]. The opposite nonequilibrium state of system, i.e. $\mathbf{T}_S < \mathbf{T}_R$ provides the opposite direction of energy/momentum flux - from [R] to [S].

The proposed mechanism of Pauli repulsion between fermions of the same spin state (Kaivarainen http://arxiv.org/abs/physics/0103031) also may play a role in repulsion between Sender and Receiver.

The length of $VirG_{SME}$, connecting elementary particles, can vary in the process of [S] *and* [R] interaction because of fast self-assembly of Bivacuum dipoles into virtual microtubules.

### 9.2 The mechanism of spin/information exchange between tuned particles of Sender and Receiver via $VirG_{\mathbf{S,M,E}}$

Most effectively the proposed mechanism of spin (information), momentum and energy exchange can work between Sender and Receiver, containing coherent molecular clusters - mesoscopic Bose condensate (mBC) (Kaivarainen, 2001, 2005). The nonlocal spin/qubit exchange between [S] and [R] via $\mathbf{VirG}_{SME}^i$ does not need the radius pulsation, but only the instantaneous polarization change of Bivacuum bosons $(\mathbf{BVB}^+ \rightleftharpoons \mathbf{BVB}^-)^i$, which is interrelated with instant spin state change of both Bivacuum fermions, forming virtual Cooper pairs in the double virtual microtubule:

$$[\mathbf{BVF}^\uparrow \bowtie \mathbf{BVF}^\downarrow]^i \overset{(S=+1/2)\to(S=-1/2)}{\rightleftharpoons} [\mathbf{BVF}^\downarrow \bowtie \mathbf{BVF}^\uparrow]^i$$

The instantaneous angular momentum (spin) exchange between The frequency of this flip-flop process is determined by frequency of reorientation of semi-integer spin of fermions of Sender and counterphase spin state change of corresponding fermion of Receiver.

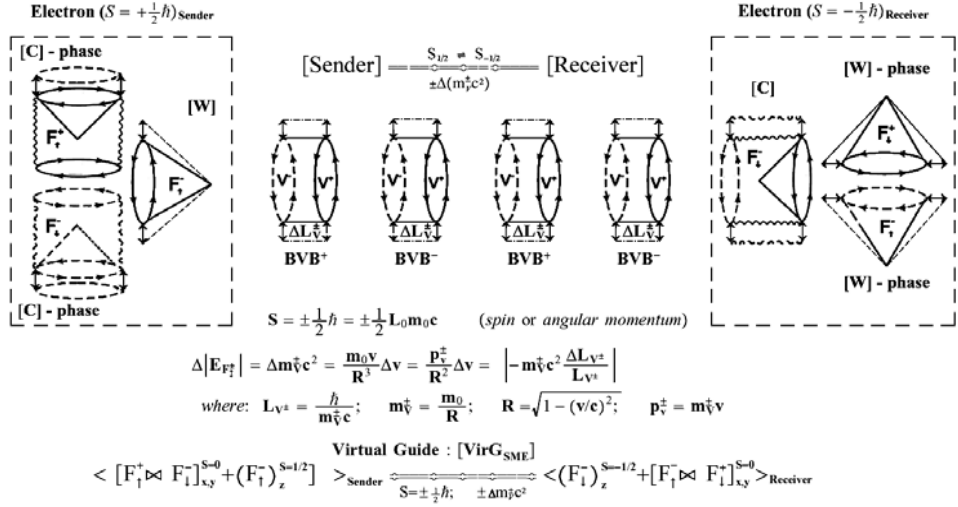

**Fig**. **4**. The mechanism of nonlocal Bivacuum mediated interaction (entanglement) between two distant unpaired sub-elementary fermions of 'tuned' elementary triplets (particles) of the opposite spins $< [\mathbf{F}_\uparrow^+ \bowtie \mathbf{F}_\downarrow^-] + \mathbf{F}_\uparrow^- >_{\text{Sender}}^i$ and $< [\mathbf{F}_\uparrow^+ \bowtie \mathbf{F}_\downarrow^-] + \mathbf{F}_\downarrow^- >_{\text{Receiver}}^i$, with close frequency of $[\mathbf{C} \rightleftharpoons \mathbf{W}]$ pulsation and close de Broglie wave length ($\lambda_\mathbf{B} = \mathbf{h}/\mathbf{m}_V^+\mathbf{v}$) of particles. The tunnelling of momentum and energy increments: $\Delta|\mathbf{m}_V^+\mathbf{c}^2|$ ~$\Delta|\mathbf{VirP}^+| \pm \Delta|\mathbf{VirP}^-|$ from Sender to Receiver and vice-verse via Virtual spin-momentum-energy Guide [$\mathbf{VirG}_{\mathbf{SME}}^i$] is accompanied by instantaneous pulsation of diameter ($2\Delta\mathbf{L}_V^\pm$) of this virtual microtubule, formed by Bivacuum bosons $\mathbf{BVB}^\pm$ or double microtubule, formed by Cooper pairs of Bivacuum fermions: [$\mathbf{BVF}^\uparrow \bowtie \mathbf{BVF}^\downarrow$]. The spin state exchange between [S] and [R] can be realized by the instantaneous change polarization of Cooper pairs: [$\mathbf{BVF}^\uparrow \bowtie \mathbf{BVF}^\downarrow$] $\rightleftharpoons$ [$\mathbf{BVF}^\downarrow \bowtie \mathbf{BVF}^\uparrow$] and Bivacuum bosons: $\mathbf{BVB}^+ \rightleftharpoons \mathbf{BVB}^-$.

The described above mechanisms of nonlocal/instant transmission of spin/information, momentum and energy between coherent clusters of elementary particles and atoms of Sender and Receiver, connected by Virtual Guides, may describe a lot of unconventional experimental results, like Kozyrev ones (http://arxiv.org/abs/physics/0103031) and Psi phenomena.

In virtual microtubules ($\mathbf{VirG}_{SME}^i$)$^i$ the time and its 'pace' are uncertain: $\mathbf{t} = \mathbf{0/0}$, if the external velocities ($\mathbf{v}$) and accelerations ($\mathbf{dv/dt}$) of Bivacuum dipoles, composing such virtual Bose condensate, are zero (see eq.7.1 and http://arxiv.org/abs/physics/0103031).

### 9.3 The role of tuning force ($\mathbf{F}_{\mathbf{VPW}^\pm}$) of virtual pressure waves $\mathbf{VPW}_q^\pm$ of Bivacuum in entanglement

The tuning between **two similar elementary** particles: 'sender (S)' and 'receiver (R)' via $\mathbf{VirG}_{SME}^i$ may be qualitatively described, using well known model of *damped harmonic oscillators,* interacting with all-pervading virtual pressure waves ($\mathbf{VPW}_{q=1}^\pm$) of Bivacuum with fundamental frequency ($\boldsymbol{\omega}_0 = \mathbf{m}_0\mathbf{c}^2/\hbar$). The criteria of tuning - synchronization of [S] and [R] is the equality of the amplitude probability of resonant energy exchange of Sender and Receiver with virtual pressure waves ($\mathbf{VPW}_{q=1}^\pm$): $\mathbf{A}_{C \rightleftharpoons W}^S = \mathbf{A}_{C \rightleftharpoons W}^R$, resulting from minimization of frequency difference $(\boldsymbol{\omega}_S - \boldsymbol{\omega}_0) \to 0$ and $(\boldsymbol{\omega}_R - \boldsymbol{\omega}_0) \to 0$:

$$\mathbf{A}_{C \rightleftharpoons W}^{S} \sim \left[ \frac{1}{(\mathbf{m}_V^+)_S} \frac{\mathbf{F}_{\mathbf{VPW}^{\pm}}}{(\boldsymbol{\omega}_S^2 - \boldsymbol{\omega}_0^2) + \mathrm{Im}\,\boldsymbol{\gamma}\boldsymbol{\omega}_S} \right] \qquad 9.5$$

$$[\mathbf{A}_{C \rightleftharpoons W}^{R}]_{x,y,z} \sim \left[ \frac{1}{(\mathbf{m}_V^+)_R} \frac{\mathbf{F}_{\mathbf{VPW}^{\pm}}}{(\boldsymbol{\omega}_R^2 - \boldsymbol{\omega}_0^2) + \mathrm{Im}\,\boldsymbol{\gamma}\boldsymbol{\omega}_R} \right] \qquad 9.5a$$

where the frequencies of $\mathbf{C} \rightleftharpoons \mathbf{W}$ pulsation of particles of Sender $(\boldsymbol{\omega}_S)$ and Receiver $(\boldsymbol{\omega}_R)$ are:

$$\boldsymbol{\omega}_R = \boldsymbol{\omega}_{\mathbf{C} \rightleftharpoons \mathbf{W}} = \mathbf{R}\,\boldsymbol{\omega}_0^{in} + (\boldsymbol{\omega}_B^{ext})_R \qquad 9.6$$

$$\boldsymbol{\omega}_S = \boldsymbol{\omega}_{\mathbf{C} \rightleftharpoons \mathbf{W}} = \mathbf{R}\,\boldsymbol{\omega}_0^{in} + (\boldsymbol{\omega}_B^{ext})_S \qquad 9.6a$$

$\boldsymbol{\gamma}$ is a damping coefficient due to *decoherence effects*, generated by local fluctuations of Bivacuum deteriorating the phase/spin transmission via $\mathbf{VirG}_{SME}$; $(\mathbf{m}_V^+)_{S,R}$ are the actual mass of (S) and (R); $[\mathbf{F}_{\mathbf{VPW}}]$ is a *tuning force of virtual pressure waves* $\mathbf{VPW}^{\pm}$ *of Bivacuum with tuning energy* $\mathbf{E}_{VPW} = \mathbf{q}\,\mathbf{m}_0\mathbf{c}^2$ *and wave length* $\mathbf{L}_{VPW} = \hbar/\mathbf{m}_0\mathbf{c}$

$$\mathbf{F}_{\mathbf{VPW}^{\pm}} = \frac{\mathbf{E}_{VPW}}{\mathbf{L}_{VPW}} = \frac{\mathbf{q}}{\hbar}\mathbf{m}_0^2\mathbf{c}^3 \qquad 9.7$$

The most probable Tuning Force (TF) has a minimum energy, corresponding to $\mathbf{q} = \mathbf{j} - \mathbf{k} = \mathbf{1}$.

The influence of *virtual pressure force* $(\mathbf{F}_{\mathbf{VPW}})$ stimulates the synchronization of [S] and [R] pulsations, i.e. $\omega_R \to \omega_S \to \omega_0$. This fundamental frequency $\boldsymbol{\omega}_0 = \mathbf{m}_0\mathbf{c}^2/\hbar$ is the same in any space volume, including those of Sender and Receiver.

The $\mathbf{VirG}_{SME}$ represent quasi 1D macroscopic virtual Bose condensate with a configuration of single microtubules, formed by Bivacuum bosons $(\mathbf{BVB}^{\pm})$ or with configuration of double microtubules, composed from Cooper pairs as described in previous section.

The effectiveness of entanglement between two or more similar elementary particles is dependent on synchronization of their $[\mathbf{C} \rightleftharpoons \mathbf{W}]$ pulsation frequency and 'tuning' the phase of these pulsations via nonlocal virtual guide $(\mathbf{VirG}_{SME})^{S,R}$ between Sender and Receiver under the action of the virtual pressure waves $\mathbf{VPW}_{q=1}^{\pm}$ and Tuning energy of Bivacuum.

The mechanism proposed may explain the experimentally confirmed nonlocal interaction between coherent elementary particles (Aspect and Gragier, 1983), atoms and remote coherent clusters of molecules.

### 9.4 The stochastic jumps of Cumulative Virtual Cloud (CVC) between primary $VR_0$ and one of
### secondary Virtual Replicas $(VR_n)$ of elementary particles, as a possible explanation of two slit experiments

In accordance to proposed mechanism of dynamics of sub-elementary particles - Bivacuum interaction, forming the photons, electrons, etc. (Fig.1 and Fig.3), their primary and secondary virtual replicas are existing. The properties of VR and their multiplication VRM(r,t) of elementary particles, described in section 8, are dependent on their de Broglie wave length, frequency and phase.

The frequency of de Broglie wave and its length can be expressed from eq.7.3 as:

$$\nu_B = \frac{(\mathbf{m}_V^+ \mathbf{v}^2)_{tr}^{ext}}{h} = \frac{\mathbf{v}}{\lambda_B} = \mathbf{v}_{\mathbf{C} \rightleftharpoons \mathbf{W}} - \mathbf{R}\nu_0 \qquad 9.8$$

$$or: \ \nu_B = \frac{\mathbf{m}_V^+ \mathbf{c}^2}{h} - \mathbf{R}\nu_0 \qquad 9.8a$$

where: $\nu_0 = \mathbf{m}_0 \mathbf{c}^2/h = \omega_0/2\pi; \ \lambda_B = h/\mathbf{m}_V^+ \mathbf{v}$

In nonrelativistic case for fermions, like electrons, when $\mathbf{v} << \mathbf{c}$ and the relativistic factor $\mathbf{R} = \sqrt{1 - (\mathbf{v}/\mathbf{c})^2} \simeq 1$, the energy of de Broglie wave is close to Tuning energy (**TE**) of Bivacuum (section 7):

$$\mathbf{E}_B = h\nu_B \simeq \mathbf{m}_V^+ \mathbf{c}^2 - \mathbf{m}_0 \mathbf{c}^2 = \mathbf{TE} \qquad 9.9$$

The fundamental phenomenon of de Broglie wave is a result of modulation of the carrying internal frequency of $[\mathbf{C} \rightleftharpoons \mathbf{W}]$ pulsation ($\omega_{in} = \mathbf{R}\omega_0 = \mathbf{R}\mathbf{m}_0 \mathbf{c}^2/h$) by the angular frequency of the de Broglie wave: $\omega_B = \mathbf{m}_V^+ \mathbf{v}_{tr}^2/h = 2\pi \mathbf{v}/\lambda_B$, equal to the frequency of beats between the actual and complementary torus and antitorus of the *anchor* Bivacuum fermion ($\mathbf{BVF}_{anc}^\updownarrow$) of unpaired $\mathbf{F}_\updownarrow^\pm$. The Broglie wave length $\lambda_B = h/(\mathbf{m}_V^+ \mathbf{v})$ and mass symmetry shift of $\mathbf{BVF}_{anc}^\updownarrow$ is determined by the external translational momentum of particle: $\vec{\mathbf{p}} = \mathbf{m}_V^+ \vec{\mathbf{v}}$. For nonrelativistic particles $\omega_B << \omega_0$. For relativistic case, when $\mathbf{v}$ is close to $\mathbf{c}$ and $\mathbf{R} \simeq 0$, the de Broglie wave frequency is close to resulting frequency of $[\mathbf{C} \rightleftharpoons \mathbf{W}]$ pulsation: $\omega_B \simeq \omega_{\mathbf{C} \rightleftharpoons \mathbf{W}}$.

Introduced in our theory notion of *Virtual replica (VR) multiplication (VRM)* of any material object in Bivacuum is a result of interference of basic Virtual Pressure Waves ($\mathbf{VPW}_{q=1}^\pm$) and Virtual Spin Waves ($\mathbf{VirSW}_{q=1}^{\pm 1/2}$) of Bivacuum (reference waves), with primary VR of the object.

The feedback reaction of one of copies of *Virtual replica* of VRM on its original and corresponding translational momentum exchange may induce the self-interference, displaying itself like wave - like behavior of even a *singe* elementary fermion $\langle [\mathbf{F}_\uparrow^- \bowtie \mathbf{F}_\downarrow^+] + \mathbf{F}_\updownarrow^\pm \rangle^{e,p}$ (Fig.1) or boson, like the photon (Fig. 3).

For free elementary particles the notion of *secondary virtual replica*, as one of multiplicated primary $VR_0$ coincides with notion of one of possible '*anchor sites*' (eq.6.36), as a conjugated dynamic complex of three Cooper pair of asymmetric fermions. The in-phase pulsation of Cooper pairs of asymmetric Bivacuum fermions of the *anchor site or secondary VR*, like the pairs $[\mathbf{F}_\uparrow^- \bowtie \mathbf{F}_\downarrow^+]$ of particles themselves, are the source of positive and negative basic Virtual pressure waves: $[\mathbf{VPW}_{q=1}^+ \bowtie \mathbf{VPW}_{q=1}^-]$. As far as the frequency and length of $\sum \mathbf{VR}$ or AS are the same, as exited by paired sub-elementary fermion and anifermion of particle in triplets $\langle [\mathbf{F}_\uparrow^- \bowtie \mathbf{F}_\downarrow^+] + \mathbf{F}_\updownarrow^\pm \rangle^{e,p}$, the interference pattern displays itself, when the both slits are open. It is important to note, that, if only one of two slit is open, the photon or electron can be registered in points of screen, far from the strait direction of particles propagation, where the interference make this registration impossible. This confirms not only the self-interference effects in case of single particle, but as well broad spatial distribution of the *anchor sites*, preexisting in the process of particle propagation in space. See eq.6.36 in section 6:

$$\mathbf{AS}(\mathbf{r}, \mathbf{t}) = \sum^N 3[\mathbf{BVF}_\downarrow^- \bowtie \mathbf{BVF}_\uparrow^+]_n \longrightarrow \sum^N 3[\mathbf{F}_\downarrow^- \bowtie \mathbf{F}_\uparrow^+]_n \qquad 9.10$$

We can see from the above analysis, that our model of duality does not need the Bohmian "quantum potential" (Bohm and Hiley, 1993) or de Broglie's "pilot wave" for explanation of wave-like behavior of elementary particles and two-slit experiment.

Scattering of the photon on a free electron will affect its velocity, momentum, mass, wave B frequency, length, its primary and secondary virtual replica (VR$_S$) and feedback influence of VR$_S$ on the electron, following by change of the interference picture. Our theory predicts that applying of the EM field to *singe electrons* with frequency resonant to their de Broglie frequency, should be accompanied by alternative acceleration of the electrons and modulation of their Virtual Replica. This can be accompanied by 'washing out' the interference pattern in two-slit experiment as a result of induced decoherence between particle and its virtual replica. This consequence of our theory of two-slit experiment can be easily verified.

## 10 New kind of Bivacuum mediated nonlocal interaction between macroscopic objects and possible temporal effects

In accordance to our approach, the remote interaction between macroscopic Sender [S] and Receiver [R] can be realized, as a result of *Bivacuum mediated interaction (BMI)*, like superposition of distant and nonlocal components of their Virtual Replicas Multiplication ($\mathbf{VRM}_S \leftrightarrows \mathbf{VRM}_R$), described in previous sections.

Nonequilibrium processes in [Sender], accompanied by acceleration of particles, like evaporation, heating, cooling, melting, boiling etc. may stimulate the *nonelastic effects* in the volume of [Receiver] and increments of modulated virtual pressure and spin waves ($\mathbf{\Delta VPW_m^{\pm}}$ and $\mathbf{\Delta VirSW_m^{\pm 1/2}}$), accompanied [$\mathbf{C \rightleftharpoons W}$] pulsation of triplets $[\mathbf{F_\uparrow^+ \bowtie F_\downarrow^-}] + \mathbf{F_\updownarrow^{\pm}} >^i$ , formed by sub-elementary fermions of different generation, representing electrons, protons and neutrons.

The following unconventional kinds of effects of nonelectromagnetic and non-gravitational nature can be anticipated in the remote interaction between **macroscopic** nonequilibrium [Sender] and sensitive detector [Receiver] via multiple Virtual spin and energy guides $\mathbf{VirG}_{SME}$ (Fig.4), if our theory of nonlocal spin, momentum and energy exchange between [S] and [R], described above is correct:

**I**. Weak repulsion and attraction between 'tuned' [S] and [R] and rotational momentum in [R] induced by [S], as a result of transmission of momentum/kinetic energy and angular momentum (spin) between elementary particles of [S] and [R]. The probability of such 'tuned' interaction between [S] and [R] is dependent on dimensions of coherent clusters of atoms and molecules of condensed matter in state of mesoscopic Bose condensation (**mBC**) (Kaivarainen, 1995; 2001; 2003; 2004). The number of atoms in such clusters is related to number of $\mathbf{VirG}_{SME}$ in the bundles, connecting **mBC** of in [S] and [R] and may be regulated by temperature, ultrasound, etc. The kinetic energy distant transmission from atoms of [S] to atoms of [R] may be accompanied by the temperature and local pressure/sound effects in [R];

**II**. Increasing the probability of thermal fluctuations in the volume of [R] due to decreasing of Van der Waals interactions, because of charges screening effects, induced by overlapping of distant virtual replicas of [S] and [R] and increasing of dielectric permittivity of Bivacuum. In water the variation of probability of cavitational fluctuations should by accompanied by the in-phase variation of pH and electric conductivity due to shifting the equilibrium: $H_2O \rightleftharpoons H^+ + HO^-$ to the right or left;

**III**. Small changing of mass of [R] in conditions, changing the probability of the inelastic recoil effects in the volume of [R] under influence of [S];

**IV**. Registration of metastable virtual particles (photons, electrons, positrons), as a result of Bivacuum symmetry perturbations.

*The first kind* (I) of new class of interactions between coherent fermions of [S] and [R] is a result of huge number (bundles) of correlated virtual spin-momentum-energy guides $\mathbf{VirG}_{SME} \equiv [\mathbf{VirSW}_S^{\circlearrowleft} =\diamond \diamond= \mathbf{VirSW}_R^{\circlearrowright}]$ formation by standing spin waves ($\mathbf{VirSW}_{S,R}$).

These guides can be responsible for:

a) virtual signals (phase/spin), momentum and kinetic energy instant transmission between [S] and [R], meaning the nonlocal information and energy exchange;

b) the regulation of Pauli repulsion effects between fermions of [S] and [R] with parallel spins;

c) the transmission of macroscopic rotational momentum from [S] of [R]. The macroscopic rotational momentum, exerted by $\mathbf{VirSW}_S$, is dependent on the difference between the *external* angular momentums of elementary fermions of [S] and [R].

*The second kind* (II) of phenomena: influence of [S] on probability of thermal fluctuations in [R], - is a consequence of the additional symmetry shift in Bivacuum fermions ($\mathbf{BVF}^{\updownarrow}$), induced by superposition of distant and nonlocal Virtual Replicas of [S] and [R]: $\mathbf{VR}^S \bowtie \mathbf{VR}^R$, which is accompanied by increasing of Bivacuum fermions ($\mathbf{BVF}^{\updownarrow} = [\mathbf{V}^+ \Updownarrow \mathbf{V}^-]$) virtual charge: $\Delta \mathbf{e} = (\mathbf{e}_{V^+} - \mathbf{e}_{V^-}) << \mathbf{e}_0$ in the volume of [R]. Corresponding increasing of Bivacuum permittivity ($\varepsilon_0$) and decreasing magnetic permeability ($\mu_0$) : $\varepsilon_0 = 1/(\mu_0 \mathbf{c}^2)$ is responsible for the charges screening effects in volume of [R], induced by [S]. This weakens the electromagnetic Van der Waals interaction between molecules of [R] and increases the probability of defects origination and cavitational fluctuations in solid or liquid phase of Receiver.

*The third kind of phenomena (III)*: reversible decreasing of mass of rigid [R] can be a result of reversible lost of energy of Corpuscular phase of particles, as a consequence of inelastic recoil effects, following the in-phase $[\mathbf{C} \rightarrow \mathbf{W}]$ transition of $\mathbf{N}_{coh}$ coherent nucleons in the volume of [R].

The probability of recoil effects can be enhanced by heating the rigid object or by striking it by another hard object. This effect can be registered directly - by the object mass decreasing. In conditions, close to equilibrium, the Matter - Bivacuum energy exchange relaxation time, following the process of coherent $[\mathbf{C} \rightleftharpoons \mathbf{W}]$ pulsation of macroscopic fraction of atoms is very short and corresponding mass defect effect is undetectable. *Such collective recoil effect of coherent particles* could be big in superconducting or superfluid systems of macroscopic Bose condensation or in good crystals, with big domains of atoms in state of Bose condensation.

*The fourth kind of the above listed phenomena* - increasing the probability of virtual particles and antiparticles origination in asymmetric Bivacuum in condition of forced resonance with exciting Bivacuum virtual waves was discussed earlier (Kaivarainen http://arxiv.org/abs/physics/0103031).

It was demonstrated (Kaivarainen: http://arxiv.org/abs/physics/0207027), that the listed nontrivial consequences of our Unified theory (I - IV) are consistent with unusual data, obtained by groups of Kozyrev (1984; 1991) and Korotaev (1999; 2000). It is important to note, that these experiments are incompatible with existing today paradigm. It means that the current paradigm is timed out and should be replaced by the new one.

### 10.1 Temporal effects in evolution of complex Virtual Replica Multiplication: VRM(t)

The 3D net of $\mathbf{VirG}_{SME}$, unifying the Universe, may transmit not only information, but also kinetic energy and momentum from one cosmic objects to another via huge distance. For such super-entanglement the remote interacting particles in regions should be 'tuned' to each other. The latter can be achieved by forced resonance of Bivacuum Virtual Pressure Waves ($\mathbf{VPW}^{\pm}$) with pulsing elementary particles (see section 9.3).

The general formula for time, discussed in section (7), is:

$$\left[ \mathbf{t} = -\frac{\vec{\mathbf{v}}}{\mathbf{d}\vec{\mathbf{v}}/\mathbf{dt}} \frac{1-(\mathbf{v/c})^2}{2-(\mathbf{v/c})^2} = -\frac{\mathbf{dt}}{\mathbf{dT}_k} \mathbf{T}_k \right]_{x,y,z} \qquad 10.1$$

The pace of time and time itself are positive ($\mathbf{t} > \mathbf{0}$), if the particles motion is slowing down ($\mathbf{dv/dt} < \mathbf{0}$ and $\mathbf{dv} < \mathbf{0}$) and negative, if particles are accelerating. In the absence of acceleration ($\mathbf{dv/dt} = \mathbf{0}$ and $\mathbf{dv} = \mathbf{0}$), the time is infinitive and its pace zero:

$$\mathbf{t} \rightarrow \infty \qquad \text{and} \qquad \frac{\mathbf{dt}}{\mathbf{t}} \rightarrow \mathbf{0} \qquad 10.2$$

$$at \quad \mathbf{dv/dt} \rightarrow \mathbf{0} \quad \text{and} \qquad \mathbf{v} = \mathbf{const} \qquad 10.2a$$

Standing waves satisfy this condition, as well, as postulated principle of *internal* kinetic energy of torus ($\mathbf{V}^+$) and antitorus ($\mathbf{V}^-$) of asymmetric Bivacuum fermions/antifermions: $\left(\mathbf{BVF}_{as}^{\updownarrow}\right)^{\phi} \equiv \mathbf{F}_{\mp}^{\pm}$ conservation (eq. 3.1).

As far the Universe expands with acceleration: $\mathbf{dv/dt} > \mathbf{0}$ and $\mathbf{v} > \mathbf{0}$, it follows from 12.11, that the Universal time - "TIME ARROW" is negative: $\mathbf{t} < \mathbf{0}$. However, the 'local' time of each cooling planet and star is positive as far $\mathbf{dv/dt} < \mathbf{0}$ and $\mathbf{v} > \mathbf{0}$ in 12.11.

The permanent ($\mathbf{t} \rightarrow \infty$) collective motion of the electrons and atoms of $^4\mathbf{He}$ in superconducting and superfluid rings, correspondingly, with constant velocity ($\mathbf{v} = \mathbf{const}$), i.e. in the absence of collisions and accelerations, is a good example of dynamic system with infinitive life-time.

*In Bivacuum with superfluid properties the excitations, like quantized vortices and rotating virtual filaments, like asymmetric Virtual Guides (VirG) with $\mathbf{T}_k = const$, are stable ($\mathbf{t} = \infty$). Such long - living Bivacuum excitations may represent at the pointed conditions the primary and secondary Virtual Replicas (VR), generated in process of their multiplication (VRM).*

At conditions: $[\mathbf{v} = \mathbf{c} = \mathbf{const}]$, valid for the case of photons, we get from (12.11) the uncertainty for time, like: $\mathbf{t} = \mathbf{0/0}$. The similar result we get for state of virtual Bose condensate (**VirBC**) in Bivacuum, when the *external translational* velocity of Bivacuum fermions ($\mathbf{BVF}^{\updownarrow}$) and Bivacuum bosons ($\mathbf{BVB}^{\pm}$) is equal to zero ($\mathbf{v} = \mathbf{0} = \mathbf{const}$). The latter condition corresponds to totally symmetric $\mathbf{BVF}^{\updownarrow}$ and $\mathbf{BVB}^{\pm}$, when their torus ($\mathbf{V}^+$) and antitorus ($\mathbf{V}^-$) mass, charge and magnetic moments compensate each other.

The uncertainty of time is valid also in part of Universe Virtual Replica (UVR), presented by bundles of filaments of Virtual Guides (VirG$_{SME}$), formed by polymerized symmetric $\sum \mathbf{BVB}^{\pm}$ or Cooper pairs of Bivacuum fermions $\sum[\mathbf{BVF}^{\uparrow} \bowtie \mathbf{BVF}^{\downarrow}]$, when their external kinetic energy is permanent and equal to zero: $\mathbf{T}_k = 0 = const$. *In such conditions the past and future are indistinguishable.*

**In any kind of virtual systems, like VR and VRM(r,t), the simulation of past and future events is possible by Quantum supercomputer or Superconsciousness (see Table 2 in section 12), because by the definition, the relativistic mechanics do not work for them and the causality principle is absent.**

Shifting of Bivacuum dipoles symmetry by physical fields (electric or gravitational), accompanied by their acceleration/deceleration and acquiring the external kinetic energy, may influence the value and even sign of pace of time of virtual replicas, i.e. their future or past.

The time-dependent superposition of individual virtual replicas multiplication of inorganic objects $\mathbf{VRM_{Ob}}(t)$ and living organisms $\mathbf{VRM_L}(t)$ of each star (solar) system forms the Star system Virtual Replica (**SVR**). In turn, superposition of all stars systems virtual replica SVR of Galactic can be responsible for formation of Galactical Virtual

Replica.

We put forward a conjecture, that as a result of ability of complex virtual replicas, like human ones or *Galactical virtual replica (GVR) with active medium properties* to evolution/self-organization and informational processing, this process can be considered as a corresponding supercomputer or Galactical Virtual Consciousness (**GVC**), as a function of **GVR(t)**:

$$\mathbf{GVC(t) = F[GVR(t)]} = \sum \mathbf{SSC(t)} \qquad 10.3$$

$$where: \quad \mathbf{SVR = F}\left\{\sum \; [\mathbf{VRM_{Ob}}(t) \Rightarrow \Leftarrow \mathbf{VRM_L}(t)]\right\}$$

where: $\mathbf{VRM_{Ob}}(t)$ and $\mathbf{VRM_L}(t)$ are the selected/individual virtual replicas multiplication of inorganic objects and living organisms not only in space, but also in time, in the case of their existence in given star system.

It follows from our approach, that Superconsciousness of the Universe is a huge hierarchical system, resulting from Bivacuum mediated interaction (BMI) between Virtual Consciousness of individual planets (like Earth Noosphere), stars, star systems, Galactical virtual consciousness, etc.

Our description of the Earth Virtual Replicas of the Universe, Solar system and Earth, their interrelation with corresponding levels of Superconsciousness, have some common features with Henry Stapp (1982) hypothesis of interrelation between collapsing of superimposed mental states and actual world.

In each selected current moment of time $\mathbf{t = t}_C$, as a result of *Virtual consciousness (VC) activity*, a big number of discreet time - *evolution* versions $VR_F^i(t_F)$ of current $VR^0$ with different probability ($P_i$) of realization appears. The most probable future virtual replica $\mathbf{VR}_{Future}$ can be calculated as:

$$\mathbf{VR}_{Future} = \frac{\sum P_i \; [VR_F^i(t_F)]}{\sum P_i} \qquad 10.4$$

In similar way we have the most probable virtual replica of the past infinitive number of possible metastable Virtual replicas $VR_P^i(t_P)$ of past, as a result of time - *involution* of current $VR^0$ :

$$\mathbf{VC}_{Past} = \frac{\sum P_P^i \; [VR_P^i(t_P)]}{\sum P_P^i} \qquad 10.5$$

*Clairvoyance or anticipation* is a result of ability of gifted psychic [Sender] to 'search', using specific 'key images' at first stage - the right $\mathbf{VR}^0$ at current time and then select from this future or past set of virtual replicas the most probable one. This complex process includes very 'tuned' interaction of the astral and mental bodies (distant VR) and ether (local VR) bodies of [Sender] with $\sum^{\infty} SVR_F(t)$. Similar mechanism works, in accordance to our approach, in extra-perception by Sender - psychic of the past of some individual $VR_P(t)$, as a selected component of $\sum_{\infty} SVR_P(t)$.

The proposed in our work mechanism of interaction of human consciousness of Receptor and related virtual replica with virtual replicas of Sender consciousness, corresponding to certain notions and images, means *telepathy*. Similar mechanism can be responsible for possibility of sharing of well formulated by individuals new *IDEAS* with certain population of mankind, sensitive enough for perception of specific VRM, corresponding to certain mental state of this individuals, induced by such ideas or notions.

*The 'phantom' effects* where revealed in a system of interacting 'charged' by intention vessel of water and few other distant vessels with aqueous solutions, surrounding the 'charged' vessel (Tiller, Dibble and Kohane, 2001 and www.tiller.org). After replacing the 'charged' vessel far out of system, the *'memory' of its presence* remains for a long time. The presence and orientation of large quartz crystal strongly affected the amplitude of 'phantom' effect. In experiment, described, screening of the target [R] from electromagnetic fields by Faraday's cage did not influence on the distant interaction between [S] and [R] and the phantom effect.

*Consequently, there are a lot of experimental evidence already, confirming the existence of new fundamental remote Bivacuum Mediated Interaction (BMI) between Sender and Receiver, following from our Unified Theory.*

## 11 Hierarchic Model of Consciousness:
## From Molecular Bose Condensation to Synaptic Reorganization

Our Hierarchic Model of Consciousness (Kaivarainen: http://arxiv.org/abs/physics/0003045) is based on Hierarchic Theory of Matter (Kaivarainen, 2000; 2001; http://arxiv.org/abs/physics/0102086). The idea of Karl Pribram of his book: 'Languages of the Brain' (1977) of holographic principles of memory and braining is very popular in quantum models of consciousness. Our model also supports this general idea and try to transform it in concrete shape. The code way of keeping information in the form of the effectons and deformons as 3D standing waves (de Broglie waves, electromagnetic, acoustic and vibro-gravitational), generated by microtubules, containing water in state of mesoscopic Bose condensation (mBC) - looks very effective and may be used in quantum computer technology.

### 11.1 Stages of Hierarchic Model of Consciousness

In accordance with our HMC, the sequence of following interrelated stages is necessary for elementary act of perception and memory, resulted from simultaneous excitation and depolarization of big enough number of neurons, forming cooperative ensemble:

1. The change of the electric component of neuron's body internal electromagnetic field as a result of cells depolarization;

2. Opening the potential - dependent $Ca^{2+}$ channels and increasing the concentration of these ions in cytoplasm. Activation of $Ca^{2+}$ - dependent protein gelsolin, which stimulate fast disassembly of actin filaments;

3. Shift of $A \rightleftharpoons B$ equilibrium between the closed (A) and open to water (B) states of cleft, formed by $\alpha$ and $\beta$ tubulins in tubulin pairs of microtubules (MT) to the right as a consequence of piezoelectric effect, induced by depolarization of membrane of nerve cell;

4. Increasing the life-time and dimensions of coherent "flickering" water clusters in MT, representing the 3D superposition of de Broglie standing waves of $H_2O$ molecules in hollow core of MT. It is a result of the water molecules immobilization in the 'open' nonpolar clefts of $(\alpha\beta)$ dimers of MT;

5. Increasing the superradiance of coherent IR photons induced by synchronization of quantum transitions of the *effectons* between *acoustic* and *optic* like states. Corresponding increasing of probability of superdeformons (cavitational fluctuations) excitation in water of cytoplasm;

6. The *disassembly of actin filaments system to huge number of subunits, [gel→sol] transition and increasing of water fraction in hydration shell of proteins in cytoplasm.* This transition is a result of cavitational fluctuations and destabilization of actin filaments by $Ca^{2+}$. Corresponding decreasing the water activity in cytoplasm - increases strongly the passive osmotic diffusion of water from the external volume to the cell;

7. As a consequence of previous stage, a jump-way increasing of the nerve cell body volume (pulsation), accompanied by disrupting the (+) ends of MTs with cytoplasmic membranes occur. This stage makes it possible for MTs to change their orientation inside neuron's body;

8. Spatial "tuning" - collective reorientation of MTs of simultaneously excited neurons to geometry, corresponding to minimum potential energy of distant (but not nonlocal) electromagnetic and vibro-gravitational interaction between MTs and centrioles twisting;

9. Decreasing the concentration of $Ca^{2+}$ to the critical one, when disassembly of actin filaments is stopped and [gel $\rightleftharpoons$ sol] equilibrium shifts to the left again, stabilizing the new MTs system spatial configuration and corresponding nerve cell body volume and geometry. This new geometry of nerve cells after fixation of (+) ends of MTs back to plasmatic membrane - determine the new distribution of ionic channels activity and reorganization of synaptic contacts in all excited ensemble of neurons after relaxation, i.e. *short-term and long-term memory.*

This cyclic consequence (hierarchy) of quantum mechanical, physicochemical and classical nonlinear events can be considered as elementary acts of memorizing and consciousness. The total period of listed above stages can be as long as 500 ms, i.e. half of second.

The resonance wave number of excitation of superdeformons, leading from our model, is equal to 1200 $(1/cm)$. The experiments of Albrecht-Buehler (1992) revealed that just around this frequency the response of surface extensions of 3T3 cells to weak IR irradiation is maximum. Our model predicts that IR irradiation of microtubules system *in vitro* with this frequency will dramatically increase the probability of microtubules catastrophes. It's one of the way to verify our model experimentally.

Except superradiance, two other known cooperative optic effects could be involved in supercatastrophe realization: *self-induced bistability and pike regime* of IR photons radiation. Self-induced bistability is light-induced phase transition. It could be related to nonlinear shift of [$a \Leftrightarrow b$] equilibrium of primary librational water effectons in MT to the right, as a result of saturation of IR (lb)-photons absorption. As far the molecular polarizability and dipole moments in (a) and (b) states of the primary effectons - differs, such shifts of [$a \Leftrightarrow b$] equilibrium should be accompanied by periodic jumps of dielectric permeability and stability of coherent water clusters. These shifts may be responsible for the pike regime of librational IR photons absorption and radiation. *As far the stability of b-states of lb effectons is less than that of a-states, the characteristic frequency of pike regime can be correlated with frequency of MTs - supercatastrophe activation.*

The Brownian effects, which influence reorientation of MTs system and probability of cavitational fluctuations, stimulating [gel - sol] transition in elementary act of consciousness - represent in our model the *non-computational element of consciousness.* Other models relate this element to wave function collapsing (Hameroff and Penrose, 1996).

## 11.2 Special features of Virtual Replica (VR), generated by neurons

Our Hierarchic Model of Consciousness - HMC (Kaivarainen, 2000b; 2001) is based on Hierarchic theory of condensed matter (Kaivarainen, 1995; 2000a; 2001). In accordance to this theory, coherent properties of water clusters in *microtubules (MT)* and distant exchange by IR photons, radiated by these clusters (*mesoscopic molecular Bose condensate - mBC*) may be responsible for distant interaction between MT of different neurons and neuron ensembles with similar orientation of MT. The existences of distant coherent interaction (from 10 microns to one millimeter) in brain has been confirmed recently by magnetic resonance imaging (MRI) by Warren and colleagues in Princeton university (Rizi et al., 2000; Richter et al., 1995; 2000, see also http://www.princeton.edu/~wwarren/NMRintro).

In accordance to our HMC, each specific kind of neuron ensembles excitation - corresponds to hierarchical system of three-dimensional (3D) standing waves of following interrelated kinds: thermal de Broglie waves (waves B), produced by anharmonic vibrations of molecules; electromagnetic (IR) waves; acoustic waves and vibro-gravitational waves (Kaivarainen, 2000b; 2001). Corresponding complex hologram may be responsible for distant quantum neurodynamics regulation and for morphogenetic field.

In this aspect our approach has a common features with ideas, developed by Pribram (1991).

In our model we consider a number of quantum collective excitations, resulted from coherent anharmonic translational and librational oscillations of water in the hollow core of the microtubules (MT) with internal diameter about 120 Å. It was shown, that water fraction, corresponding to 3D standing de Broglie waves, related to librations, represent 3D mesoscopic molecular Bose condensate (mBC) in form of coherent clusters. The dimensions of water clusters (nanometers) and frequency of their IR radiation may be enhanced by interaction with walls of MT. It is most organized and orchestrated fraction of condensed matter in biological cells. The Brownian effects, which influence reorientation of MT system and probability of cavitational fluctuations, stimulating [gel - sol] transition in nerve cells - may be responsible for non-computational element of consciousness. Other models (Wigner, 1955; Penrose, 1994; Hameroff and Penrose, 1996) relate this element to wave function collapse.

Change of the ordered fraction of water in microtubules in form of mBC, leads to [gel-sol] transition, related to reversible [assembly - disassembly] of actin microfilaments, change of osmotic pressure, pulsation of cells volume and membranes deformation. Corresponding "holomovement" of Virtual replica (VR) of living organism may be responsible for mind-matter interaction, telepathy and other phenomena, related to parapsychology. The bigger is number of MTs with similar orientation of coherently interacting cells, the bigger is corresponding fraction of ordered water, very sensitive to nerve excitation.

It is shown experimentally that the frequency of [$gel \rightleftharpoons sol$] transitions in cytoplasm, regulated by $Ca^{2+}$ and enzyme gelsolin, is about 40 s$^{-1}$ (Miamoto, 1995, Muallem et al. 1995).

There are evidence, pointing that spatial properties of DNA and [MTs + microtubules associated proteins (MAP)] system follow the Golden mean or Fibonacci series rule (see related articles by Stuart Hameroff:
www.consciusness.arizona.edu/hameroff/info-processing01.htm). In accordance to our results, it is a condition, optimal for exchange interaction of matter with Bivacuum by means of Bivacuum gap oscillations. The introduced in section 3, notion of *Tuning energy of Bivacuum* looks to be responsible for adaptation of biosystems dynamic and spatial properties to those of Bivacuum in a course of long term biological evolution. It is a result of forced resonance of [$C \rightleftharpoons W$] pulsations of the electrons and nuclears of condensed matter (water, proteins, DNA, microtubules, other microfilaments, etc.) with basic Bivacuum virtual pressure waves: $VPW_{q=1}^{\pm}$, corresponding to their lower excitation state: $q = j - k = 1$. This kind of interaction is responsible also for realization of 2nd and 3d laws of thermodynamics.

Topological quantum computational/error correction in [MTs - MAP], based on Fibonacci series, related directly to Golden mean (GM) condition, suggested by Roger Penrose, may be responsible for resistance of subunits of MTs to decoherence. In turn, this means resistance to decoherence of water clusters (primary librational effectons in state of mBC) inside MTs, maintaining their stability.

It leads from our approach, that DNA, chromosomes, microtubules and bunches of MTs may serve as effective virtual jet generators (VJG), increasing virtual pressure (VirP$^{\pm}$)$_{1,2,3}$ in selected direction. It may be due to coherent [$\mathbf{C} \rightleftharpoons \mathbf{W}$] pulsation of elementary particles, composing atoms and molecules in state of mesoscopic molecular Bose condensate (mBC), existing in these structures and modulating dynamics of Bivacuum.

At the "rest" condition of cells the resulting concentration of internal anions of neurons is bigger than that of external ones, providing the difference of potentials equal to 50-100mV. As far the thickness of membrane is only about 5nm or 50Å it means that the gradient of electric tension is about:

$$100.\ 000\ V/sm$$

i.e. it is extremely high. Depolarization of membrane usually is related to penetration of $Na^{+}$ ions into the cell. The processes of depolarization, accompanied by pulsation of nerve cell body, - change the properties of membranes as Casimir chambers and, consequently, the virtual replica of cell.

*The virtual replica of cells, involved in nonequilibrium process of nerve excitation, including acupuncture points, change in-phase with corresponding elementary acts of consciousness.*

Corresponding coherent cooperative changes: increments-decrements of kinetic energy and momentum, accompanied, for example phase transitions of water clusters and lipid domains in axons and nerve cells membranes participate, in accordance to our theory, in nonlocal mind-matter and mind-mind interactions via Virtual Guides of spin, momentum and energy (see Fig.12).

## 12. Bivacuum mediated Mind-Matter and Mind-Mind Interaction, including Remote Vision, Telepathy, Remote Healing and Telekinesis

### *12.1 The stages of Psi channel formation*

Theories of Virtual Replica (**VR**) of material objects in Bivacuum and primary **VR** Multiplication (**VRM**), described in chapter 8, in combination with theory of Virtual Guides (**VirG**$_{SME}$) (see chapter 9), are the background for explanation of different kind of listed above Psi phenomena. The primary **VR** represents a three-dimensional (3D) superposition of Bivacuum virtual standing waves **VPW**$_m^{\pm}$ and **VirSW**$_m^{\pm 1/2}$, modulated by [$\mathbf{C} \rightleftharpoons \mathbf{W}$] pulsation of elementary particles and translational and librational de Broglie waves of molecules of macroscopic matter (http://arxiv.org/abs/physics/0207027).

The infinitive multiplication of VR in space and time: **VRM(r,t)** in form of 3D packets of virtual standing waves, representing *iterated* primary VR, is a result of interference of all pervading external coherent basic *reference waves* - Bivacuum Virtual Pressure Waves (**VPW**$_{q=1}^{\pm}$) and Virtual Spin Waves (**VirSW**$_{q=1}^{\pm 1/2}$) with similar kinds of modulated standing waves, forming primary VR. The latter can be considered as the *object waves,* making it possible to name the **VRM**, as **Holoiteration** by analogy with regular hologram.

We put forward a conjecture, that the dependence of complex **VRM(t)** on time is a consequence of its ability to self-organization in both directions - positive (evolution) and negative (devolution) in nonequilibrium conditions. Virtuality of such systems, means by definition, that relativistic mechanics and causality principle do not work for them.

Depending on the type modulation (section 8.1) **VR** and **VRM(r,t)** are subdivided on the:

a) frequency modulated;
b) amplitude modulated;
c) phase modulated;

d) polarization modulated.

Only their superposition contains all the information about positions and dynamics of atoms/molecules, composing object, possibilities of their evolution and devolution.

The nonlocal Virtual Guides (**VirG$_{SME}$**) of spin, momentum and energy (chapter 9), represent virtual microtubules with properties of one-dimensional virtual Bose condensate, constructed from 'head-to-tail' polymerized Bivacuum bosons (**BVB$^\pm$**) or Cooper pairs of Bivacuum fermions (**BVF$^\updownarrow$**). The bundles of **VirG$_{SME}$**, connecting coherent atoms of Sender (S) and Receiver (S) are responsible for nonlocal weak interaction between them. The introduced in our theory *Bivacuum Mediated Interaction* (**BMI**) is a new fundamental interaction due to superposition of Virtual replicas of Sender and Receiver, because of **VRM** mechanism, and connection of their coherent atoms via **VirG$_{SME}$** bundles. Just **BMI(r,t)** is responsible for remote ultraweak nonlocal interaction and different psi-phenomena. For activation of psi-channels the system: [S + R] should be in nonequilibrium state.

**After our Unified Model**, **the informational (spin)**, **momentum and energy exchange interaction between Sender [S] and Receiver [R]**, **representing Psi channel formation**, **involves following three stages**:

**1**. Superposition of nonlocal (informational/spin) components of [S] and [R] Virtual Replicas Multiplication:

$$\mathbf{VRM}_S^{nl} \bowtie \mathbf{VRM}_R^{nl}$$ 12.1

formed by modulated by the objects de Broglie waves virtual spin waves of Sender and Receiver: **VirSW$_S^{S=\pm1/2}$** and **VirSW$_R^{S=\pm1/2}$**;

**2**. Formation of bundles of nonlocal Virtual guides of spin, momentum and energy, connecting coherent nucleons and electrons of [S] and [R]:

$$\sum[\mathbf{VirG}_{S,R}]^i = \sum\left[\ \mathbf{VirSW}_S^{S=+1/2}\ \underset{\mathbf{BVF}^\uparrow\bowtie\mathbf{BVF}^\downarrow}{\overset{\mathbf{BVB}^\pm}{=\diamond=}}\ \mathbf{VirSW}_R^{S=-1/2}\ \right]^i$$ 12.2

**VirG$_{S,R}^i$** is quasi-1D virtual microtubule (quasi one-dimensional virtual Bose condensate), formed primarily by standing **VirSW$_S^{S=+1/2}$** $\overset{\mathbf{BVB}^\pm}{\underset{\mathbf{BVF}^\uparrow\bowtie\mathbf{BVF}^\downarrow}{=\diamond=}}$ **VirSW$_R^{S=-1/2}$** of opposite spins, following by self-assembly of Cooper pairs of [**BVF$^\uparrow$** $\bowtie$ **BVF$^\downarrow$**]$^i$ or Bivacuum bosons (**BVB$^\pm$**)$^i$;

**3**. Superposition of distant components of Virtual Replicas Multiplication of [S] and [R], formed by standing virtual pressure waves $[\mathbf{VPW}_m^+ \bowtie \mathbf{VPW}_m^-]_S^i =\diamond= [\mathbf{VPW}_m^+ \bowtie \mathbf{VPW}_m^-]_R^i$, modulated by [S] and [R]:

$$\mathbf{VRM}_S^{dis} \bowtie \mathbf{VRM}_R^{dis} = \sum\{[\mathbf{VPW}_m^+ \bowtie \mathbf{VPW}_m^-]_S^i =\diamond= [\mathbf{VPW}_m^+ \bowtie \mathbf{VPW}_m^-]_R^i\}$$ 12.3

The described above three stage of [S] and [R] Bivacuum mediated interaction (BMI) provide formation of *psi channel*. For activation of this channel, the whole system: ([S] + [R]) should be in nonequilibrium state.

**We put forward a conjecture**, **that even teleportation or spatial exchange of macroscopic number of coherent atoms between very remote regions of the Universe is possible via coherent Psi-channels**. **If this consequence of our theory will be confirmed**, **we get a new crucial method of the instant inter-stars propulsion**.

For special case if Sender [*S*] or Receiver [*R*] is psychic, the double conducting membranes of the coherent nerve cells (like in axons) may provide the cumulative Casimir effect, contributing Virtual Replica of [S] and [R].

The quantum neurodynamics processes in Sender (Healer) may be accompanied by radiation of electromagnetic waves or magnetic impulses, propagating in Bivacuum via virtual guides: $\mathbf{VirG}_{SME}$. Such kind of radiation from different regions of Sender/Healer has been revealed experimentally.

The important role in Bivacuum mediated Mind-Matter and Mind-Mind interaction, plays the coherent fraction of water in **microtubules** of neurons in state of *mesoscopic molecular Bose condensate (mBC)* (Kaivarainen: http://arxiv.org/abs/physics/0102086). This fraction of **mBC** is a variable parameter, dependent on structural state of microtubules and number of simultaneous elementary acts of consciousness (Kaivarainen: http://arxiv.org/abs/physics/0003045). It can be modulated not only by excitation of nerve cells, but also by *specific interaction with virtual replica of one or more chromosomes $(VR^{DNA})$ of the same or other cells.*

*The change of frequency of selected kind of thermal fluctuations, like cavitational ones, in the volume of receiver [R], including cytoplasm water of nerve cells, is accompanied by reversible disassembly of microtubules and actin' filaments, i.e. [**gel** ⇌ **sol**] transitions. These reactions, responsible for elementary act of consciousness,* are dependent on the changes of corresponding activation barriers.

*The mechanisms of macroscopic quantum entanglement, proposed in our work, is* responsible for change of intermolecular Van der Waals interaction in the volume of [R] and probability of selected thermal fluctuations (i.e. cavitational fluctuations), induced by [S]. In this case, realization of certain series of elementary acts of consciousness of [S] will induce similar series in nerve system of [R]. This means informational exchange between $VR^R$ and $VR^S$ of two psychics via external Virtual Guides ($\mathbf{VirG}_{SME}^{ext}$), and *their bundles, forming Psi-channels, i.e. telepathy.*

The *specific character* of telepathic signal transmission from [S] to [R] may be provided by modulation of $\mathbf{VRM}_{MT}^{S}$ of microtubules by $\mathbf{VRM}_{DNA}^{S}$ of DNA of sender's chromosomes in neuron ensembles, responsible for subconsciousness, imagination and consciousness. The resonance - most effective remote informational/energy exchange between two psychics is dependent on corresponding 'tuning' of their nerve systems. As a background of this tuning can be the described Bivacuum mediated interaction (BMI) between regulative cells components of [S] to [R], like:

$$\sum \left[ \mathbf{2\ centrioles + chromosomes} \right]_S \overset{\mathbf{BMI}}{\Longleftrightarrow} \sum \left[ \mathbf{2\ centrioles + chromosomes} \right]_R \qquad 12.4$$

In accordance to our theory of elementary act of consciousness and *three stages of BMI mediated Psi channel formation*, described above, the modulation of dynamics of [assembly ⇌ disassembly] of microtubules by influence on probability of cavitational fluctuations in the nerve cells and corresponding [*gel* ⇌ *sol*] transitions by directed mental activity of [Sender] can provide **telepathic contact and remote viewing** between [Sender] and [Receiver].

The mechanism of **remote healing** could be the same, but the local targets in the body of patient [R] are not necessarily the MTs and chromosomes of the nerve cells, but **centrioles + chromosomes** of the ill organs (heart, liver, etc.).

The **telekinesis**, as example of mind-matter interaction, should be accompanied by significant nonequilibrium process in the nerve system of Sender, related to increasing of kinetic energy of coherent molecules in neurons of Sender, like cumulative momentum of water clusters, coherently melting in microtubules of centrioles and inducing their disassembly. Corresponding momentum and kinetic energy are transmitted to 'receiver - target' via multiple correlated bundles of $\mathbf{VirG}_{SME}^{ext}$ in superimposed $\mathbf{VRM}_{S,R}$ (Psi-channels).

The specific magnetic potential exchange between [S] and [R] via Psi-channels can be generated by the nerve impulse regular propagation along the axons and depolarization of nerve cells membranes (i.e. electric current) in the 'tuned' ensemble of neuron cells of psychic - [Sender], accompanied by magnetic flux. These processes are accompanied by $\mathbf{BVF}^{\uparrow} \rightleftharpoons \mathbf{BVB}^{\pm} \rightleftharpoons \mathbf{BVF}^{\downarrow}$ equilibrium shift to the right or left, representing magnetic field excitation.

The evidence are existing, that *Psi-channel* between [S] and [R] works better, if the frequencies of geomagnetic Schumann waves - around 8 Hz (close to brain waves frequency) are the same in location of [S] and [R]. However, the main coherence factor in accordance to our theory, are all-pervading Bivacuum virtual pressure waves ($\mathbf{VPW}_{q=1}^{\pm}$), with basic Compton frequency $[\omega_0 = \mathbf{m}_0\mathbf{c}^2/\hbar]^i$, equal to carrying frequency of [Corpuscle $\rightleftharpoons$ Wave] pulsations of the electrons, protons, neurons, composing real matter and providing entanglement. The macroscopic Bivacuum flicker fluctuation, activated by nonregular changes/jumps in properties of complex Hierarchical Virtual replica of Solar system and even galactic, related to *sideral time*, also may influence on quality of Psi-chanells between Sender and Receiver.

Formation of the different kinds of virtual standing waves, representing nonlocal and distant fractions of Virtual Replicas (VR)$_{S,R}$ of Sender [S] and Receiver [R], necessary for $\mathbf{VirG}_{SME}$ and *Psi-channel* formation, are presented in Table 1

**TABLE 1**

## The role of paired and unpaired sub-elementary particles of the electron's [Corpuscle $\rightleftharpoons$ Wave] pulsation:

$$\langle [\mathbf{F}_\uparrow^+ \bowtie \mathbf{F}_\downarrow^-]_W + (\mathbf{F}_\uparrow^-)_C \rangle \rightleftharpoons \langle [\mathbf{F}_\uparrow^+ \bowtie \mathbf{F}_\downarrow^-]_C + (\mathbf{F}_\uparrow^-)_W \rangle$$

## in Bivacuum mediated interaction between sender [S] and receiver [R]

| **Pair of sub-elementary particle and antiparticle pulsation**: $$[\mathbf{F}_\uparrow^+ \bowtie \mathbf{F}_\downarrow^-]_W \rightleftharpoons \langle [\mathbf{F}_\uparrow^+ \bowtie \mathbf{F}_\downarrow^-]_C$$ | **Unpaired sub-elementary fermion pulsation**: $$(\mathbf{F}_{\uparrow\downarrow}^\pm)_C \overset{\mathbf{BvSO}}{\rightleftharpoons} (\mathbf{F}_{\uparrow\downarrow}^\pm)_W \overset{\mathbf{CVC}^{\circlearrowright\circlearrowleft}}{\Longleftrightarrow} \mathbf{VirSW}^{\circlearrowright\circlearrowleft}$$ |
|---|---|
| 1. Virtual Pressure Waves: $\left[\mathbf{VPW}^+ \bowtie \mathbf{VPW}^-\right]$ | 1. Electromagnetic potential: $\mathbf{E}_{EM} = \boldsymbol{\alpha}\, \mathbf{m}_V^+ \mathbf{c}^2 \sim$ $$\sim \frac{1}{2}\left|\mathbf{VirP}_{\mathbf{F}_\uparrow^+}^+ - \mathbf{VirP}_{\mathbf{F}_\downarrow^-}^-\right|^{[\mathbf{F}_\uparrow^+ \bowtie \mathbf{F}_\downarrow^-]}$$ |
| 2. Total Virtual Pressure energy increment, equal to that of total and unpaired ($\Delta\mathbf{E}_{\mathbf{F}_\uparrow^+}$): $$\Delta\mathbf{E}_{\mathbf{F}_\uparrow^+} \sim \Delta\mathbf{VirP}_{\mathbf{F}_\uparrow^+}^+ = \frac{1}{2}\left|\mathbf{VirP}_{\mathbf{F}_\uparrow^+}^+ - \mathbf{VirP}_{\mathbf{F}_\downarrow^-}^-\right|^{[\mathbf{F}_\uparrow^+ \bowtie \mathbf{F}_\downarrow^-]} +$$ $$+ \frac{1}{2}\left|\mathbf{VirP}_{\mathbf{F}_\uparrow^+}^+ + \mathbf{VirP}_{\mathbf{F}_\downarrow^-}^-\right|^{[\mathbf{F}_\uparrow^+ \bowtie \mathbf{F}_\downarrow^-]}$$ | 2. Gravitational potential: $\mathbf{E}_G = \boldsymbol{\beta}\,[\mathbf{m}_V^+ + |\mathbf{m}_V^-|]\mathbf{c}^2 \sim$ $$\sim \frac{1}{2}\left|\mathbf{VirP}_{\mathbf{F}_\uparrow^+}^+ + \mathbf{VirP}_{\mathbf{F}_\downarrow^-}^-\right|^{[\mathbf{F}_\uparrow^+ \bowtie \mathbf{F}_\downarrow^-]}$$ |
| where the kinetic and potential energy increments: $$\Delta\mathbf{T}_k = \frac{1}{2}\left|\mathbf{VirP}_{\mathbf{F}_\uparrow^+}^+ - \mathbf{VirP}_{\mathbf{F}_\downarrow^-}^-\right|^{[\mathbf{F}_\uparrow^+ \bowtie \mathbf{F}_\downarrow^-]}$$ $$\Delta\mathbf{V} = \frac{1}{2}\left|\mathbf{VirP}_{\mathbf{F}_\uparrow^+}^+ + \mathbf{VirP}_{\mathbf{F}_\downarrow^-}^-\right|^{[\mathbf{F}_\uparrow^+ \bowtie \mathbf{F}_\downarrow^-]}$$ | 3. Virtual Spin Waves (**VirSW**): $\mathbf{I}_S = \mathbf{I}_{\mathbf{VirSW}^{\pm 1/2}} \sim \mathbf{K}_{BVF^\uparrow \rightleftharpoons BVF^\downarrow}(\mathbf{t}) =$ $(\mathbf{K}_{BVF^\uparrow \rightleftharpoons BVF^\downarrow})_0\,[\sin(\omega_0^t t) + \gamma\omega_B^{lb}\sin(\omega_B^{lb}t)]$ |
| 3. Virtual Replica of the Object (VR=VR$^{in}$+VR$^{sur}$) | 4. The bundles of Virtual Guide (**VirG**$_{SME}^{ext}$)$^i$ formation between remote [S] and [R]: |
| 4. Virtual Replicas of [S] and [R] Multiplication: $$\mathbf{VRM}_S = \sum \mathbf{VR}_S \Longrightarrow \sum \mathbf{VR}_R = \mathbf{VRM}_R$$ | $\mathbf{VirSW}_S^{S=+1/2} \overset{\mathbf{BVB}^\pm}{\underset{BVF^\uparrow \bowtie BVF^\downarrow}{\Longleftrightarrow}} \mathbf{VirSW}_R^{S=-1/2}$ |

Pauli attraction (Cooper pairs formation) or repulsion between **BVF**$^{\updownarrow}$

of the opposite or similar spins

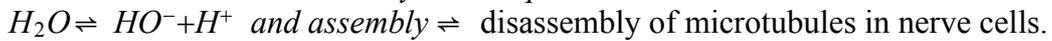

*One of the result of Psi-channel formation, as superposition of VRM$_{S,R}$ and bundles of VirG$_{SME}^{ext}$,*
*is a change of permittivity $\varepsilon_0$ and permeability $\mu_0$ of Bivacuum [$\varepsilon_0 = n_0^2 = 1/(\mu_0 c^2)$].*
*In turn, ($\pm\Delta\varepsilon_0$) influence Van-der-Waals interactions in condensed matter,*
*changing the probability of defects origination in solids and cavitational fluctuations in liquids.*

*Bidirectional change of pH of water under Psi field action can be a consequence*
*of $\pm\Delta VP^\pm$ and $\pm\Delta\varepsilon_0$ influence on cavitational fluctuations, accompanied by shift of dynamic equilibrium:*
*$H_2O \rightleftharpoons HO^- + H^+$ and assembly $\rightleftharpoons$ disassembly of microtubules in nerve cells.*

The coherency of all components of Virtual wave guide between [S] and [R], formed by nonlocal virtual spin waves (**VirSW**$^\circlearrowright$ and **VirSW**$^\circlearrowleft$) of two opposite angular momentums and virtual pressure waves ($\mathbf{VPW}_q^+$ and $\mathbf{VPW}_q^-$) of two opposite energies, corresponds to finest tuning of mind-matter and mind-mind interaction. The coherency between signals of [S] and [R] can be provided by *Tuning Force (TF) of Bivacuum* and modulation of nonlocal Virtual Guides ($\mathbf{VirG}_{SME}$) by cosmic and geophysical magnetic flicker noise.

The [*dissociation* ⇌ *association*] equilibrium oscillation of coherent water clusters in state of molecular Bose condensate (mBC) in microtubules of nerve cells, modulating ($\mathbf{VirSW}^{\circlearrowleft,\circlearrowright}$) and $\mathbf{VPW}^{\pm}$, is a crucial factor for realization of quantum Psi phenomena. The virtual replica (VR) of microtubules and its multiplication (VRM) can be modulated also by VRM of DNA.

### 12.2 The Bivacuum mediated interaction between remote biopolymers, cells and different organisms

In accordance to our Unified Model, all material objects, composed from elementary particles, including biosystems: proteins, DNA, membranes, microtubules, cells, organs, whole animals, etc. has a complex hierarchical virtual replicas multiplication (VRM) in superfluid Bivacuum, described in previous sections.

The main factors of Bivacuum Mediated Interaction (**BMI**) are:

a) the carriers of information and spin, like Virtual spin waves ($\mathbf{VirSW}^{\pm 1/2}$);

b) the carriers of the momentum and kinetic energy via 3D system of nonlocal bundles of Virtual Guides $\sum[\mathbf{VirG}_{SME}^{i}(S <==> R)]$, connecting big number of coherent atoms of Sender (S) and Receiver (R);

c) the distant and nonlocal components of $VRM^{dis}$ and $VRM^{nl}$ are responsible for Virtual Replicas and spatial images exchange between Sender and Receiver.

The virtual replica ($\mathbf{VR}_{DNA}^{S,R}$) of highly ordered parts of the eukaryote's chromosomes (introns), not participating directly in biosynthesis and including more than 95% of total DNA, may play the active and very specific role in the [Mind-Matter] and [Mind-Mind] interaction. The so-called *nucleosomes* represent long parts of DNA, tuned around the histone octamer. The diameter of such nucleosomes (stable nuclear-protein complexes) is about 110 Å. The histones are able to influence the DNA structure and functions. In accordance to our Hierarchic theory of condensed matter (Kaivarainen, 1995; 2001: http://arxiv.org/abs/physics/0102086), the high density and stability of nucleosomes and native chromosomes provides condition for mesoscopic Bose condensation [**mBC**] of their atoms at ambient temperature. The [**mBC**] may exist in form of so-called *primary translational and librational effectons (the latter are bigger) with dimension of tens of angstroms*, representing 3D standing de Broglie waves of atoms/molecules of biopolymers, microtubules, chromosomes and membranes.

The existence of [**mBC**] means coherent [$\mathbf{C} \rightleftharpoons \mathbf{W}$] pulsation of elementary particles of atoms, composing these clusters. The same is true for water clusters in state of **mBC** in composition of hydration shells of DNA and in the holes of microtubules with internal diameter about 140 Å.

The huge 3D net of $\mathbf{VirG}_{SME}$, connecting coherent hydrogens, oxygen, carbons, nitrogen and other atoms in similar [**mBC**] of (S) and (R), allows the *macroscopic quantum entanglement between chromosomes, microtubules, ribosomes, etc.* of distant cells and even between different organisms. The nonlocal interaction 'in-vitro' between the remote (separated by tens of kilometers) samples of DNA from the same source, ribosomes and collagen - was demonstrated by team of Peter Gariaev in 1992. The activation of this interaction by irradiation of Sender samples with two laser visible beams with orthogonal polarization, was used. The immediate remote interaction effect 'in vitro' was registered by change of light scattering of the Receiver(R) samples after Sender(S) activation.

**We put forward a conjecture**, that in natural conditions the activation and *synchronization of atoms de Broglie waves, stimulating the interaction between distant chromosomes and ribosomes of the same or different organisms, can be realized by coherent IR photons scattering on Sender structures. These IR photons are* radiated by water clusters (mBC), named primary librational effectons (Kaivarainen, 1995; 2001) in microtubules of pairs of orthogonal centrioles. Corresponding nonequilibrium states should enhance the effect of quantum entanglement via bundles of Virtual Guides ($\mathbf{VirG}_{SME}$), connecting atoms of 'tuned' parts of the remote DNA(S)<==>DNA(R). The orthogonal direction of IR photons radiation by microtubules may provide conditions of 3D effects of the Sender treatment. The interference of one pair of orthogonal IR photons with orthogonal IR photons of *other* cells can form the pattern of 3D standing waves, regulating the morphogenesis.

It is possible also, that the IR photons, radiated by two orthogonal centrioles, each of them containing about 50 microtubules, can be normally polarized as respect to each other, like in described experiments with two laser beams.

The librational IR superradiation of coherent water clusters in microtubules can activate the librational dynamics of water in DNA (Sender) hydration shell. In turn, this activation is responsible for information, energy and momentum transmission from chromosomes of Sender to Receiver, mediated by nonlocal bundles $\sum \mathbf{VirG}_{SME}^{i}(S <==> R)$ and $\mathbf{VRM}_{S} \bowtie \mathbf{VRM}_{M}$ superposition.

The frequency of such coherent librational photons, radiated by water in microtubules of pairs of centrioles, activating $\mathbf{VirG}_{SME}^{i}$, is about:

$$\nu_{lb} = \widetilde{\nu}\,\mathbf{c} \simeq \mathbf{750cm}^{-1} \times 3 \cdot 10^{10}\,cm/s = 2,25 \cdot 10^{13}\,s^{-1}$$

The vibration frequencies of different fractions of vicinal water of chromosomes also may correspond to wide range of radio-frequency: $(\mathbf{10^{4} \div 10^{9}})\,s^{-1}$ (Kaivarainen, 1995; 2001). It means that corresponding collective excitations of model [water + DNA (chromosomes)] system *in vitro* can be activated by artificial ultrasound and radio-waves of resonant frequencies.

In similar way, *in vivo* in cells of eucaryotes the 3D net of $\mathbf{VRM}_{S} \bowtie \mathbf{VRM}_{M}$ and Virtual Guides $\sum[\mathbf{VirG}_{SME}^{i}(S <==> R)]$ between 'tuned' (S) and (R), providing entanglement, is activated by system [$\mathbf{two\ centriols + chromosoms}$].

The molecules of water, contain two hydrogen and one oxygen atoms, which provide their tendency to clusterization and mesoscopic Bose condensation (mBC). It is a reason, why $H_2O$ is very good molecules for formation of coherent, stable and effective coaxial pairs of Virtual Guides. The correlated pair of VirG$_{SME}$ can be formed by both hydrogen atoms of $\mathbf{H_2O}$, connecting the electrons and protons of remote water molecules of S and R:

$$[\mathbf{VirG}_{SME}^{e} + \mathbf{VirG}_{SME}^{p}]_{H}^{S} <===> [\mathbf{VirG}_{SME}^{e} + \mathbf{VirG}_{SME}^{p}]_{H}^{R} \qquad 12.5$$

The virtual guides with Compton radius, determined by protons [$\mathbf{VirG}_{SME}^{p}(S <==> R)$], are smaller than that, determined by the electrons [$\mathbf{VirG}_{SME}^{p}(S <==> R)$] of hydrogen atoms. Consequently, $\mathbf{VirG}_{SME}^{p}$ is localized inside $\mathbf{VirG}_{SME}^{e}$, forming a double coaxial virtual microtubule between water hydrogens of S and R.

It is important to point out, that the basic virtual pressure waves $(\mathbf{VPW}_{q=1}^{\pm})^{i}$ of Bivacuum, responsible for synchronization of [$\mathbf{C} \rightleftharpoons \mathbf{W}$] pulsations of the electrons and nucleons (protons) of $\mathbf{DNA - S}$ and $\mathbf{DNA - R}$, necessary for entanglement (chapter 9) can be very effective in case of just these biopolymers. It follows from our theory, that the evolution of matter and especially biosystems, occur under permanent influence of Bivacuum virtual pressure waves, driving matter to Golden mean (GM) conditions on each hierarchical level of its organization.

The nonequilibrium conditions, following from competition between Tuning Energy (TE) of $(\mathbf{VPW}_{q=1}^{\pm})^i$, driving the mass/energy of elementary particles to the rest mass ($m_0^i$), i.e. to temperature of the absolute zero from one side and between thermal energy of living organisms from other side, provide self-organization of matter and evolution of biosystems.

In the case of DNA macromolecule there are convincing evidences, that its large-scale geometry obey the Golden mean proportions (see Dan Winter site: www.soulinvitation.com/magneticx). For example: one 360 degree turn of DNA measures 34 Å in the direction of the main axis. The width of the molecule is 20 Å. These dimensions ratio $\mathbf{34 : 20 = 1.7 \simeq 1.618 = 1/\phi}$ is close to Golden Mean within the limits of the accuracy of the measurements.

The proposed in this work mechanism of $\mathbf{VRM}^{dis,nl}$ and $\mathbf{VirG}_{SME}$ − mediated quantum entanglement between synchronized similar atoms of chromosomes of Sender and Receiver, may stimulate the high-energy cavitational fluctuations of water near specific sites of DNA/chromosomes. The energy of these thermal fluctuations is enough for breaking the chemical bonds in DNA and can be accompanied by *remote genetic transmutation (RGT)*. Such a process can occur between cells (donors/senders) to cells (acceptors/receivers) inside the same multicell organism or between different organisms.

The remote genetic transmission (RGT) was discovered by Dzang Kangeng in 1981 and patented in 1992. Later it was confirmed by Gariaev's group in 1994 - 2004. The impressive results of this group are online: http://www.self-managing.net/genetica/.

The coherent bundles of $\sum \mathbf{VirG}_{SME}(S <==> R)$, mediating the transmission and exchange of momentum and kinetic energy between complementary parts of remote *'healthy'* DNA/chromosomes of cells in system [Sender<==>Receiver] do not induce any damage in structure of both. In this *correct* case the S and R 'dance' in the same dynamic rhythm, coinciding by frequency, amplitudes and phase.

However, in the case of remote *healthy* DNA- S and *ill* DNA- R structures are dynamically "incompatible" 'because of mutations/mistakes in the 'ill' DNA-R. Similar situation takes a place if DNA- S and DNA- R belong to different organisms. In both cases the cumulative effect of $\sum \mathbf{VirG}_{SME}(S <==> R)$ bundles may stimulate a high-energy cavitational fluctuations in the hydration shell near regions of structural incompatibility of DNA - R. The thermal energy of corresponding microbubbles collapsing corresponds to 6000K. It is much more than energy, necessary for covalent bonds breach between nucleotides of DNA and amino acids of proteins.

Such selective Bivacuum mediated selected destruction of DNA-Receiver on fragments and their assembly in new combinations are controlled by the Virtual Replica Multiplication (VRM$_S$) of DNA-Sender targeting DNA-Receiver by means of $\sum \mathbf{VirG}_{SME}(\mathbf{S} <==> \mathbf{R})$. The proposed in our work mechanism of Remote Genetic Transmutation (RGT) may be realized in two cases:

1) DNA-S and DNA-R belong to the two similar population of cells of the same genetic origin and one of population (DNA-R) is damaged (poisoned). In this case the induced by DNA-S substitution of damaged regions of DNA-R can be realized by 'jumps' of selected *transposons* or *plasmids* in bacteria, localized in *'silent' or 'junk'* part of chromosomes (R), representing about 95% of genetic apparatus of any cell. We consider this 'silent' DNA, as a storage of very mobile genes fragments, duplicating the fragments of active genes, participating in biosynthesis. In some sense one may said, that 'silent' part of chromosomes is a multiple 'backup' of the active genes;

2) DNA-S and DNA-R belong to *different* organisms with different genes. In such case of the induced *Remote Genetic Transmutation (RGT)*, the DNA-S via bundles of $\sum \mathbf{VirG}_{SME}(\mathbf{S} <==> \mathbf{R})$ and superposition of $\mathbf{VRM}_S \bowtie \mathbf{VRM}_M$ may stimulate conversion of part of *silent* genes or their fragments in DNA-R, which are *close by primary structure* to DNA-S, to the active for RNA and proteins biosynthesis state. The new assembly of activated fragments of silent part of DNA-R to coding genes, pertinent for DNA-S can be realized under the control of $\mathbf{VRM}_S \bowtie \mathbf{VRM}_M$ by the action of $\sum \mathbf{VirG}_{SME}(\mathbf{S} \Longleftrightarrow \mathbf{R})$. As a consequence, the organism of Receiver start to be able for synthesis of RNA, proteins, cells and their spatial organization (morphogenesis), pertinent for organism of Sender.

The Kangeng's (1981-1992) and Gariaev's (1992-2004) experiments with induced by the *artificial* electromagnetic polarized radiation of Sender are in total accordance with our mechanism of Remote Genetic Transmutation (RGT). As a result of this kind of interaction between remote *different species* of animals or animals and plants, the organism of Receiver gain a strong morphological features, pertinent for organism of Sender, but not for Receiver.

These genetic transmutations where not as stable, as the primary genetic code of these species. This is confirmed by disappearance of induced visible transmutations after 2-3 generation of hybrid animals or plants. Such instability can be explained by the action of known mechanisms of genetic mistakes reparation in the process of transformed cells division.

Dzang Kangeng (1981-1992) used hexahedron, cone, sphere and a parabolic-reflector aerial, as a kind of forms, providing specific spinning (polarization) of the electromagnetic (EM) field. In the Kangeng's equipment the high-frequency generator of *orthogonally-polarized* electromagnetic beams has been used, which repeatedly pass throw the donor (Sender) and the accepting (Receptor) organisms, necessary for informational exchange between them. Both organisms were remote, but interconnected by the waveguide of EM polarized beams.

All known biopolymers are optically active, sometimes in very specific way. Consequently, polarization of photons should increase the effectiveness of their scattering on DNA and other biopolymers, dependent also on their frequency and density. The *polarization modulation* of electromagnetic beams in device of Kangeng may influence the small-scale librational dynamics of *both* DNA: Sender and Receiver, stimulating the formation of bundles of very effective $\mathbf{VirG}_{SME}$ between hydration shell and DNA of (S) and (R).

Similar principles were used in Gariaev's group device. *Each of two orthogonally polarized laser visible beams* ($\lambda = 632\,\mathrm{nm}$) transforms to wide band of *radio waves (RW)* of frequency range: $(10^3 - 10^6)\,s^{-1}$ after repeatedly scattering on biosystem - Sender and laser mirrors. Most probable interpretation of RW origination, in my opinion, is a quantum beats between the primary visible photons of laser and secondary photons - scattered on the object or laser mirror after multiple passing to and fro. As far as the frequencies of secondary - scattered photons are slightly shifted as respect to primary ones with $\lambda = 632\,\mathrm{nm}$ due to inelastic Raman - like scattering, their interference and quantum beats are followed by generation of radio-frequency EM radiation of wide range:

$$\omega_{RW} = \pm\Delta\omega = \pm[\omega_1 - (\omega_2^0 \pm \delta\omega_2)] \qquad 12.6$$

The wideness of radio-frequency range is determined by wideness of scattered frequency: $\omega_2 = (\omega_2^0 \pm \delta\omega_2)$.

However, *no visible photons, no radiowaves* can serve themselves, as a carriers of genetic or other information (signals) on distances of many kilometers, used in Gariaev' group experiments.

In accordance to our theory, their role is only the activation of nonlocal Virtual Guides of spin, momentum and kinetic energy $\sum[\mathbf{VirG}_{SME}\,(\mathbf{S} \Longleftrightarrow \mathbf{R})]$ and nonequilibrium state of superimposed $\mathbf{VRM}_S \bowtie \mathbf{VRM}_M$ between coherent atoms of Sender and Receiver.

We suppose, that the hierarchical system of $\mathbf{VRM}_S \bowtie \mathbf{VRM}_M$ is a crucial part of *morphogenetic field,* which can be energetically realized/induced via
$\mathbf{VirG}_{SME}$ ($\mathbf{S} <\Longrightarrow \mathbf{R}$).

The spontaneous radiation of coherent 'biophotons' by DNA of wide frequency range has been revealed earlier by Gurvich (1977) and Popp (2000).

In few experiments the *'DNA phantom effects'* was revealed by Gariaev's group. This effect display itself, as keeping the same shape of auto-correlation function of light scattering in device chamber even after replacement of DNA gel from chamber. This unusual 'ghost' effect confirms our theory of stable Ether Body - Virtual Replica (VR) and Astral Body (VRM) (see section 8.3) of the object in Bivacuum and its ability influence the interaction and dynamics of the air molecules and water vapor via $\mathbf{VirG}_{SME}$. The influence of VR on the gas molecules interaction (for example, as a result of molecules polarizability change due to variation of Bivacuum dielectric constant in volume of VR) was confirmed by nitrogen jet washing out (destroying) the DNA phantom. Few minutes after nitrogen 'wind' was stopped, the DNA phantom restores again.

Consequently, the 'phantoms' are the reflection of stable 'Virtual Replicas' of the object. The mechanism of feedback reaction between VR and VRM of any object and object itself is mediated by the internal $\mathbf{VirG}_{SME}$, connecting atoms of the same object/organism.

The proposed in our work mechanisms of the Induced Remote Genetic Transmutation (RT) and Induced Remote Morphogenesis (RM) of one kind of plants and animals by another, mediated by superposition $\mathbf{VRM}_S \bowtie \mathbf{VRM}_R$ and bundles of Virtual Guides between chromosomes of Sender and Receiver:

$$\sum_{n \to \infty} \left[\mathbf{VirG}_{SME}(\mathbf{S} <\Longrightarrow \mathbf{R})\right]_n \qquad 12.7$$

activated by coherent IR radiation of microtubules (MTs) of orthogonal centrioles, makes it possible the directed manipulation of RT, for example, by modulating the MTs properties of Sender cells by nerve excitation or external resonant electromagnetic or acoustic fields influence on the cells membranes.

The same mechanism explains the well known phenomena of *remote healing.* It follows from our approach, that the "Remote Healer" as a Sender, should be healthy himself and be able to activate the interaction in system (pair of centrioles + chromosomes) in own cells, corresponding to diseased organs of "Patient - Receiver" by specific nerve system of "Healer" excitation. *Self-healing* and regeneration of the damaged organs of the same animals, like liver, kidney, heart and even head brain can be also realized via VRM and $\mathbf{VirG}_{SME}$.

### *12.3 The mechanism of microbes adaptation to antibiotics, following from proposed model of Remote Genetic Transmutation*

In the framework of our approach, such important in medicine phenomena, like *adaptation of microbes to antibiotics* can be explained in following sequence of stages:

1. The antibiotics, killing the microbes (for example, vancomycin) stimulate the *chaotic* 'jumps' of plasmids in their DNA, making the chromosome system nonequilibrium and unstable;

2. Finally, as a result of such casual chromosomes reorganization, the small population of microbes emerge, stable/immune to this antibiotics;

3. Such stable microbes population increases due to their regular successful division;

4. This stable population stimulates the similar genetical rearrangement in sensitive to antibiotic microbes population via described above mechanism of Induced Remote Genetic Transmutation (RGT), provided by superposition of $\mathbf{VRM}_S \bowtie \mathbf{VRM}_R$ and bundles of Virtual Guides $\sum^{n\to\infty}$ $[\mathbf{VirG}_{SME}(\mathbf{S} <==> \mathbf{R})]_n$ between DNA of of stable and unstable microbes.

The chromosomes of Senders - stable microbes are stimulated for RGT by coherent IR photons, radiated by water clusters inside microtubules:

$$Microtubules <\overset{IR\,photons}{=====}> Chromosoms$$

The another kind of data, pointing to possibility of 'healing' of different species of the *same* populations, like bacteria and insects, are existing. In such experiments one group of population, poisoned by antibiotics survived, if it was close enough to healthy population of the same organisms. Our theory may explain this effect by resonant interaction between Virtual Replicas Multiplications $\mathbf{VRM}_S \bowtie \mathbf{VRM}_R$ of healthy and poisoned populations of bacteria (Parsons and Heal, 2002) and insects (Agadjanian, 2003) via $\mathbf{VirG}_{SME}$ beam.

The Kaznacheyev' team (1986 - 2002) also was conducted the extensive and successful studies of interaction between remote living cells, unexplainable in the framework of existing paradigm, but easily compatible with proposed mechanism of RGT.

The Chinese notion of *meridians*, connecting a big number of *acupuncture points* on the surface of human body may correspond to multiple beams of $\mathbf{VirG}_{SME}$, connecting groups of cells, containing similar by their resonant properties systems of:

$$[chromosomes + pair\,of\,orthogonal\,centrioles + water\,clusters(mBC)] \qquad 12.8$$

not only on the surface, but also in the internal organs of human's body. The external influence on one or few of the surface *acupuncture points* should change the properties of all other such points, connected by meridians. It means, that the Virtual Guides - mediated interaction $\sum^{n\to\infty}$ $[\mathbf{VirG}_{SME}(\mathbf{S} <==> \mathbf{R})]_n$ between human body's organs and its modulation by needles or otherwise (for example, using Audio-Video Skin Transmitter: http://www.karelia.ru/~alexk/innovations/skin-transmitter.html) is a background of Chinese medicine.

The water molecules ($\mathbf{H_2O}$), containing two hydrogen atoms, tending to mesoscopic Bose condensation at physiological temperature, are the best molecules for creation of stable and effective double: (electron + proton)$_H$ virtual guides ($\mathbf{VirG}^e_{SME} + \mathbf{VirG}^p_{SME}$)$_H$. Taking this into account, we put forward a conjecture, that in volume of meridians the *water contents* is higher, than outside meridians. If so, the dielectric properties of meridians should differ from those of surrounding tissues. It is a way for experimental verification of our idea.

The *Homeopathic drugs*, changing the water properties in cells in specific way, including water in hydration shell of chromosomes, also may influence on Bivacuum - mediated nonlocal interaction between remote cells of the same organism, shifting [coherence $\rightleftharpoons$ decoherence] equilibrium in meridians to the right or left. Our Comprehensive Analyzer of Matter Properties (CAMP) could be a powerful tool for understanding the role of water perturbations in homeopathy (Kaivarainen: http://arxiv.org/abs/physics/0207114).

### 12.4 The examples of Bivacuum mediated remote mental action on different physical targets and their possible explanation

A big number of examples, collected by Savva (2000), demonstrate how Bivacuum mediated interaction (BMI), generated by gifted psychic - 'Sender' [S], can interfere with real physical fields and targets [R] of nonbiological (like water) and biological nature:

1. Speeding up and slowing down the rate of americium 241Am nuclear decay - point to

influence of **BMI** of psychic on the energy exchange of quarks and gluons with Bivacuum Virtual Pressure waves $(\mathbf{VPW}_q^{\pm})^i$ in the process of quarks (sub-elementary fermions) $[\mathbf{C} \rightleftharpoons \mathbf{W}]$ pulsation;

2. Rotation of the plane of polarization of laser beam by $7\text{-}30^0$ points to perturbation of Bivacuum optical isotropic properties due to change of dynamic equilibrium between Bivacuum fermions and antifermions to the left or right:

$$\mathbf{BVF}^{\uparrow} \rightleftharpoons \mathbf{BVB}^{\pm} \rightleftharpoons \mathbf{BVF}^{\downarrow} \qquad\qquad 12.9$$

under the psychic influence, accompanied by magnetic field excitation in Bivacuum (the Faraday - like optical effect);

3. Deviation of the electrical resistance of a thermoresistor. This can be explained as a result of **BMI** influence on resonance exchange interaction of the electrons with Bivacuum, as a result of deviation of frequency of $\mathbf{VPW}_q^{\pm}$, induced by psychic, from frequency of $[\mathbf{C} \rightleftharpoons \mathbf{W}]$ pulsation of the electrons, increasing or decreasing their kinetic energy (see section 7);

4. Induction of a periodic electrical signal from a piezoelectric sensor. It can be the result of effect, like described above, combined with influence of BMI on Van der Waals interaction between atoms/ions of lattice, changing it rigidness and probability of thermal fluctuations, accompanied by defects origination in lattice, via resonance exchange of $(\mathbf{VPW}_q^{\pm})^i$ and $[\mathbf{C} \rightleftharpoons \mathbf{W}]$ pulsation of the electrons, protons and neutrons;

5. Induction of a pulse magnetic field (100 nT and up to 27x106 nT) by psychic, accompanied by rotation of a compasses needle. This effect is similar to that, described in item 2. *Our theory predict, that psychic-excited magnetic field will induce the rotation or oscillation of the plane of polarization of laser beam;*

6. Moving the plate of an encased precise analytical balance equivalent to 100 mg force point to influence of BMI of psychic on gravitation. This can be a result of momentum and kinetic energy transmission from the nerve cells of psychic to selected coherent atoms of the balance via Virtual Guides of spin, momentum and energy $(\mathbf{VirG}_{SME})$;

7. Induction of a temporary peak in the Raman spectrum of tap water at 2200 cm$^{-1}$. This can be a consequence of mechanism described above, confirming the existence of $\mathbf{VirG}_{SME}$, connecting atoms (protons) of water in microtubules of nerve cells and in water - target (Receiver) and the ability of bundles of $\sum\limits^{N} [\mathbf{VirG}_{SME}(\mathbf{S} <==> \mathbf{R})]_n$ influence the internal and external properties of water molecules;

8. Temporary changes in the microstructure of water as observed through scattering of laser beam ($\lambda = 632.8$ nm) at various angles. The same mechanism, as described above;

9. Deviation of UV adsorption spectra of DNA - water solution in the area of 220- 280 nm in three independent observations. The same mechanism as above;

10. Predetermined by operator deviation from randomness with high probability of various random number generators has been revealed in Princeton group. *This can be a result of asymmetric influence of BMI of the operator on elementary particles of electronics, responsible for randomization of results;*

11. Increasing of the concentration of dislocations (missing atoms in microcrystalline structure) in "metal bending" experiments with local increase of surface hardness. The same mechanism, as in the item 4 listed above, describing the BMI influence on piezoelectric materials.

It is interesting to note, that water treated with a magnetic field, like the psychic intent-imprinted water, has stimulating effect on plant growth. The IR spectra of water, its surface tension and crystallization patterns are similar for both types of water treatment (Benor, 1992). The detailed study of magnetic field action on water was performed in our work (Kaivarainen: http://arxiv.org/abs/physics/0207114).

Marcel Vogel (after Savva, 2000) claimed that *quartz crystals of certain shape could amplify the mental intent* action on water. He demonstrated that in water, circulating around an intent-charged crystal, the following changes are revealed:

    -decreasing in surface tension, increasing of conductivity;

    - a significant drop in the freezing point (as low as -30 degrees);

    - bidirectional alterations in the pH up to 3 points;

    - the appearance of two new bands in the IR and UV absorption spectrum, etc.

These important experiments point to ability of certain materials (crystal in this case) for taping the properties of multiplied Virtual Replica (i.e. secondary VR) of psychic. This process and the following processing/reading of taped properties, have a similarity with fixation and reproduction of regular holograms from photomaterials. *The role of 'reference waves' in reproducing of taped Virtual Replica in crystal play basic Virtual pressure waves $(VPW_q^\pm)$ of Bivacuum. The same mechanism may be responsible for the 'ghost' or phantom phenomena.*

Dean and Brame found that healer - treated water demonstrated changes with both IR spectrophotometry (indicating altered hydrogen bonding) and specific peaks with UV spectrophotometry. The half-life for these effects lasted from three days to as long as three years. *For explanation of induction effects see items 6 and 7.*

The long relaxation time (memory of water) is explained by our Hierarchical theory of condensed matter (http://arxiv.org/abs/physics/0207114), as a result of very big resulting activation barrier, separating the induced by BMI shift of dynamic equilibrium between 24 collective excitations in water from the equilibrium one.

It was demonstrated, that mental influence, as a carrier of information, cannot be significantly blocked by any physical screening and that the effect does not depend on the distance. This point to nonelectromagnetic nature of mind-matter interaction. *The nature of Bivacuum mediated interaction (BMI) satisfy this condition.*

The remote influence of psychic on direct magnetic field sensor (ferroprobe magnetometer), screened from the external regular EM influence was revealed by Ageev, Dulnev, Kolmakov, etc. (2003). The detector of alternating magnetic field did not revealed any significant changes, induced by remote 'sender/psychic'. This in accordance with mechanism, described by equilibrium shift between Bivacuum fermions and antifermions with opposite spins, mentioned above: $\mathbf{BVF}^\uparrow \rightleftharpoons \mathbf{BVB}^\pm \rightleftharpoons \mathbf{BVF}^\downarrow$.

The active Psi-channel seems to be non-isotropic, i.e. strictly spatially directed. The same group revealed that good psychic (sender) is able selectively change the electric potential of the electrodes in the aqueous solution of sodium chloride (100 ml of 0.9% NaCl solution) in one of two vessels, separated from each other for 2 meters only. The distance between psychic and vessels was about 500 kilometers.

Well registered phenomena of remote viewing (RV) and precognition by Hal Puthoff and Rassel Targ at Stanford Research Institute (1996) are confirmed in other laboratories. Our theory of BMI explains the RV phenomena as a result of superposition of Virtual replicas multiplication of Sender (psychic) and Receiver (target): $\mathbf{VRM}_S \bowtie \mathbf{VRM}_R$ and feedback reaction between this superimposed interference pattern and the nerve system of the psychic, exciting visual centers of his brain.

Numerous studies have demonstrated (Targ & Katra, 1998) that size of the target (down to 1 mm square) and distance between sender [S] and target (up to 10,000 miles) do not appear to significantly impair signal perception. The electromagnetic shielding by Faraday cage or sea water does not negatively impact remote viewing ability. *These data, pointing to anisotropic component of BMI, are in accordance with proposed mechanism of nonlocal* *interaction via bundles of* $\sum\limits_{}^{N}$ $[\mathbf{VirG}_{SME}(\mathbf{S} \Longleftrightarrow \mathbf{R})]_n$, *connecting coherent elementary particles of Sender and Receiver, as a part of BMI. We can see, that the proposed in this work new resonant Bivacuum mediated interaction (BMI) between psychic and target can explain all above described phenomena. The old, currently existing paradigm, failed to do this.*

### 12.5 The Biological and Biochemical effects of Bivacuum Mediated Interaction between Sender and Receiver

There are two classes of remote healing (RH):

1) when the target (Receiver) is found by the healer (Sender) on the basis of a name, location, birth date, etc. (in remote viewing terms, this is "coordinate"). This process involves superposition of Virtual Replica Multiplicators of $[S]$ and $[R]$: $\mathbf{VRM}_S \bowtie \mathbf{VRM}_R$, necessary for *recognition and targeting and formation of Virtual guides bundles* $\sum\limits_{}^{N} [\mathbf{VirG}_{SME}(\mathbf{S} \Longleftrightarrow \mathbf{R})]_n$, connecting coherent atoms of $[S]$ and $[R]$;

2) when the *adjunct* (an object previously treated by the healer, such as water, cloth, a crystal, etc.) is used by the patient with or without the healer's knowledge. In this case the *imprinting* of VR of healer in *adjunct* object should take a place. The possibility of *imprinting/taping of secondary* VR becomes possible because of primary VR multiplication process. In this case the role of *primary* VR of the of the *healer/sender* takes the *secondary VR*, imprinted in the adjunct matter.

In a 1991 Chien & al. report that they found the following biochemical effects when studying the influence of a qigong master, generating Psi-field, on a culture of human fibroblasts: a 1.8% increase in cell growth rate in 24 hrs; 10-15% increase in DNA synthesis and 3-5% increase in cell protein synthesis in a 2 h period.

When the master emitted "inhibiting" intention and corresponding VRM, the cell growth decreased by 6%, while DNA and protein synthesis decreased by 20-23%, respectively 35-48%.

Intent-modulated emission of *biophotons* from the hands of qigong practitioners is a well-known phenomenon that has often been reported in the scientific literature. Eugene Wallace reported measuring up to 100 time stronger emissions from the hands of gifted persons compared to controls.

A study by Nakamura & al. (2000) reports an increase in subject's hand biophotons intensity associated with a drop in skin surface temperature during Qigong practice. The significantly higher (up to 105nT) magnetic signals during Qi emission from hands of qigong practitioners, as compared to the controls were revealed (Lin & Chen, 2001).

It was proposed by Oschman (2000) that under special conditions, resonant brainwaves may entrain the body's neural system to deliver healing frequencies to diseased tissues, or become coupled to the Schumann resonance and thus transmit distant healing effects to the target. But even accepting the Schumann resonance as a non-dissipative mechanism of information transmission, the electromagnetism based approach still fail to explain such effects, as influence of mental activity on internuclear and gravitational forces, described above. Only certain Bivacuum symmetry and dynamics perturbations, proposed in our theory, may be responsible for like phenomena. The positive role of Schumann resonance in psi phenomena we may explain by its influence on the nerve system of Sender/psychic, activating psi-abilities.

### 12.6 Superconsciousness of the Universe and its Evolution

The stability of each new synaptic distribution in the ensemble of nerve cells of human's brain, participating in elementary act of consciousness, is responsible partly for long-term memory. Each redistribution in neuron's system of brain and central nerve system is accompanied by modulation of *internal (ether) virtual replica* $VR_{in}$. This part of our approach is close to idea of Karl Pribram about holographic nature of memory.

The *ether body* $\mathbf{VR}^{in} + \mathbf{VR}^{sur}$ (see section 8.3) is generated by set of biological structures, including synaptic configuration, responsible for memory on neurodynamics level. After creation of such series of $\mathbf{VR}^{in,sur}$, in form of stable hierarchical superpositions of 3D virtual standing waves, the corresponding primary *ether body (VR)* and its multiplication ($\mathbf{VRM}$) or *holoiteration become independent on material object*. It is true not only for *ether* VR of biological objects humans, animals, trees, etc.), but for any inorganic objects as well (rocks, water, oil, etc.). Such mechanism explains the *ether memory* formation of *phantoms, as a system of secondary VR*.

The primary virtual replica multiplications: distant ($\mathbf{VRM}^{dis}$) and nonlocal ($\mathbf{VRM}^{nl}$) (see section 8) of human's brain and nerve system provide the *astral and mental memories*, interrelated with *ether memory*.

In turn, the combination of the individual *astral and mental memories* can be incorporated into complex Virtual Replica Multiplication ($\mathbf{VRM}$) of the whole Earth/Planet: $\mathbf{P\text{-}VRM}$ .

It means that, the individual human's $\mathbf{VRM(r, t)}$, related to elementary act of memorization and consciousness, can be incorporated in holographic structure of $\mathbf{P\text{-}VRM}$. Such mechanism stands for a imprinting of the individual mental activity to $\mathbf{P\text{-}VRM}$, i.e. the NOOSPHERE formation.

*A kind of filters and thresholds*, preventing 'downloading' to $\mathbf{Earth\ VRM}$ (Noosphere) the information in form of standing $\mathbf{VPW}^{\pm}$, of already existing or 'destructive' info, should exist. The principle of selection of "valuable" new information/perturbation can be based, for example, on criteria of Golden mean or Hidden harmony of primary VRM construction, as a background principle of [Bivacuum +Matter] dynamic and spatial self-organization.

The notion of "Collective Informational Bank" in such approach means formation of **Universe Virtual Replica** ($\mathbf{UVR}$) from huge number of planets ($\mathbf{P\text{-}VRM}$) and stars ($\mathbf{S\text{-}VRM}$) of the Universe. The ability of processing of information may be a consequence of ability of Universe Virtual Replica:

$$\mathbf{UVR} = \sum(\mathbf{P} - \mathbf{VRM}) + \sum(S - \mathbf{VRM}) \qquad 12.10$$

to self-organization, evolution, devolution and formation of metastable states. The feedback reaction between different components of Hierarchical $\mathbf{UVR}$, can be considered, as a Superconsciousness ($\mathbf{SC}$) of the Universe".

Such Superconsciousness ($\mathbf{SC}$) have a properties of Quantum Supercomputer, able to processing of information and organizing it by certain principles. This can be considered as the evolution of Superconsciousness.

The conjecture of Universe with properties of "Even Bigger Computer" with ability to simulate future and memorizing past, was proposed also by Tom Campbell (2003). Earlier similar ideas were discussed by Dean Radin in his book: 'The conscious universe: the scientific truth of psychic phenomena', HarperEdge, 1998, 362 pp, ISBN 0-06-251502-0.

**TABLE 2**

**The Stages of Self-Organization and Evolution of the Universe,
Mediated by Virtual Pressure Waves ($\mathbf{VPW}_q^{\pm}$)**

(I)  **BIVACUUM** $\overset{\text{Bivacuum Symmetry Breach}}{<==============>}$ [**MATTER + FIELDS** + $\sum$ **VRM OF MATTER**] $\rightleftharpoons$

(II)

**BIOSYSTEMS** $\xleftarrow{\text{Consciousness}}$ **[COMPLEX ORGANISMS (CO) + $\sum$ VRM(r, t) of CO]** $\rightleftharpoons$

(III) **[$\sum$ VRM OF MATTER + $\sum$ VRM of CO]** $\xleftarrow{\text{self–organization of Bivacuum + Matter}}$ **(IV)**

(IV) **UNIVERSE VR (UVR)** $\xleftarrow{\text{self–organization of UVR to Quant Supercomp}}$ **SUPERCONSCIOUSNESS (SC)**

---

where: **VRM(r,t)** is virtual replicas multiplication in space and time; **CO** means complex organisms; **UVR** means the Universe Virtual Replica, self-organizing to Superconsciousness (SC) with properties of Quantum Supercomputer.

The important contribution to realization of Superconsciousness, as a self-organizing quantum supercomputer, is related to superposition of nonlocal and temporal components of hierarchical systems of Virtual Replicas Multiplications VRM(r,t) and Virtual Guides (**VirG**$_{SME}$) formation, providing different kind of Bivacuum mediated interaction (BMI).

The Hierarchical system of bundles of virtual microtubules, like the axons in humans body, exchanging the information, momentum and energy between remote complex objects of the Universe, containing coherent regions, i.e. neutron stars, black holes in centers of galactics, etc. is a crucial part of the Universe Quantum Supercomputer or Superconsciousness. The mechanism of possible time effects in huge domains of virtual Bose Condensate (VirBC) in the Universe, formed by the bundles of nonlocal **VirG**$_{SME}$ was described in section 10.1.

The possibility of feedback reaction between *Superconsciusness* and *Planet Noosphere*, following from our concept, means that at proper conditions of Universe Virtual Replica (UVR) high instability (bifurcation point), even small perturbation of planet **P-VRM** may influence Superconsciousness.

### 12.7 The effects of virtual replica of asymmetric constructions, like pyramids, on the matter

It looks, that Virtual Replica in Bivacuum, generated by psychic or by the [Earth-Moon-Sun] dynamic system, can be imitated and modulated by some asymmetric inorganic constructions, like pyramids, rings, etc. In work of Adamenko, Levchook (1994), Narimanov (2001) and Miakin (2002) such effects has been demonstrated on examples of following test-systems, placed inside pyramids: the cultures of microbes (dynamic behavior), water (pH, $O_2$ concentration), polymers solution (optical density), benzene acid (UV absorption).

The Virtual Replicas of the pyramids or cones should be much more asymmetric, than VR generated by cube. The effects of different virtual replicas on test systems, like water and aqueous solutions, generated by such two hollow or filled structures, are anticipated to be different also. This consequence of our model is confirmed experimentally by Narimanov (2001). Keeping a flask with water under the *pyramid* during few days, makes pH of water lower, than in control flask, placed under *cube* in the same room and temperature. The ice, formed from the 'pyramid - treated water' melts about 10% faster, than the control ice. These results point to decreasing of intermolecular interaction in pyramid - treated water.

*The sharpening of the razor blades* after their keeping inside pyramids, revealed experimentally, may be a consequence of increasing probability of virtual charged particles + antiparticles pairs origination in the internal VR of pyramid due to its asymmetry (i.e. Bivacuum polarization). Consequently, the dielectric permittivity ($\varepsilon_0$) of Bivacuum increases. In turn, this induces the decreasing of ion-ion, ion-dipole and dipole-dipole interactions in condensed matter (blade) inside the pyramid. As a result, the small structural irregularities with bigger relative interface, interacting with perturbed Bivacuum, on the top of blade, responsible for its sharpness, became unstable and gradually destroyed under the effect of thermal fluctuations. The blade becomes sharper.

The dependence of internal VR of cavity on its shape, leading from our theory, is confirmed by the different Lamb shifts in atomic spectra of samples in cavities of different shape. It is known, that the Lamb shift is determined by screening of the electrons and nuclears charges by the charged virtual vacuum particles and antiparticles. In our model such a particles/antiparticles may be represented by $\mathbf{BVF}^\uparrow = [\mathbf{V}^+\uparrow\uparrow \ \mathbf{V}^-]$ and $\mathbf{BVF}^\downarrow = [\mathbf{V}^+\downarrow\downarrow \ \mathbf{V}^-]$, acquiring nonzero charge, as a result of their torus - antitorus small asymmetry.

### 12.8 The experimental program for verification of the proposed mechanism of Bivacuum - Mediated Mental Interaction

Such a program should include the following directions:

1. Finding a way for increasing the psychic abilities to make the results of experiments more reliable and reproducible with aid of special devices, like our Audio-Video Skin Transmitter: www.karelia.ru/~alexk (see 'Innovations');

2. Confirmation of correlation of properties of the Virtual Replica of any object, including human's body with dielectric permittivity ($\varepsilon_0$) and magnetic permeability ($\mu_0$) of Bivacuum, using Kirlian effect or its developed computerized version (Korotkov, et all), named Gas Discharge Visualization (GDV);

Our theory explains high sensitivity of Kirlian effect to nerve and physiological state of man by corresponding variation of threshold of air molecules excitation and ionization, following their thermal collisions, accompanied the variation of ($\varepsilon_0$) and Coulomb interaction between polarized air molecules in strong gradients of electric field. The photons radiation, registered by Kirlian effect and GDV is, therefore directly related to density of virtual pressure waves ($VPW^\pm$) and charge symmetry shift of Bivacuum dipoles, induced by Ether and Astral Virtual replicas of the object under study;

2. The monitoring of behavior of microtubules in nerve cells, in cells culture and isolated systems of microtubules 'in vitro', like their cooperative assembly/disassembly, providing interaction between such systems, may confirm the role of systems [centrioles + chromosomes] in telepathic contacts and remote healing;

3. Confirmation of data, pointing to ability of crystals, like quartz, and aqueous systems for imprinting of Virtual replica multiplication of living organisms: $\mathbf{VRM(r,t)}$ or Biofield and elucidation of mechanism of imprinting, using our Comprehensive Analyzer of Matter Properties (CAMP) (Kaivarainen: http://arxiv.org/abs/physics/0207114);

4. Systematic study of Bivacuum perturbation by biofield - $\mathbf{VRM(r,t)}$, using Casimir attraction force between conducting plates and spectral Lamb shift in the same set of experiments, as a criteria of Virtual Replica Multiplication properties;

5. Confirmation of interrelation between the shape of the object and properties of virtual replica, generated by this object. For example, the influence of hollow pyramid and cube on water parameters (pH, conductivity, etc.) in vessel under them, is proved to be different (Narimanov, 2001);

6. Confirmation of consequence of our theory, that any nonequilibrium processes, like melting or boiling, are accompanied by replacement of virtual replica of the previous stable phase of condensed matter by the new one. The monitoring of mechanism of long relaxation time of virtual replica (phantom effect) of the object after its moving away, using Kirlian effect of our Comprehensive Analyzer of Matter Properties (CAMP), with water as a test system.

## Main Conclusions

1. A new Bivacuum model, as the infinite dynamic superfluid matrix of virtual dipoles, named Bivacuum fermions ($\mathbf{BVF}^{\updownarrow}$)$^i$ and Bivacuum bosons ($\mathbf{BVB}^{\pm}$)$^i$, formed by correlated torus ($\mathbf{V}^+$) and antitorus ($\mathbf{V}^-$), as a collective excitations of subquantum particles and antiparticles of opposite energy, charge and magnetic moments and separated by energy gap, is developed. In primordial non polarized Bivacuum, i.e. in the absence of matter and fields, these parameters of torus and antitorus totally compensate each other. Their spatial and energetic properties correspond to three generations of electrons, muons and tauons ($i = e, \mu, \tau$). The positive and negative Virtual Pressure Waves ($\mathbf{VPW}_q^{\pm}$) and Virtual Spin Waves ($\mathbf{VirSW}_q^{S=\pm 1/2}$) are the result of emission and absorption of positive and negative energy Virtual Clouds ($\mathbf{VC}_q^{\pm}$), resulting from transitions of $\mathbf{V}^+$ and $\mathbf{V}^-$ between different states of excitation, symmetrical in realms of positive and negative energy: $j - k = q$;

2. The symmetry shift between $\mathbf{V}^+$ and $\mathbf{V}^-$ actual and complementary mass and charge to the left or right, opposite for Bivacuum fermions $\mathbf{BVF}^{\uparrow}$ and antifermions $\mathbf{BVF}^{\downarrow}$, has the relativistic and reverse to that dependence on these dipoles external rotational-translational velocity. This shift is accompanied by sub-elementary fermion and antifermion formation. The formation of sub-elementary fermions/antifermions and their fusion to stable triplets of elementary fermions, like electrons or protons $\langle[\mathbf{F}_{\uparrow}^- \bowtie \mathbf{F}_{\downarrow}^+] + \mathbf{F}_{\updownarrow}^{\pm}\rangle^{e,p}$, following by the *rest mass and charge* origination, become possible at the certain rotation velocity ($\mathbf{v}$) of Cooper pairs of $[\mathbf{BVF}^{\uparrow} \bowtie \mathbf{BVF}^{\downarrow}]$ around their common axis. It is shown, that this rotational-translational velocity value is determined by Golden Mean condition: $(\mathbf{v/c})^2 = \phi = 0.618$;

3. The fundamental physical roots of Golden Mean condition: $(\mathbf{v/c})^2 = \mathbf{v}_{gr}^{ext}/\mathbf{v}_{ph}^{ext} = \phi$ are revealed, as the equality of internal and external group and phase velocities of torus and antitorus of sub-elementary fermions, correspondingly: $\mathbf{v}_{gr}^{in} = \mathbf{v}_{gr}^{ext}$; $\mathbf{v}_{ph}^{in} = \mathbf{v}_{ph}^{ext}$. These equalities are named 'Hidden Harmony Conditions';

4. The new expressions for total, potential and kinetic energies of de Broglie waves of elementary particles were obtained. The former represents the extended basic Einstein and Dirac formula for free particle:

$$\mathbf{E}_{tot} = \mathbf{m}_V^+ \mathbf{c}^2 = \sqrt{1 - (\mathbf{v/c})^2}\, \mathbf{m}_0 \mathbf{c}^2 + \mathbf{h}^2/\mathbf{m}_V^+ \boldsymbol{\lambda}_B^2$$

The new formulas take into account the contributions of the actual mass/energy of torus ($\mathbf{V}^+$) and those of complementary antitorus ($\mathbf{V}^-$), correspondingly, of asymmetric sub-elementary fermions to the total ones. The shift of symmetry between the inertial and inertialess mass and other parameters of torus and antitorus of sub-elementary fermions are dependent on their *internal* rotational-translational dynamics in composition of triplets and the *external* translational velocity of the whole triplets. The latter determines the external translational momentum and the empirical de Broglie wave frequency: $\mathbf{v}_B = \mathbf{m}_V^+ \mathbf{v}^2/\mathbf{h}$ and length: $\boldsymbol{\lambda}_B = \mathbf{h}/\mathbf{m}_V^+ \mathbf{v}$;

5. A dynamic mechanism of [corpuscle ($\mathbf{C}$) $\rightleftharpoons$ wave ($\mathbf{W}$)] duality is proposed. It involves the modulation of the internal (hidden) quantum beats frequency between the asymmetric 'actual' (torus) and 'complementary' (antitorus) states of sub-elementary fermions or antifermions by the external - empirical de Broglie wave frequency of the whole particles (triplets), equal to beats of similar states of the 'anchor' Bivacuum fermion. In nonrelativistic conditions such modulation stands for the wave packets origination;

6. The photon is a result of fusion (annihilation) of two triplets of particle and antiparticle. It represents a rotating sextet of sub-elementary fermions and antifermions with axial structural symmetry and minimum energy $\mathbf{2m}_0^e \mathbf{c}^2$. The electromagnetic field, is a result of Corpuscle - Wave pulsation of photon, exciting $\mathbf{VPW^+ \bowtie VPW^-}$ and its fast rotation with angle velocity ($\omega_{rot}$), equal to pulsation frequency. The clockwise or anticlockwise direction of photon rotation, as respect to direction of its propagation, corresponds to its spin sign: $s = \pm\hbar$;

9. It is demonstrated, that the dimensionless 'pace of time' ($\mathbf{dt/t} = -\mathbf{dT}_k/\mathbf{T}_k$) and time itself for each closed system are determined by the change of this system kinetic energy. They are positive, if the particles of the system are slowing down under the influence of basic virtual pressure waves $\mathbf{VPW}_{q=1}^{\pm}$ of Bivacuum. The ($\mathbf{dt/t}$) and ($\mathbf{t}$) are negative in the opposite case. This new concept of time extends the Einstein relativistic theory. For example, our formula for time includes not only velocity, but also acceleration of the particles;

10. Theory of Virtual Replica ($\mathbf{VR}$) of material objects in Bivacuum and $\mathbf{VR}$ Multiplication: $\mathbf{VRM}$ ($\mathbf{r,t}$). The $\mathbf{VR}$ represents a three-dimensional (3D) superposition of Bivacuum virtual standing waves $\mathbf{VPW}_m^{\pm}$ and $\mathbf{VirSW}_m^{\pm 1/2}$, modulated by [$\mathbf{C \rightleftharpoons W}$] pulsation of elementary particles and translational and librational de Broglie waves of molecules of macroscopic object (http://arxiv.org/abs/physics/0207027). The infinitive multiplication of primary $\mathbf{VR}$ in space in form of 3D packets of virtual standing waves, representing set of *secondary* $\mathbf{VR}$: $\mathbf{VRM(r)}$, is a result of interference of all pervading external coherent basic *reference waves* - Bivacuum Virtual Pressure Waves ($\mathbf{VPW}_{q=1}^{\pm}$) and Virtual Spin Waves ($\mathbf{VirSW}_{q=1}^{\pm 1/2}$) with similar kinds of modulated standing waves, forming $\mathbf{VR}$. The $\mathbf{VR}$ plays the role of the *object waves*. This phenomena may stand for *remote vision* of psychic. The ability of enough complex system of $\mathbf{VRM(r,t)}$ to self-organization in nonequilibrium conditions, make it possible multiplication of primary $\mathbf{VR}$ not only in space but as well, in time in both time direction - positive (evolution) and negative (devolution). The feedback reaction between most probable/stable $\mathbf{VRM(r,t)}$ and nerve system of psychic, including visual centers of brain, can by responsible for *clairvoyance*. The $\mathbf{VR}$ of elementary particles coincides with notion of their *anchor site*, representing two or three conjugated Cooper pairs [$\mathbf{BVF^{\uparrow} \bowtie BVF^{\downarrow}}$]$_{as}$ of asymmetric Bivacuum fermions. The stochastic jumps of $\mathbf{CVC^{\pm}}$ of [W] phase of particle from one anchor site to another and the ability of interference of single particle with its own *anchor site* explains two slit experiment;

11. The new general presentation of wave function, based on our wave-corpuscle duality model, takes into account not only the external *translational* dynamics of particle, but also the internal *rotational* one, responsible for the rest mass and charge origination;

12. The *eigen wave functions*, as a solutions of Shrödinger equation, describe the linear superposition of multiple *anchor site,* as a possible alternatives for realization of particle's [C] phase;

13. A possible Mechanism of Quantum entanglement between remote coherent elementary particles: electrons and nuclears of atoms of Sender(S) and Receiver(R) via Virtual Guides of spin, momentum and energy ($\mathbf{VirG_{S,M,E}}$) is proposed. The single $\mathbf{VirG_{S,M,E}^{BVB^{\pm}}}$ can be assembled from Bivacuum bosons ($\mathbf{BVB^{\pm}}$)$^i$ by 'head-to-tail' principle. The doubled $\mathbf{VirG_{S,M,E}^{BVF^{\uparrow}\bowtie BVF^{\downarrow}}}$ from the adjacent microtubules, rotating in opposite directions, can be formed by Cooper pairs of Bivacuum fermions [$\mathbf{BVF^{\uparrow}\bowtie BVF^{\downarrow}}$]$^i$, polymerized by the same principle. The spin/information transmission via Virtual Guides is accompanied by reorientation of spins of tori and antitori of Bivacuum dipoles. The momentum and energy transmission from S to R is realized by the instant pulsation of diameter of such virtual microtubule with frequency of beats, equal to difference between frequencies of $\mathbf{C \rightleftharpoons W}$ pulsation of S and R. The length of $\mathbf{VirG_{S,M,E}}$, connecting fluctuating in space particles of (S) and (R), also can correspondingly vary, because of immediate self-assembly/disassembly of $\mathbf{VirG_{S,M,E}}$ from the infinitive source of Bivacuum dipoles.

The Virtual Guides of both kinds represent the quasi 1D virtual Bose condensate with nonlocal properties, similar to that of 'wormholes'. The bundles of $\mathbf{VirG}_{SME}$, connecting coherent atoms of Sender (S) and Receiver (S), as well as nonlocal component of $\mathbf{VRM(r,t)}$, determined by interference pattern of Virtual Spin Waves, are responsible for nonlocal weak interaction;

14. The introduced *Bivacuum Mediated Interaction* (**BMI**) is a new fundamental interaction, resulting from superposition of Virtual replicas of Sender and Receiver, because of $\mathbf{VRM(r,t)}$ mechanism, and connection of their coherent atoms via $\sum \mathbf{VirG}_{SME}(\mathbf{S} <==> \mathbf{R})$ bundles. Just **BMI** is responsible for remote ultraweak nonlocal interaction. The system: [S + R] should be in nonequilibrium state;

15. The mechanism of Remote Genetic Transmutation (RGT), Remote Morphogenesis (RM) and Remote Healing (RH) is proposed. It is based on conjecture, that the system:

[**pair of orthogonal Centrioles + Chromosomes**]

stands for *sending* and *receiving* of specific genetic information via bundles of $\sum \mathbf{VirG}_{SME}^{i}(\mathbf{S} <==> \mathbf{R})$, connecting coherent elementary particles of [S] and [R];

16. Different Psi phenomena, like remote vision, telepathy, telekinesis, clairvoyance, etc. where considered. The telepathic signal transmission from Sender [S] to Receiver [R] may be provided by multiplication of virtual replicas of microtubules $\mathbf{VRM}_{MT}^{S}(\mathbf{r,t})$ and virtual replica of DNA $\mathbf{VRM}_{DNA}^{S}(\mathbf{r,t})$, and their superposition with corresponding $\mathbf{VRM}_{MT,DNA}^{R}(\mathbf{r,t})$ of the Target/Receiver. The modulation of dynamics of [assembly $\rightleftharpoons$ disassembly] of microtubules and corresponding [*gel* $\rightleftharpoons$ *sol*] transitions in the 'tuned' nerve cells ensembles in [Receiver] by directed mental activity of [Sender] can provide *telepathic contact and remote viewing* between [Sender] and [Receiver]. The resonance remote informational/energy exchange between two living organisms or psychics is dependent on 'tuning' of their [Centrioles + Chromosomes] systems in complementary neuron ensembles via $\sum \mathbf{VirG}_{SME}(\mathbf{S} <==> \mathbf{R})$ bundles;

17. The *telekinesis and remote healing*, as example of mind-matter interaction, should be accompanied by strong collective nonequilibrium process (excitation) in the nerve system of Sender. The excessive momentum and kinetic energy are transmitted from Sender to Receiver or 'Target' due to superposition of $\mathbf{VRM(r,t)}_{S} \bowtie \mathbf{VRM(r.t)}_{R}$ and multiple bundles of Virtual Guides, connecting 'tuned' elementary particles (electrons, protons, neutrons) of [S] and [R]:

$$\sum \mathbf{VirG}_{SME}^{e,p}(\mathbf{S} <==> \mathbf{R}) \equiv Psi - channels$$

*We put forward a conjecture, that even teleportation or spatial exchange of macroscopic number of coherent atoms between very remote regions of the Universe is possible via coherent Psi-channels. If this consequence of our theory will be confirmed, we get a new crucial method of the instant inter-stars propulsion.*

**The correctness of our Unified Theory (UT) follows from its ability to explain a lot of unconventional experimental data, like Kozyrev ones, remote genetic transmutation, remote vision, mind-matter interaction, etc. without contradictions with fundamental laws of nature. For details see: http://arxiv.org/abs/physics/0103031.**